# A Tutorial on the Classical Theories of Electromagnetic Scattering and Diffraction

Masud Mansuripur

College of Optical Sciences, The University of Arizona, Tucson

**Abstract**. Starting with Maxwell's equations, we derive the fundamental results of the Huygens-Fresnel-Kirchhoff and Rayleigh-Sommerfeld theories of scalar diffraction and scattering. These results are then extended to cover the case of vector electromagnetic fields. The famous Sommerfeld solution to the problem of diffraction from a perfectly conducting half-plane is elaborated. Far-field scattering of plane waves from obstacles is treated in some detail, and the well-known optical cross-section theorem, which relates the scattering cross-section of an obstacle to its forward scattering amplitude, is derived. Also examined is the case of scattering from mild inhomogeneities within an otherwise homogeneous medium, where, in the first Born approximation, a fairly simple formula is found to relate the far-field scattering amplitude to the host medium's optical properties. The related problem of neutron scattering from ferromagnetic materials is treated in the final section of the paper.

**1. Introduction**. The classical theories of electromagnetic (EM) scattering and diffraction were developed throughout the nineteenth century by the likes of Augustine Jean Fresnel (1788-1827), Gustav Kirchhoff (1824-1887), John William Strutt (Lord Rayleigh, 1842-1919), and Arnold Sommerfeld (1868-1951).[1-3] A thorough appreciation of these theories requires an understanding of the Maxwell-Lorentz electrodynamics[4-11] and a working knowledge of vector calculus, differential equations, Fourier transformation, and complex-plane integration techniques.[12] The relevant physical and mathematical arguments have been covered (to varying degrees of clarity and completeness) in numerous textbooks, monographs, and research papers.[13-23] The goal of this tutorial is to present the core concepts of the classical theories of scattering and diffraction by starting with Maxwell's equations and deriving the fundamental results using mathematical arguments that should be accessible to students of optical sciences as well as practitioners of modern optical engineering and photonics technologies. A consistent notation and uniform terminology is used throughout the paper. To maintain the focus on the main results and reduce the potential for distraction, some of the longer derivations and secondary arguments have been relegated to the appendices.

The organization of the paper is as follows. After a brief review of Maxwell's equations in Sec.2, we provide a detailed analysis of an all-important Green function in Sec.3. The Huygens-Fresnel-Kirchhoff scalar theory of diffraction is the subject of Sec.4, followed by the Rayleigh-Sommerfeld modification and enhancement of that theory in Sec.5. These scalar theories are subsequently generalized in Sec.6 to arrive at a number of formulas for vector scattering and vector diffraction of EM waves under various settings and circumstances. Section 6 also contains a few examples that demonstrate the application of vector diffraction formulas in situations of practical interest. The famous Sommerfeld solution to the problem of diffraction from a perfectly electrically conducting half-plane is presented in some detail in Sec.7.

Applying the vector formulas of Sec.6 to far-field scattering, we show in Sec.8 how the forward scattering amplitude for a plane-wave that illuminates an arbitrary object relates to the scattering cross-section of that object. This important result in the classical theory of scattering is formally known as the optical cross-section theorem (or the optical theorem).[1,9,24]

An alternative approach to the problem of EM scattering when the host medium contains a region of weak inhomogeneities is described in Sec.9. Here, we use Maxwell's macroscopic equations in conjunction with the Green function of Sec.3 to derive a fairly simple formula for the far-field scattering amplitude in the first Born approximation. The related problem of slow neutron scattering from ferromagnetic media is treated in Sec.10. The paper closes with a few conclusions and final remarks in Sec.11.

**2. Maxwell's equations**. The standard equations of the classical Maxwell-Lorenz theory of electrodynamics relate four material sources to four EM fields in the Minkowski spacetime $(\boldsymbol{r}, t)$.[4-11] The sources are the free charge density $\rho_{\text{free}}$, free current density $\boldsymbol{J}_{\text{free}}$, polarization $\boldsymbol{P}$, and magnetization $\boldsymbol{M}$, while the fields are the electric field $\boldsymbol{E}$, magnetic field $\boldsymbol{H}$, displacement $\boldsymbol{D}$, and magnetic induction $\boldsymbol{B}$. In the $SI$ system of units, where the free space (or vacuum) has permittivity $\varepsilon_0$ and permeability $\mu_0$, the displacement is defined as $\boldsymbol{D} = \varepsilon_0\boldsymbol{E} + \boldsymbol{P}$, and the magnetic induction as $\boldsymbol{B} = \mu_0\boldsymbol{H} + \boldsymbol{M}$. Let the total charge-density $\rho(\boldsymbol{r}, t)$ and total current-density $\boldsymbol{J}(\boldsymbol{r}, t)$ be defined as

$$\rho(\boldsymbol{r}, t) = \rho_{\text{free}}(\boldsymbol{r}, t) - \boldsymbol{\nabla} \cdot \boldsymbol{P}(\boldsymbol{r}, t). \tag{1}$$

$$\boldsymbol{J}(\boldsymbol{r}, t) = \boldsymbol{J}_{\text{free}}(\boldsymbol{r}, t) + \partial_t\boldsymbol{P}(\boldsymbol{r}, t) + \mu_0^{-1}\boldsymbol{\nabla} \times \boldsymbol{M}(\boldsymbol{r}, t). \tag{2}$$

The charge-current continuity equation, $\boldsymbol{\nabla} \cdot \boldsymbol{J} + \partial_t\rho = 0$, is generally satisfied by the above densities, irrespective of whether their corresponding sources are free (i.e., $\rho_{\text{free}}$ and $\boldsymbol{J}_{\text{free}}$), or bound electric charges within electric dipoles (i.e., $-\boldsymbol{\nabla} \cdot \boldsymbol{P}$ and $\partial_t\boldsymbol{P}$), or bound electric currents within magnetic dipoles (i.e., $\mu_0^{-1}\boldsymbol{\nabla} \times \boldsymbol{M}$). Invoking Eqs.(1) and (2), Maxwell's macroscopic equations are written as follows:

$$\varepsilon_0\boldsymbol{\nabla} \cdot \boldsymbol{E}(\boldsymbol{r}, t) = \rho(\boldsymbol{r}, t). \tag{3}$$

$$\boldsymbol{\nabla} \times \boldsymbol{B}(\boldsymbol{r}, t) = \mu_0\boldsymbol{J}(\boldsymbol{r}, t) + \mu_0\varepsilon_0\partial_t\boldsymbol{E}(\boldsymbol{r}, t). \tag{4}$$

$$\boldsymbol{\nabla} \times \boldsymbol{E}(\boldsymbol{r}, t) = -\partial_t\boldsymbol{B}(\boldsymbol{r}, t). \tag{5}$$

$$\boldsymbol{\nabla} \cdot \boldsymbol{B}(\boldsymbol{r}, t) = 0. \tag{6}$$

Taking the curl of both sides of the third equation, using the vector identity $\boldsymbol{\nabla} \times \boldsymbol{\nabla} \times \boldsymbol{V} = \boldsymbol{\nabla}(\boldsymbol{\nabla} \cdot \boldsymbol{V}) - \boldsymbol{\nabla}^2\boldsymbol{V}$, and substituting from the first and second equations, we find

$$\boldsymbol{\nabla} \times \boldsymbol{\nabla} \times \boldsymbol{E} = -\mu_0\partial_t\boldsymbol{J} - \mu_0\varepsilon_0\partial_t^2\boldsymbol{E} \quad \rightarrow \quad (\boldsymbol{\nabla}^2 - c^{-2}\partial_t^2)\boldsymbol{E} = \mu_0\partial_t\boldsymbol{J} + \varepsilon_0^{-1}\boldsymbol{\nabla}\rho. \tag{7}$$

Here, $c = (\mu_0\varepsilon_0)^{-½}$ is the speed of light in vacuum. Similarly, taking the curl of both sides of Eq.(4) and substituting from Eqs.(5) and (6), we find

$$\boldsymbol{\nabla} \times \boldsymbol{\nabla} \times \boldsymbol{B} = \mu_0\boldsymbol{\nabla} \times \boldsymbol{J} - \mu_0\varepsilon_0\partial_t^2\boldsymbol{B} \quad \rightarrow \quad (\boldsymbol{\nabla}^2 - c^{-2}\partial_t^2)\boldsymbol{B} = -\mu_0\boldsymbol{\nabla} \times \boldsymbol{J}. \tag{8}$$

The scalar and vector potentials, $\psi(\boldsymbol{r}, t)$ and $\boldsymbol{A}(\boldsymbol{r}, t)$, are defined such that $\boldsymbol{B} = \boldsymbol{\nabla} \times \boldsymbol{A}$ and $\boldsymbol{E} = -\boldsymbol{\nabla}\psi - \partial_t\boldsymbol{A}$. With these definitions, Maxwell's third and fourth equations are automatically satisfied. In the Lorenz gauge, where $\boldsymbol{\nabla} \cdot \boldsymbol{A} + c^{-2}\partial_t\psi = 0$, Maxwell's second equation yields

$$\boldsymbol{\nabla} \times \boldsymbol{\nabla} \times \boldsymbol{A} = \mu_0\boldsymbol{J} - \mu_0\varepsilon_0\boldsymbol{\nabla}(\partial_t\psi) - \mu_0\varepsilon_0\partial_t^2\boldsymbol{A} \quad \rightarrow \quad (\boldsymbol{\nabla}^2 - c^{-2}\partial_t^2)\boldsymbol{A} = -\mu_0\boldsymbol{J}. \tag{9}$$

Similarly, substitution into Maxwell's first equation for the $E$-field in terms of the potentials yields

$$-\varepsilon_0\boldsymbol{\nabla} \cdot (\boldsymbol{\nabla}\psi + \partial_t\boldsymbol{A}) = \rho \quad \rightarrow \quad (\nabla^2 - c^{-2}\partial_t^2)\psi = -\rho/\varepsilon_0. \tag{10}$$

Since monochromatic fields oscillate at a single frequency $\omega$, their time-dependence factor is generally written as $\exp(-\mathrm{i}\omega t)$. Consequently, the spatiotemporal dependence of all the fields and all the sources can be separated into a space part and a time part. For example, the $E$-field may now be written as $\boldsymbol{E}(\boldsymbol{r})e^{-\mathrm{i}\omega t}$, the total electric charge-density as $\rho(\boldsymbol{r})e^{-\mathrm{i}\omega t}$, and so on. Defining the free-space wavenumber $k_0 = \omega/c$, the Helmholtz equations (7)-(10) now become

$$(\boldsymbol{\nabla}^2 + k_0^2)\boldsymbol{E}(\boldsymbol{r}) = -\mathrm{i}\omega\mu_0 \boldsymbol{J}(\boldsymbol{r}) + \varepsilon_0^{-1}\boldsymbol{\nabla}\rho(\boldsymbol{r}). \tag{11}$$

$$(\boldsymbol{\nabla}^2 + k_0^2)\boldsymbol{B}(\boldsymbol{r}) = -\mu_0\boldsymbol{\nabla}\times\boldsymbol{J}(\boldsymbol{r}). \tag{12}$$

$$(\boldsymbol{\nabla}^2 + k_0^2)\boldsymbol{A}(\boldsymbol{r}) = -\mu_0 \boldsymbol{J}(\boldsymbol{r}). \tag{13}$$

$$(\nabla^2 + k_0^2)\psi(\boldsymbol{r}) = -\rho(\boldsymbol{r})/\varepsilon_0. \tag{14}$$

In regions of free space, where $\rho(\boldsymbol{r}) = 0$ and $\boldsymbol{J}(\boldsymbol{r}) = 0$, the right-hand sides of Eqs.(11)-(14) vanish, thus allowing one to replace $k_0^2\boldsymbol{E}(\boldsymbol{r})$ with $-\boldsymbol{\nabla}^2\boldsymbol{E}(\boldsymbol{r})$, and similarly for $\boldsymbol{B}(\boldsymbol{r})$, $\boldsymbol{A}(\boldsymbol{r})$, and $\psi(\boldsymbol{r})$, whenever the need arises. These substitutions will be used in the following sections.

**3. The Green function**. In the spherical coordinate system $(r,\theta,\varphi)$, the Laplacian of the spherically symmetric function $G(\boldsymbol{r}) = e^{\mathrm{i}k_0 r}/r$ equals $\partial^2(rG)/r\partial r^2 = -k_0^2 G(\boldsymbol{r})$ everywhere except at the origin $r = 0$, where the function has a singularity. Thus, $(\nabla^2 + k_0^2)G(\boldsymbol{r}) = 0$ at all points $\boldsymbol{r}$ except at the origin. A good way to handle the singularity at $r = 0$ is to treat $G(\boldsymbol{r})$ as the limiting form of another function that has no such singularity, namely,

$$G(\boldsymbol{r}) = lim_{\varepsilon\to 0}\big(e^{\mathrm{i}k_0 r}/\sqrt{r^2+\varepsilon}\big). \tag{15}$$

The Laplacian of our well-behaved, non-singular function is readily found to be

$$r^{-2}\partial_r\big[r^2\partial_r\big(e^{\mathrm{i}k_0 r}/\sqrt{r^2+\varepsilon}\,\big)\big] = -\left[\frac{3\varepsilon}{(r^2+\varepsilon)^{5/2}} - \frac{2\mathrm{i}k_0\varepsilon}{r(r^2+\varepsilon)^{3/2}} + \frac{k_0^2}{(r^2+\varepsilon)^{1/2}}\right]e^{\mathrm{i}k_0 r}. \tag{16}$$

The first two functions appearing inside the square brackets on the right-hand side of Eq.(16) are confined to the vicinity of the origin at $r = 0$; they are tall, narrow, symmetric, and have the following volume integrals (see Appendix A for details):

$$\int_0^\infty 4\pi r^2(r^2+\varepsilon)^{-5/2}\mathrm{d}r = 4\pi/3\varepsilon. \tag{17}$$

$$\int_0^\infty 4\pi r(r^2+\varepsilon)^{-3/2}\mathrm{d}r = 4\pi/\sqrt{\varepsilon}\cdot \tag{18}$$

Thus, in the limit of sufficiently small $\varepsilon$, the first two functions appearing on the right-hand side of Eq.(16) can be represented by $\delta$-functions,[†] and the entire equation may be written as

$$(\nabla^2 + k_0^2)(e^{\mathrm{i}k_0 r}/\sqrt{r^2+\varepsilon}) = -4\pi(1 - \mathrm{i}2k_0\sqrt{\varepsilon})\delta(\boldsymbol{r}). \tag{19}$$

This is the sense in which we can now state that $(\nabla^2 + k_0^2)G(\boldsymbol{r}) = -4\pi\delta(\boldsymbol{r})$ in the limit when $\varepsilon \to 0$. Shifting the center of the function to an arbitrary point $\boldsymbol{r}_0$, we will have

$$(\nabla^2 + k_0^2)G(\boldsymbol{r},\boldsymbol{r}_0) = -4\pi\delta(\boldsymbol{r}-\boldsymbol{r}_0). \tag{20}$$

The gradient of the Green function $G(\boldsymbol{r},\boldsymbol{r}_0)$, which plays an important role in our discussions of the following sections, is now found to be

$$\boldsymbol{\nabla}G(\boldsymbol{r},\boldsymbol{r}_0) = \boldsymbol{\nabla}\left(\frac{e^{\mathrm{i}k_0|\boldsymbol{r}-\boldsymbol{r}_0|}}{|\boldsymbol{r}-\boldsymbol{r}_0|}\right) = (\mathrm{i}k_0 - |\boldsymbol{r}-\boldsymbol{r}_0|^{-1})\left(\frac{e^{\mathrm{i}k_0|\boldsymbol{r}-\boldsymbol{r}_0|}}{|\boldsymbol{r}-\boldsymbol{r}_0|}\right)\frac{\boldsymbol{r}-\boldsymbol{r}_0}{|\boldsymbol{r}-\boldsymbol{r}_0|}\cdot \tag{21}$$

Appendix B provides an analysis of Eq.(20), an inhomogeneous Helmholtz equation, via Fourier transformation.

---

[†] When multiplied by $\varepsilon$, as required by Eq.(16), the integrand in Eq.(17) peaks at $\sim\varepsilon^{-1/2}$ at $r = \sqrt{2/3\,\varepsilon}$, then drops steadily to $\sim\sqrt[4]{\varepsilon}$ at $r = \sqrt[4]{\varepsilon}$. Similarly, the integrand in Eq.(18), again multiplied by $\varepsilon$, peaks at $\sim 1$ at $r = \sqrt{1/2\,\varepsilon}$, then drops steadily to $\sim\sqrt{\varepsilon}$ at $r = \sqrt[4]{\varepsilon}$.

**4. The Huygens-Fresnel-Kirchhoff theory of diffraction.**[1-3,9] Consider a scalar function $\psi(\boldsymbol{r})$ that satisfies the homogeneous Helmholtz equation $(\nabla^2 + k_0^2)\psi(\boldsymbol{r}) = 0$ everywhere within a volume $V$ of free space enclosed by a surface $S$. Two examples of the geometry under consideration are depicted in Fig.1.[9] In general, $\psi(\boldsymbol{r})$ can represent the scalar potential or any Cartesian component of the monochromatic $E$-field, $B$-field, or $A$-field associated with an EM wave propagating in free space with frequency $\omega$ and wave-number $k_0 = \omega/c$. In Fig.1(a), the EM wave arrives at the surface $S_1$ from sources located on the left-hand side of $S_1$, and enters the volume $V$ contained within the closed surface $S = S_1 + S_2$. In Fig.1(b), the EM waves emanate from inside the closed surface $S_1$ and permeate the volume $V$ enclosed by $S_1$ on one side and by a second closed surface $S_2$ that defines the outer boundary of $V$. In both figures, the point $\boldsymbol{r}_0$, where the field is being observed, is located inside the volume $V$, and the surface normals $\hat{\boldsymbol{n}}$ everywhere on the closed surface $S$ are unit-vectors that point inward (i.e., into the volume $V$).

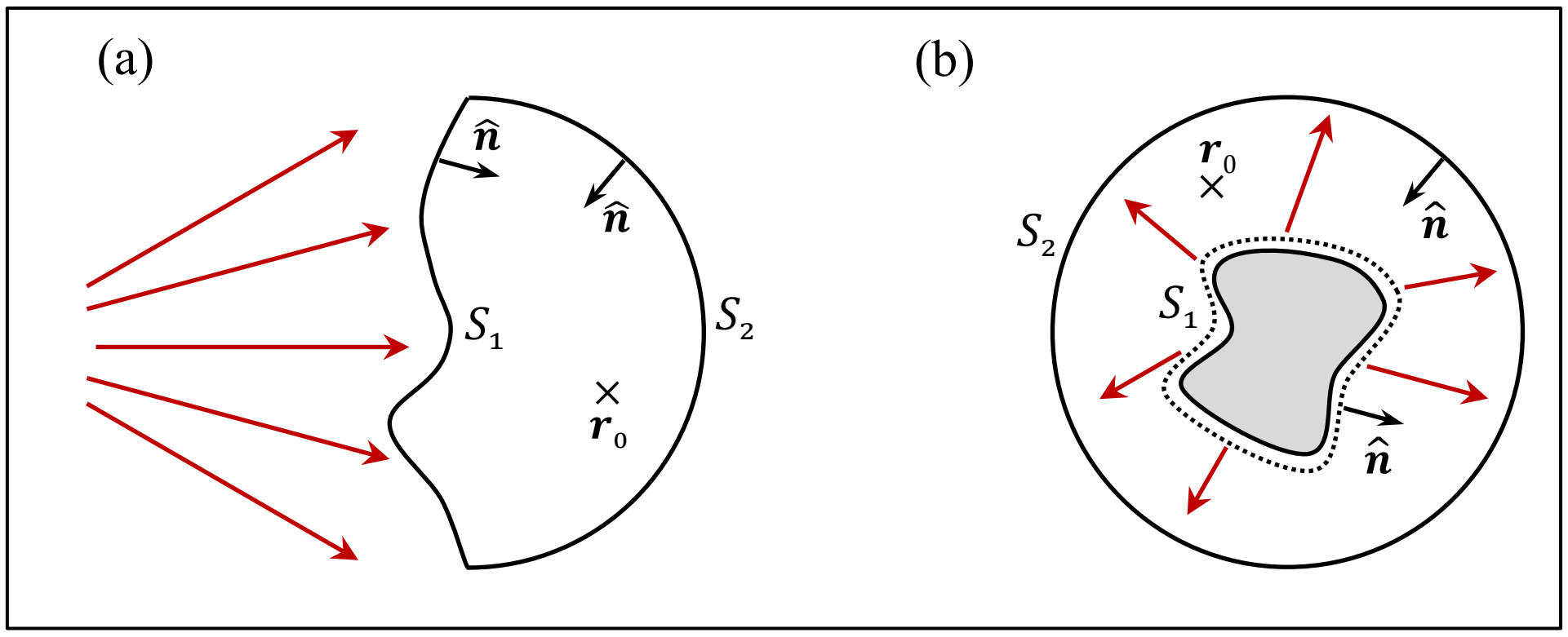

**Fig.1**. Two surfaces $S_1$ and $S_2$ bound the scattering region, which is assumed to be free of sources and material bodies. All the radiation within the scattering region comes from the outside. In (a) the sources of radiation are on the left-hand side of $S_1$, while in (b) the radiation emanates from the sources inside the closed surface $S_1$. The observation point $\boldsymbol{r}_0$ is an arbitrary point within the scattering region. All the points located on $S_2$ are assumed to be far away from $\boldsymbol{r}_0$, so that the fields that reach $S_2$ do not contribute to the fields observed at $\boldsymbol{r}_0$. At each and every point on $S_1$ and $S_2$, the surface normals $\hat{\boldsymbol{n}}$ point *into* the scattering region.

The essence of the theory developed by G. Kirchhoff in 1882 (building upon the original ideas of Huygens and Fresnel)[1] is an exact mathematical relation between the observed field $\psi(\boldsymbol{r}_0)$ and the field $\psi(\boldsymbol{r})$ that exists everywhere on the closed surface $S$. This relation is derived below using the sifting property of Dirac's $\delta$-function, the relation between the $\delta$-function and the Green function given in Eq.(20), some well-known identities in standard vector calculus, and Gauss's famous theorem of vector calculus, according to which $\int_V \boldsymbol{\nabla}\cdot\boldsymbol{V}\mathrm{d}v = \oint_S \boldsymbol{V}\cdot\mathrm{d}\boldsymbol{s}$. We have

$$\psi(\boldsymbol{r}_0) = \int_V \psi(\boldsymbol{r})\delta(\boldsymbol{r}-\boldsymbol{r}_0)\mathrm{d}\boldsymbol{r} = -(4\pi)^{-1}\int_V \psi(\boldsymbol{r})(\nabla^2 + k_0^2)G(\boldsymbol{r},\boldsymbol{r}_0)\mathrm{d}\boldsymbol{r} \quad \leftarrow \boxed{\text{See Eq.(20)}}$$

$$= -(4\pi)^{-1}\int_V [\psi(\boldsymbol{r})\boldsymbol{\nabla}\cdot\boldsymbol{\nabla}G(\boldsymbol{r},\boldsymbol{r}_0) - G(\boldsymbol{r},\boldsymbol{r}_0)\boldsymbol{\nabla}\cdot\boldsymbol{\nabla}\psi(\boldsymbol{r})]\mathrm{d}\boldsymbol{r} \quad \leftarrow \boxed{\text{Replace } k_0^2\psi \text{ with } -\nabla^2\psi.}$$

$$= -(4\pi)^{-1}\int_V [\boldsymbol{\nabla}\cdot(\psi\boldsymbol{\nabla}G) - \cancel{\boldsymbol{\nabla}G\cdot\boldsymbol{\nabla}\psi} - \boldsymbol{\nabla}\cdot(G\boldsymbol{\nabla}\psi) + \cancel{\boldsymbol{\nabla}\psi\cdot\boldsymbol{\nabla}G}]\mathrm{d}\boldsymbol{r} \quad \leftarrow \boxed{\boldsymbol{\nabla}\cdot(\phi\boldsymbol{V}) = \boldsymbol{V}\cdot\boldsymbol{\nabla}\phi + \phi\boldsymbol{\nabla}\cdot\boldsymbol{V}}$$

$$= (4\pi)^{-1}\int_S [\psi(\hat{\boldsymbol{n}}\cdot\boldsymbol{\nabla}G) - G(\hat{\boldsymbol{n}}\cdot\boldsymbol{\nabla}\psi)]\mathrm{d}s \quad \leftarrow \boxed{\text{Gauss' theorem; } \hat{\boldsymbol{n}} \text{ points into the volume } V.}$$

$$= (4\pi)^{-1}\int_S [\psi(\boldsymbol{r})\partial_n G(\boldsymbol{r},\boldsymbol{r}_0) - G(\boldsymbol{r},\boldsymbol{r}_0)\partial_n\psi(\boldsymbol{r})]\mathrm{d}s. \qquad (22)$$

**Example**. In the extreme situation where $S$ is a small sphere of radius $\varepsilon$ centered at $\boldsymbol{r}_0$, we will have $\psi(\boldsymbol{r}) \cong \psi(\boldsymbol{r}_0)$, $G(\boldsymbol{r},\boldsymbol{r}_0) = e^{\mathrm{i}k_0\varepsilon}/\varepsilon$, and $\partial_n G(\boldsymbol{r},\boldsymbol{r}_0) = -(\mathrm{i}k_0 - \varepsilon^{-1})e^{\mathrm{i}k_0\varepsilon}/\varepsilon$. Given that the surface area of the sphere is $4\pi\varepsilon^2$, the second term in Eq.(22) makes a negligible contribution to the overall integral when $\varepsilon \to 0$. The first term, however, contains $e^{\mathrm{i}k_0\varepsilon}/\varepsilon^2$, which integrates to $4\pi$ in the limit of $\varepsilon \to 0$, yielding $\psi(\boldsymbol{r}_0)$ as the final result.

---

Taking the spherical (or hemi-spherical) surface $S_2$ in Fig.1 to be far away from the region of interest, the field $\psi(\boldsymbol{r})$ everywhere on $S_2$ should have the general form of $f(\theta,\varphi)e^{\mathrm{i}k_0 r}/r$ and, consequently, $\partial_n\psi(\boldsymbol{r}) \sim (\mathrm{i}k_0 - r^{-1})\psi(\boldsymbol{r})$. Similarly, across the surface $S_2$, the Green function has the asymptotic form $G(\boldsymbol{r},\boldsymbol{r}_0) \sim e^{\mathrm{i}k_0 r}/r$ and, therefore, $\partial_n G(\boldsymbol{r},\boldsymbol{r}_0) \sim (\mathrm{i}k_0 - r^{-1})G(\boldsymbol{r},\boldsymbol{r}_0)$. Thus, on the faraway surface $S_2$, the integrand in Eq.(22) must decline faster than $1/r^2$, which means that the contribution of $S_2$ to $\psi(\boldsymbol{r}_0)$ as given by Eq.(22) should be negligible. Kirchhoff's diffraction integral now relates the field at the observation point $\boldsymbol{r}_0$ to the field distribution on $S_1$, as follows:

$$\psi(\boldsymbol{r}_0) = (4\pi)^{-1}\int_{S_1}[\psi(\boldsymbol{r})\partial_n G(\boldsymbol{r},\boldsymbol{r}_0) - G(\boldsymbol{r},\boldsymbol{r}_0)\partial_n\psi(\boldsymbol{r})]\mathrm{d}s. \tag{23}$$

To make contact with the Huygens-Fresnel theory of diffraction, Kirchhoff suggested that both $\psi(\boldsymbol{r})$ and $\partial_n\psi(\boldsymbol{r})$ vanish on the opaque areas of the screen $S_1$, whereas in the open (or transparent, or unobstructed) regions, they retain the profiles they would have had in the absence of the screen. These suggestions, while reasonable from a practical standpoint and resulting in good agreement with experimental observations under many circumstances, are subject to criticism for their mathematical inconsistency, as will be elaborated in the next section.

**5. The Rayleigh-Sommerfeld theory**. In the important special case where the surface $S_1$ coincides with the $xy$-plane at $z = 0$, one can adjust the Green function in such a way as to eliminate either the first or the second term in the integrand of Eq.(23). These situations would then correspond, respectively, to the so-called Neumann and Dirichlet boundary conditions.[1,9] Let the observation point, which we assume to reside on the right-hand side of the planar surface $S_1$, be denoted by $\boldsymbol{r}_0^+ = (x_0, y_0, z_0)$, while its mirror image in $S_1$ (located on the left-hand side of $S_1$) is denoted by $\boldsymbol{r}_0^- = (x_0, y_0, -z_0)$. If we use $G(\boldsymbol{r},\boldsymbol{r}_0^+)$ in Eq.(23), we obtain $\psi(\boldsymbol{r}_0)$ on the left-hand side of the equation, but if we use $G(\boldsymbol{r},\boldsymbol{r}_0^-)$ instead, the integral will yield zero — simply because the peak of the corresponding $\delta$-function now resides outside the integration volume. Thus, we can replace $G(\boldsymbol{r},\boldsymbol{r}_0)$ in Eq.(23) with either of the two functions $G(\boldsymbol{r},\boldsymbol{r}_0^+) \pm G(\boldsymbol{r},\boldsymbol{r}_0^-)$. On the $S_1$ plane, where $z = 0$, we have $G(\boldsymbol{r},\boldsymbol{r}_0^-) = G(\boldsymbol{r},\boldsymbol{r}_0^+)$ and $\partial_n G(\boldsymbol{r},\boldsymbol{r}_0^-) = -\partial_n G(\boldsymbol{r},\boldsymbol{r}_0^+)$. The resulting diffraction integrals, respectively satisfying the Neumann and Dirichlet boundary conditions, will then be

$$\psi(\boldsymbol{r}_0) = -\frac{1}{2\pi}\int_{S_1}\frac{\exp(\mathrm{i}k_0|\boldsymbol{r}-\boldsymbol{r}_0|)}{|\boldsymbol{r}-\boldsymbol{r}_0|}\partial_n\psi(\boldsymbol{r})\mathrm{d}s. \tag{24}$$

$$\psi(\boldsymbol{r}_0) = \frac{1}{2\pi}\int_{S_1}\frac{(\mathrm{i}k_0 - |\boldsymbol{r}-\boldsymbol{r}_0|^{-1})\exp(\mathrm{i}k_0|\boldsymbol{r}-\boldsymbol{r}_0|)}{|\boldsymbol{r}-\boldsymbol{r}_0|^2}[(\boldsymbol{r}-\boldsymbol{r}_0)\cdot\hat{\boldsymbol{n}}]\,\psi(\boldsymbol{r})\mathrm{d}s. \tag{25}$$

Note that, while Eq.(23) applies to any arbitrary surface $S_1$, the Rayleigh-Sommerfeld equations (24) and (25) are restricted to distributions that are specified on a flat plane. Given the scalar field profile $\psi(\boldsymbol{r})$ and/or its gradient on a flat plane, all three equations are exact consequences of Maxwell's equations. To apply these equations in practice, one must resort to some form of approximation to estimate the field distribution on $S_1$. The conventional approximation is that, in the opaque regions of the screen $S_1$, either $\psi(\boldsymbol{r})$ or $\partial_n\psi(\boldsymbol{r})$ or both are vanishingly small and, therefore, negligible, whereas in the transparent (or unobstructed) regions of $S_1$, the field $\psi(\boldsymbol{r})$ and/or its gradient $\partial_n\psi(\boldsymbol{r})$ (along the surface-normal) retain the profile they would have had in the absence of the screen. In this way, one can proceed to evaluate the integral on the open (or transparent, or unobstructed) apertures of $S_1$ in order to arrive at a reasonable estimate of $\psi(\boldsymbol{r}_0)$ at the desired observation location.

As a formula for computing diffraction patterns from one or more apertures in an otherwise opaque screen, the problem with Eq.(23) is that, when combined with Kirchhoff's assumption that *both* $\psi$ and $\partial_n\psi$ vanish on the opaque regions of the physical screen at $S_1$, it becomes mathematically inconsistent. This is because an analytic function such as $\psi(\boldsymbol{r})$ vanishes everywhere if both $\psi$ and $\partial_n\psi$ happen to be zero on any patch of the surface $S_1$. In contrast, Eq.(24), when applied to a physical screen, requires only the assumption that $\partial_n\psi$ be zero on the opaque regions of the screen. While still an approximation, this is a much more mathematically palatable condition than the Kirchhoff requirement.[9] Similarly, Eq.(25) requires only the approximation that $\psi$ be zero on the opaque regions. Thus, on the grounds of mathematical consistency, there is a preference for either Eq.(24) or Eq.(25) over Eq.(23). However, given the aforementioned approximate nature of the values chosen for $\psi$ and $\partial_n\psi$ across the screen at $S_1$, it turns out that scalar diffraction calculations based on these three formulas yield nearly identical results, rendering them equally useful in practical applications.[1,9]

An auxiliary consequence of Eqs.(24) and (25) is that, upon subtracting one from the other, the integral of $\partial_n(G\psi)$ over the entire flat plane $S_1$ is found to vanish; that is,

$$\int_{S_1}\partial_n[G(\boldsymbol{r},\boldsymbol{r}_0)\psi(\boldsymbol{r})]\mathrm{d}s = 0. \tag{26}$$

We will have occasion to use this important identity in the following section.

**6. Vector diffraction**. Applying the Kirchhoff formula in Eq.(22), where the integral is over the closed surface $S = S_1 + S_2$, and the function $\psi(\boldsymbol{r})$ is any scalar field that satisfies the Helmholtz equation, to a Cartesian component of the $E$-field, say, $E_x$, we write

$$\begin{aligned}4\pi E_x(\boldsymbol{r}_0) &= \int_S[E_x(\boldsymbol{r})\partial_n G(\boldsymbol{r},\boldsymbol{r}_0) - G(\boldsymbol{r},\boldsymbol{r}_0)\partial_n E_x(\boldsymbol{r})]\mathrm{d}s \qquad \boxed{\boldsymbol{\nabla}(\phi\psi)=\phi\boldsymbol{\nabla}\psi+\psi\boldsymbol{\nabla}\phi}\\ &= \int_S[(\hat{\boldsymbol{n}}\cdot\boldsymbol{\nabla}G)E_x - (\hat{\boldsymbol{n}}\cdot\boldsymbol{\nabla}E_x)G]\mathrm{d}s = \int_S[2(\hat{\boldsymbol{n}}\cdot\boldsymbol{\nabla}G)E_x - \hat{\boldsymbol{n}}\cdot\boldsymbol{\nabla}(GE_x)]\mathrm{d}s\\ &= 2\int_S(\hat{\boldsymbol{n}}\cdot\boldsymbol{\nabla}G)E_x\mathrm{d}s + \int_V\boldsymbol{\nabla}\cdot\boldsymbol{\nabla}(GE_x)\mathrm{d}\boldsymbol{r}. \leftarrow \boxed{\text{Gauss' theorem; } \hat{\boldsymbol{n}} \text{ points into the volume } V}\end{aligned} \tag{27}$$

Given that Eq.(27) is similarly satisfied by the remaining components $E_y, E_z$ of the $E$-field, the vectorial version of Kirchhoff's formula may be written down straightforwardly. Algebraic manipulations (using standard vector calculus identities described in Appendix C) simplify the final result, yielding the following expression for the $E$-field at the observation point:[9]

$$\boldsymbol{E}(\boldsymbol{r}_0) = (4\pi)^{-1}\int_S[(\hat{\boldsymbol{n}}\times\boldsymbol{E})\times\boldsymbol{\nabla}G + (\hat{\boldsymbol{n}}\cdot\boldsymbol{E})\boldsymbol{\nabla}G + \mathrm{i}\omega(\hat{\boldsymbol{n}}\times\boldsymbol{B})G]\mathrm{d}s. \tag{28}$$

Once again, it is easy to show that the contribution of the spherical (or hemi-spherical) surface $S_2$ to the overall integral in Eq.(28) is negligible. This is because, in the far field, $\hat{\boldsymbol{n}}\cdot\boldsymbol{E}\to 0$ and $\hat{\boldsymbol{n}}\times\boldsymbol{B}\to\boldsymbol{E}/c$, while $|\boldsymbol{E}|\sim e^{\mathrm{i}k_0 r}/r$, $G\sim e^{\mathrm{i}k_0 r}/r$ and $\boldsymbol{\nabla}G\sim -(\mathrm{i}k_0 - r^{-1})e^{\mathrm{i}k_0 r}\hat{\boldsymbol{n}}/r$. Consequently,

$$\boldsymbol{E}(\boldsymbol{r}_0) = (4\pi)^{-1}\int_{S_1}[(\hat{\boldsymbol{n}}\times\boldsymbol{E})\times\boldsymbol{\nabla}G + (\hat{\boldsymbol{n}}\cdot\boldsymbol{E})\boldsymbol{\nabla}G + \mathrm{i}\omega(\hat{\boldsymbol{n}}\times\boldsymbol{B})G]\mathrm{d}s. \tag{29}$$

A similar argument can be used to arrive at the vector Kirchhoff formula for the $B$-field, namely,

$$\boldsymbol{B}(\boldsymbol{r}_0) = (4\pi)^{-1} \int_{S_1} [(\hat{\boldsymbol{n}} \times \boldsymbol{B}) \times \boldsymbol{\nabla} G + (\hat{\boldsymbol{n}} \cdot \boldsymbol{B}) \boldsymbol{\nabla} G - \mathrm{i}(\omega/c^2)(\hat{\boldsymbol{n}} \times \boldsymbol{E}) G] \mathrm{d}s. \tag{30}$$

Needless to say, Eq.(30) could also be derived directly from Eq.(29), or vice versa, although the algebra becomes tedious at times; see Appendix D for one such derivation.

We mention in passing that the arguments that led to Eqs.(29) and (30) could *not* be repeated for the vector potential $\boldsymbol{A}(\boldsymbol{r})$, since in arriving at Eq.(28), in the step where $\boldsymbol{\nabla} \cdot \boldsymbol{E}$ or $\boldsymbol{\nabla} \cdot \boldsymbol{B}$ are set to zero (see Appendix C), we now have $\boldsymbol{\nabla} \cdot \boldsymbol{A} = \mathrm{i}\omega\psi/c^2$.[‡]

In parallel with the arguments advanced previously in conjunction with the Rayleigh-Sommerfeld formulation for scalar fields, one may also modify Eqs.(29) and (30) by setting $G(\boldsymbol{r}, \boldsymbol{r}_0) = 0$ and $\boldsymbol{\nabla} G = 2\partial_n G(\boldsymbol{r}, \boldsymbol{r}_0)\hat{\boldsymbol{n}}$, provided that $S_1$ is a planar surface. This is equivalent to applying Eq.(25) directly to the $x, y, z$ components of the $E$-field (or the $B$-field). It is *not* permissible, however, to retain $G$ and remove $\boldsymbol{\nabla} G$ (again, in the case of a planar $S_1$), because the gradient of $G(\boldsymbol{r}, \boldsymbol{r}_0^+) + G(\boldsymbol{r}, \boldsymbol{r}_0^-)$ has a nonzero projection onto the $xy$-plane.

In those special (yet important) cases where $S_1$ coincides with the $xy$-plane at $z = 0$, one could begin by applying either Eq.(24) or Eq.(25) to the $x, y, z$ components of the field under consideration. Manipulating the resulting equation with the aid of vector-algebraic identities in conjunction with the fact that $\boldsymbol{\nabla} G = -\boldsymbol{\nabla}_0 G$, leads to vector diffraction formulas that could be useful under special circumstances. For instance, the vectorial equivalent of Eq.(25) yields

$$\begin{aligned}
2\pi \boldsymbol{E}(\boldsymbol{r}_0) &= \int_{S_1} (\hat{\boldsymbol{n}} \cdot \boldsymbol{\nabla} G) \boldsymbol{E}(\boldsymbol{r}) \mathrm{d}s = \int_{S_1} [(\hat{\boldsymbol{n}} \times \boldsymbol{E}) \times \boldsymbol{\nabla} G + (\boldsymbol{E} \cdot \boldsymbol{\nabla} G)\hat{\boldsymbol{n}}] \mathrm{d}s \quad \leftarrow \boxed{(\boldsymbol{a} \times \boldsymbol{b}) \times \boldsymbol{c} = (\boldsymbol{a} \cdot \boldsymbol{c})\boldsymbol{b} - (\boldsymbol{b} \cdot \boldsymbol{c})\boldsymbol{a}} \\
&= \int_{S_1} \boldsymbol{\nabla}_0 G \times (\hat{\boldsymbol{n}} \times \boldsymbol{E}) \mathrm{d}s + \{\textstyle\int_{S_1} [\boldsymbol{\nabla} \cdot (G\boldsymbol{E}) - \cancelto{0}{G\boldsymbol{\nabla} \cdot \boldsymbol{E}}] \mathrm{d}s\} \hat{\boldsymbol{n}} \quad \leftarrow \boxed{\boldsymbol{\nabla} \cdot (\psi \boldsymbol{V}) = \psi \boldsymbol{\nabla} \cdot \boldsymbol{V} + \boldsymbol{V} \cdot \boldsymbol{\nabla}\psi} \\
&= \int_{S_1} \boldsymbol{\nabla}_0 \times [(\hat{\boldsymbol{n}} \times \boldsymbol{E}) G] \mathrm{d}s + [\textstyle\int_{S_1} \boldsymbol{\nabla} \cdot (G\boldsymbol{E}) \mathrm{d}s] \hat{\boldsymbol{n}} \quad \leftarrow \boxed{\boldsymbol{\nabla} \times (\psi \boldsymbol{V}) = \psi \boldsymbol{\nabla} \times \boldsymbol{V} + \boldsymbol{\nabla}\psi \times \boldsymbol{V}} \\
&= \boldsymbol{\nabla}_0 \times \int_{S_1} [\hat{\boldsymbol{n}} \times \boldsymbol{E}(\boldsymbol{r})] G(\boldsymbol{r}, \boldsymbol{r}_0) \mathrm{d}s + \left[\iint_{-\infty}^{\infty} [\cancelto{0}{\partial_x(GE_x)} + \cancelto{0}{\partial_y(GE_y)} + \cancelto{0}{\partial_z(GE_z)}] \mathrm{d}x\mathrm{d}y\right] \hat{\boldsymbol{n}}.
\end{aligned} \tag{31}$$

It is easy to see that the first two terms of the second integral on the right-hand side of Eq.(31) vanish since both $GE_x$ and $GE_y$ go to zero in the far out regions of the $xy$-plane. The vanishing of the third term, however, requires invoking Eq.(26) as applied to the $z$-component of the $E$-field. Note, as a matter of consistency, that on the left-hand side of Eq.(31), $\boldsymbol{\nabla}_0 \cdot \boldsymbol{E}(\boldsymbol{r}_0) = 0$ and that, on the right-hand side, the divergence of the curl is always zero.

A similar argument can be advanced for the $B$-field and, therefore, the following vector diffraction equations are generally valid for a planar surface $S_1$:

$$\boldsymbol{E}(\boldsymbol{r}_0) = (2\pi)^{-1} \boldsymbol{\nabla}_0 \times \int_{S_1} [\hat{\boldsymbol{n}} \times \boldsymbol{E}(\boldsymbol{r})] G(\boldsymbol{r}, \boldsymbol{r}_0) \mathrm{d}s. \tag{32}$$

$$\boldsymbol{B}(\boldsymbol{r}_0) = (2\pi)^{-1} \boldsymbol{\nabla}_0 \times \int_{S_1} [\hat{\boldsymbol{n}} \times \boldsymbol{B}(\boldsymbol{r})] G(\boldsymbol{r}, \boldsymbol{r}_0) \mathrm{d}s. \tag{33}$$

---

[‡] In setting $\boldsymbol{\nabla} \cdot \boldsymbol{E} = 0$, Maxwell's first equation, $\varepsilon_0 \boldsymbol{\nabla} \cdot \boldsymbol{E} = \rho_{\text{total}} = \rho_{\text{free}} + \rho_{\text{bound}}$, has been invoked, with the caveat that the surface $S$ is slightly detached from material bodies where electric charges of one kind or another may reside. No such caveat is needed, however, when setting $\boldsymbol{\nabla} \cdot \boldsymbol{B} = 0$, which is simply Maxwell's fourth equation. The situation is quite different with the vector potential $\boldsymbol{A}$, since setting $\boldsymbol{\nabla} \cdot \boldsymbol{A} = 0$ implies working in the Coulomb gauge. While the standard relations $\boldsymbol{B} = \boldsymbol{\nabla} \times \boldsymbol{A}$ and $\boldsymbol{E} = -\boldsymbol{\nabla}\psi - \partial_t \boldsymbol{A}$ remain valid in all gauges, the equations that relate $\psi$ and $\boldsymbol{A}$ to the charge and current densities are gauge dependent. In particular, $\boldsymbol{A}$ in the Coulomb gauge depends not only on the current-density $\boldsymbol{J}_{\text{total}}$, but also on the charge-density $\rho_{\text{total}}$. While the charge-current continuity equation $\boldsymbol{\nabla} \cdot \boldsymbol{J} + \partial_t \rho = 0$ can be used to arrive at a Helmholtz equation for $\boldsymbol{A}(\boldsymbol{r}, t)$ in the Coulomb gauge, the term appearing on the right-hand side of the equation will be the transverse current-density, which does not necessarily vanish in the free-space regions of the system under consideration.

In similar fashion, the vectorial equivalent of Eq.(24) yields

$$2\pi\boldsymbol{E}(\boldsymbol{r}_0) = -\int_{S_1} G\partial_z\boldsymbol{E}\mathrm{d}s = -\int_{S_1} G[\partial_z E_x\hat{\boldsymbol{x}} + \partial_z E_y\hat{\boldsymbol{y}} - (\partial_x E_x + \partial_y E_y)\hat{\boldsymbol{z}}]\mathrm{d}s$$

$$\boxed{\boldsymbol{\nabla}\cdot\boldsymbol{E} = 0 \;\rightarrow\; \partial_z E_z = -(\partial_x E_x + \partial_y E_y)}$$

$$= \int_{S_1}[E_x(\partial_z G\hat{\boldsymbol{x}} - \partial_x G\hat{\boldsymbol{z}}) + E_y(\partial_z G\hat{\boldsymbol{y}} - \partial_y G\hat{\boldsymbol{z}})$$

$$-\partial_z(GE_x)\hat{\boldsymbol{x}} - \partial_z(GE_y)\hat{\boldsymbol{y}} + \partial_x(GE_x)\hat{\boldsymbol{z}} + \partial_y(GE_y)\hat{\boldsymbol{z}}]\mathrm{d}s$$

$$= \int_{S_1}(\hat{\boldsymbol{n}}\times\boldsymbol{E})\times\boldsymbol{\nabla}G\mathrm{d}s - \iint_{-\infty}^{\infty}[\cancelto{0}{\partial_z(GE_x)}\hat{\boldsymbol{x}} + \cancelto{0}{\partial_z(GE_y)}\hat{\boldsymbol{y}} - \cancelto{0}{\partial_x(GE_x)}\hat{\boldsymbol{z}} - \cancelto{0}{\partial_y(GE_y)}\hat{\boldsymbol{z}}]\mathrm{d}x\mathrm{d}y$$

$$= \int_{S_1}\boldsymbol{\nabla}_0 G\times(\hat{\boldsymbol{n}}\times\boldsymbol{E})\mathrm{d}s = \boldsymbol{\nabla}_0\times\int_{S_1}[\hat{\boldsymbol{n}}\times\boldsymbol{E}(\boldsymbol{r})]G(\boldsymbol{r},\boldsymbol{r}_0)\mathrm{d}s. \tag{34}$$

On the penultimate line of Eq.(34), the first two terms in the integral are seen to vanish when Eq.(26) is applied to the $x$ and $y$ components of the $E$-field; the 3rd and 4th terms go to zero due to the vanishing of $GE_x$ and $GE_y$ in the far out regions of the $xy$-plane. Equation (34) thus yields the same expression for $\boldsymbol{E}(\boldsymbol{r}_0)$ as the one reached via Eq.(31).

**Example 1**. A plane-wave $\boldsymbol{E}(\boldsymbol{r},t) = (E_{x0}\hat{\boldsymbol{x}} + E_{y0}\hat{\boldsymbol{y}} + E_{z0}\hat{\boldsymbol{z}})e^{\mathrm{i}(k_x x+k_y y+k_z z-\omega t)}$ arriving from the region $z<0$ is reflected from a perfectly conducting plane mirror located in the $xy$-plane at $z=0$. The exact cancellation of the tangential components of the $E$-field at the mirror surface means that the scattered $E$-field in the $xy$-plane at $z=0^+$ is given by[§]

$$\boldsymbol{E}_s(x,y) = -(E_{x0}\hat{\boldsymbol{x}} + E_{y0}\hat{\boldsymbol{y}} + E_{z0}\hat{\boldsymbol{z}})e^{\mathrm{i}(k_x x+k_y y)}. \tag{35}$$

Beyond the mirror in the region $z>0$, the scattered $E$-field is obtained from Eq.(32), as follows:

$$2\pi\boldsymbol{E}_s(\boldsymbol{r}_0) = \boldsymbol{\nabla}_0\times\iint_{-\infty}^{\infty}\hat{\boldsymbol{n}}\times(-E_{x0}\hat{\boldsymbol{x}} - E_{y0}\hat{\boldsymbol{y}} - E_{z0}\hat{\boldsymbol{z}})e^{\mathrm{i}(k_x x+k_y y)}\frac{\exp\{\mathrm{i}k_0[(x-x_0)^2+(y-y_0)^2+z_0^2]^{½}\}}{[(x-x_0)^2+(y-y_0)^2+z_0^2]^{½}}\mathrm{d}x\mathrm{d}y$$

$$= \boldsymbol{\nabla}_0\times(E_{y0}\hat{\boldsymbol{x}} - E_{x0}\hat{\boldsymbol{y}})e^{\mathrm{i}(k_x x_0+k_y y_0)}\iint_{-\infty}^{\infty}e^{\mathrm{i}(k_x x+k_y y)}\frac{\exp[\mathrm{i}k_0(x^2+y^2+z_0^2)^{½}]}{(x^2+y^2+z_0^2)^{½}}\mathrm{d}x\mathrm{d}y. \tag{36}$$

The 2D Fourier transform appearing in Eq.(36) is readily found to be (see Appendix E):

$$\iint_{-\infty}^{\infty}\frac{\exp\left(\mathrm{i}k_0\sqrt{x^2+y^2+z^2}\right)}{\sqrt{x^2+y^2+z^2}}e^{\mathrm{i}(k_x x+k_y y)}\mathrm{d}x\mathrm{d}y = \mathrm{i}(2\pi/k_z)e^{\mathrm{i}k_z z}. \tag{37}$$

This is a valid equation whether the incident beam is of the propagating type (i.e., homogeneous plane-wave, with real-valued $k_z$), or of the evanescent type (i.e., inhomogeneous, with imaginary $k_z$). Substitution into Eq.(36) now yields

$$\boldsymbol{E}_s(\boldsymbol{r}_0) = \boldsymbol{\nabla}_0\times\mathrm{i}(E_{y0}\hat{\boldsymbol{x}} - E_{x0}\hat{\boldsymbol{y}})e^{\mathrm{i}(k_x x_0+k_y y_0+k_z z_0)}/k_z$$

$$= [-E_{x0}\hat{\boldsymbol{x}} - E_{y0}\hat{\boldsymbol{y}} + (k_x E_{x0} + k_{y0} E_y)\hat{\boldsymbol{z}}/k_z]e^{\mathrm{i}(k_x x_0+k_y y_0+k_z z_0)}$$

$$= -(E_{x0}\hat{\boldsymbol{x}} + E_{y0}\hat{\boldsymbol{y}} + E_{z0}\hat{\boldsymbol{z}})e^{\mathrm{i}(k_x x_0+k_y y_0+k_z z_0)}. \tag{38}$$

---

[§] Whereas on opposite facets of the mirror the scattered $E_x$ (as well as the scattered $E_y$) are identical, the presence of surface charges requires that the sign of the scattered $E_z$ flip between $z=0^-$ and $z=0^+$. The scattered $E$-field amplitude is thus $(-E_{x0}, -E_{y0}, E_{z0})$ at $z=0^-$ and $(-E_{x0}, -E_{y0}, -E_{z0})$ at $z=0^+$. Similarly, the existence of surface currents requires the scattered $B$-field amplitude to be $(B_{x0}, B_{y0}, -B_{z0})$ at $z=0^-$ and $(-B_{x0}, -B_{y0}, -B_{z0})$ at $z=0^+$.

As expected, in the half-space $z > 0$, the scattered field precisely cancels out the incident field. This result is quite general and applies to any profile for the incident beam, not just plane-waves, for the simple reason that any incident beam can be expressed as a superposition of plane-waves. It should also be clear that, in the absence of the perfectly conducting reflector in the $xy$-plane, the distribution of the tangential $E$-field (or the tangential $B$-field) throughout the $xy$-plane at $z = 0$ can be used to reconstruct the entire distribution in the $z > 0$ half-space via either Eq.(32) or Eq.(33).

**Example 2**. Consider a screen in the $xy$-plane at $z = 0$ consisting of obstructing segment(s) in the form of thin sheet(s) of perfect conductors and open regions otherwise. A monochromatic incident beam creates surface charges and surface currents on the metallic segment(s) of this planar screen. The scattered fields produced by the induced surface charges and currents can be described in terms of the scattered scalar and vector potentials $\psi_s(\boldsymbol{r},t)$ and $\boldsymbol{A}_s(\boldsymbol{r},t)$. Given that the induced surface current has no component along the $z$-axis, the vector potential will likewise have only $x$ and $y$ components. Application of Eq.(24) to $A_{sx}(\boldsymbol{r})$ and $A_{sy}(\boldsymbol{r})$ yields

$$\boldsymbol{A}_s(\boldsymbol{r}_0) = -\frac{1}{2\pi}\int_{S_1} \frac{\exp(\mathrm{i}k_0|\boldsymbol{r}-\boldsymbol{r}_0|)}{|\boldsymbol{r}-\boldsymbol{r}_0|}\big(\partial_z A_{sx}\hat{\boldsymbol{x}} + \partial_z A_{sy}\hat{\boldsymbol{y}}\big)\mathrm{d}s. \tag{39}$$

Now, $\boldsymbol{B}_s(\boldsymbol{r}) = \boldsymbol{\nabla}\times\boldsymbol{A}_s(\boldsymbol{r}) = -\partial_z A_{sy}\hat{\boldsymbol{x}} + \partial_z A_{sx}\hat{\boldsymbol{y}} + (\partial_x A_{sy} - \partial_y A_{sx})\hat{\boldsymbol{z}}$. Consequently,

$$\boldsymbol{A}_s(\boldsymbol{r}_0) = -\frac{1}{2\pi}\int_{S_1} \frac{\exp(\mathrm{i}k_0|\boldsymbol{r}-\boldsymbol{r}_0|)}{|\boldsymbol{r}-\boldsymbol{r}_0|}(B_{sy}\hat{\boldsymbol{x}} - B_{sx}\hat{\boldsymbol{y}})\mathrm{d}s. \tag{40}$$

The symmetry of the scattered field ensures that $B_{sx}$ and $B_{sy}$ are zero in the open areas of the screen, so the integral in Eq.(40) need be evaluated only on the metallic surfaces of the screen. We will have

$$\boldsymbol{B}_s(\boldsymbol{r}_0) = \boldsymbol{\nabla}_0\times\boldsymbol{A}_s(\boldsymbol{r}_0) = (2\pi)^{-1}\boldsymbol{\nabla}_0\times\int_{\text{metal}}[\hat{\boldsymbol{n}}\times\boldsymbol{B}_s(\boldsymbol{r})]\frac{\exp(\mathrm{i}k_0|\boldsymbol{r}-\boldsymbol{r}_0|)}{|\boldsymbol{r}-\boldsymbol{r}_0|}\mathrm{d}s. \tag{41}$$

In situations where a thin, flat metallic object acts as a scatterer, Eq.(41) provides a simple way to compute the scattered field provided, of course, that an estimate of the magnetic field at the metal surface (or, equivalently, an estimate of the induced surface current-density) can be obtained. Needless to say, considering that, in the absence of the scatterer, the continued propagation of the incident beam into the $z > 0$ half-space can be reconstructed from the tangential component of the incident $B$-field in the $z = 0$ plane (see Example 1), one may add the incident $B$-field to the scattered field of Eq.(41) and obtain, unsurprisingly, the general vector diffraction formula of Eq.(33).

**Example 3**. Figure 2 shows a perfectly conducting thin sheet residing in the upper half of the $xy$-plane at $z = 0$. The incident beam is a monochromatic plane-wave of frequency $\omega$, wavelength $\lambda_0 = 2\pi c/\omega$, linear-polarization aligned with $\hat{\boldsymbol{y}}$, and propagation direction along the unit-vector $\boldsymbol{\sigma}_{\text{inc}} = (\sin\theta_{\text{inc}})\hat{\boldsymbol{x}} + (\cos\theta_{\text{inc}})\hat{\boldsymbol{z}}$, as follows:

$$\boldsymbol{E}(\boldsymbol{r},t) = E_{\text{inc}}\hat{\boldsymbol{y}}\exp[\mathrm{i}k_0(\sin\theta_{\text{inc}}\,x + \cos\theta_{\text{inc}}\,z - ct)]. \tag{42}$$

Using Eq.(32), we derive the diffracted $E$-field at the observation point $\boldsymbol{r}_0 = x_0\hat{\boldsymbol{x}} + z_0\hat{\boldsymbol{z}}$. (The symmetry of the problem ensures that the field profile is independent of the $y$-coordinate of the observation point.) In what follows, we use the $0^{\text{th}}$ and $1^{\text{st}}$ order Hankel functions of type 1, $H_0^{(1)}(\zeta)$ and $H_1^{(1)}(\zeta)$, as well as the (complex) Fresnel integral $F(\zeta) = \int_\zeta^\infty \exp(\mathrm{i}x^2)\,\mathrm{d}x$.[26,27] For a graphical representation of the Fresnel integral via the so-called Cornu spiral,[28] see Appendix F. The following identities will be needed further below:

$$\int_{-\infty}^{\infty} \frac{\exp(\mathrm{i}\sqrt{x^2+\zeta^2})}{\sqrt{x^2+\zeta^2}}\,\mathrm{d}x = \mathrm{i}\pi H_0^{(1)}(\zeta). \qquad \leftarrow \text{G\&R}^{27}\ \mathbf{8.421}\text{-}11 \qquad (43)$$

$$\frac{\mathrm{d}}{\mathrm{d}\zeta} H_0^{(1)}(\zeta) = -H_1^{(1)}(\zeta). \qquad \leftarrow \text{G\&R}^{27}\ \mathbf{8.473}\text{-}6 \qquad (44)$$

$$H_1^{(1)}(\zeta) \sim \sqrt{\frac{2}{\pi\zeta}}\, e^{\mathrm{i}(\zeta - \frac{3}{4}\pi)}, \qquad (\zeta \gg 1). \qquad \leftarrow \text{G\&R}^{27}\ \mathbf{8.451}\text{-}3 \qquad (45)$$

$$\begin{aligned} F(\zeta) &= \int_0^\infty \exp(\mathrm{i}x^2)\,\mathrm{d}x - \left[\int_0^\zeta \cos(x^2)\,\mathrm{d}x + \mathrm{i}\int_0^\zeta \sin(x^2)\,\mathrm{d}x\right] \\ &= \sqrt{\pi/4}\, e^{\mathrm{i}\pi/4} - \sqrt{\pi/2}\,[C(\zeta) + \mathrm{i}S(\zeta)]. \end{aligned} \qquad \leftarrow \text{G\&R}^{27}\ \mathbf{8.250}\text{-}2{,}3 \qquad (46)$$

The $E$-field at the observation point $\boldsymbol{r}_0$ is found to be

$$\begin{aligned} \boldsymbol{E}(\boldsymbol{r}_0) &\cong (2\pi)^{-1}\boldsymbol{\nabla}_0 \times \int_{x=-\infty}^{0}\int_{y=-\infty}^{\infty} (\hat{\boldsymbol{z}} \times E_{\text{inc}}\hat{\boldsymbol{y}}) e^{\mathrm{i}k_0 \sin\theta_{\text{inc}} x} \times \frac{\exp\{\mathrm{i}k_0[(x-x_0)^2+y^2+z_0^2]^{½}\}}{[(x-x_0)^2+y^2+z_0^2]^{½}}\,\mathrm{d}y\mathrm{d}x \\ &= -½\mathrm{i}\boldsymbol{\nabla}_0 \times E_{\text{inc}}\,\hat{\boldsymbol{x}} \int_{-\infty}^{0} e^{\mathrm{i}k_0 \sin\theta_{\text{inc}} x} H_0^{(1)}\left[k_0\sqrt{(x-x_0)^2+z_0^2}\right]\mathrm{d}x \\ &= ½\mathrm{i}E_{\text{inc}}\,\hat{\boldsymbol{y}} \int_{-\infty}^{0} \frac{k_0 z_0}{[(x-x_0)^2+z_0^2]^{½}} e^{\mathrm{i}k_0 \sin\theta_{\text{inc}} x} H_1^{(1)}\left[k_0\sqrt{(x-x_0)^2+z_0^2}\right]\mathrm{d}x \\ &\cong \frac{\mathrm{i}e^{-\mathrm{i}3\pi/4}E_{\text{inc}}\,\hat{\boldsymbol{y}}}{\sqrt{2\pi}} \int_{-\infty}^{0} \frac{\sqrt{k_0}z_0}{[(x-x_0)^2+z_0^2]^{3/4}} \exp(\mathrm{i}k_0 \sin\theta_{\text{inc}}\, x) \exp\left[\mathrm{i}k_0\sqrt{(x-x_0)^2+z_0^2}\right]\mathrm{d}x. \end{aligned} \qquad (47)$$

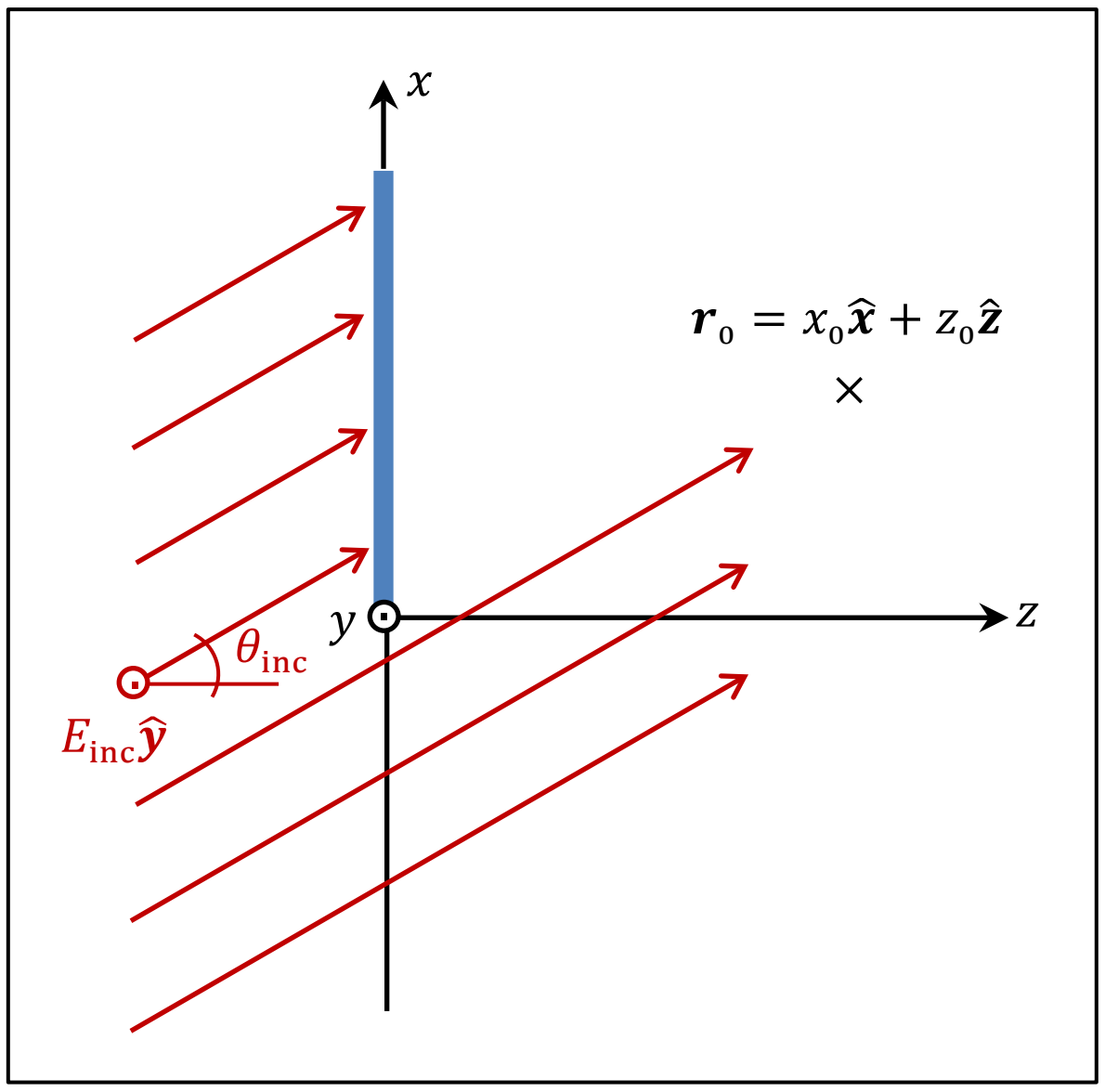

**Fig.2**. A perfectly conducting thin sheet sits in the upper half of the $xy$-plane at $z = 0$. The incident plane-wave, which is linearly polarized along the $y$-axis, has amplitude $E_{\text{inc}}$, frequency $\omega$, wave-number $k_0 = \omega/c$, and propagation direction $\boldsymbol{\sigma}_{\text{inc}} = \sin\theta_{\text{inc}}\,\hat{\boldsymbol{x}} + \cos\theta_{\text{inc}}\,\hat{\boldsymbol{z}}$. The observation point $\boldsymbol{r}_0$ is in the $xz$-plane.

Here, we have used the large-argument approximate form of $H_1^{(1)}(\zeta)$ given by Eq.(45). Assuming that $z_0 \gg |x - x_0|$, we proceed by invoking the following approximation:

$$\exp\left[\mathrm{i}k_0\sqrt{(x-x_0)^2+z_0^2}\right]\big/[(x-x_0)^2+z_0^2]^{3/4} \cong z_0^{-3/2}e^{\mathrm{i}k_0[z_0+(x-x_0)^2/2z_0]}. \qquad (48)$$

Note that this approximation is not accurate when $|x - x_0|$ acquires large values; however, the rapid phase variations of the integrand in Eq.(47) ensure that the contributions to the integral at points $x$ that are far from $x_0$ are insignificant. Substitution from Eq.(48) into Eq.(47) yields

$$\begin{aligned}
\boldsymbol{E}(\boldsymbol{r}_0) &\cong \frac{\sqrt{k_0}\,E_{\text{inc}}\hat{\boldsymbol{y}}}{\sqrt{2\pi z_0}}e^{\mathrm{i}(k_0z_0-\pi/4)}\int_{-\infty}^{0}\exp(\mathrm{i}k_0\sin\theta_{\text{inc}}\,x)\exp[\mathrm{i}k_0(x-x_0)^2/(2z_0)]\,\mathrm{d}x\\
&= \frac{\sqrt{k_0}\,E_{\text{inc}}\hat{\boldsymbol{y}}}{\sqrt{2\pi z_0}}e^{\mathrm{i}(k_0z_0-\pi/4)}\exp[\mathrm{i}k_0(x_0\sin\theta_{\text{inc}}-\tfrac{1}{2}z_0\sin^2\theta_{\text{inc}})]\\
&\qquad\times\int_{-\infty}^{0}\exp[\mathrm{i}k_0(x-x_0+z_0\sin\theta_{\text{inc}})^2/(2z_0)]\,\mathrm{d}x\\
&= \frac{E_{\text{inc}}\hat{\boldsymbol{y}}}{\sqrt{\pi}e^{\mathrm{i}\pi/4}}e^{\mathrm{i}k_0(x_0\sin\theta_{\text{inc}}+z_0-\frac{1}{2}z_0\sin^2\theta_{\text{inc}})}\int_{\sqrt{k_0/2z_0}(x_0-z_0\sin\theta_{\text{inc}})}^{\infty}\exp(\mathrm{i}x^2)\,\mathrm{d}x\\
&= \frac{F\left[\sqrt{\pi/(\lambda_0z_0)}\,(x_0-z_0\sin\theta_{\text{inc}})\right]}{\sqrt{\pi}\exp(\mathrm{i}\pi/4)}E_{\text{inc}}\,\hat{\boldsymbol{y}}\,e^{\mathrm{i}k_0[x_0\sin\theta_{\text{inc}}+z_0(1-\frac{1}{2}\sin^2\theta_{\text{inc}})]}. \qquad (49)
\end{aligned}$$

At the edge of the geometric shadow, i.e., the straight line where $x_0 = z_0\sin\theta_{\text{inc}}$, we have $F(0) = \tfrac{1}{2}\sqrt{\pi}e^{\mathrm{i}\pi/4}$. On this line, the $E$-field amplitude is $\tfrac{1}{2}E_{\text{inc}}$. Above the shadow's edge, the field amplitude steadily declines, whereas below the edge, there occur a large number of oscillations before the field settles into what is essentially the incident plane-wave.[1] Figure 3 shows a typical plot of the far-field intensity (i.e., the square of the $E$-field amplitude) versus the distance (along the $x$-axis at a fixed value of $z_0$) from the edge of the geometric shadow.

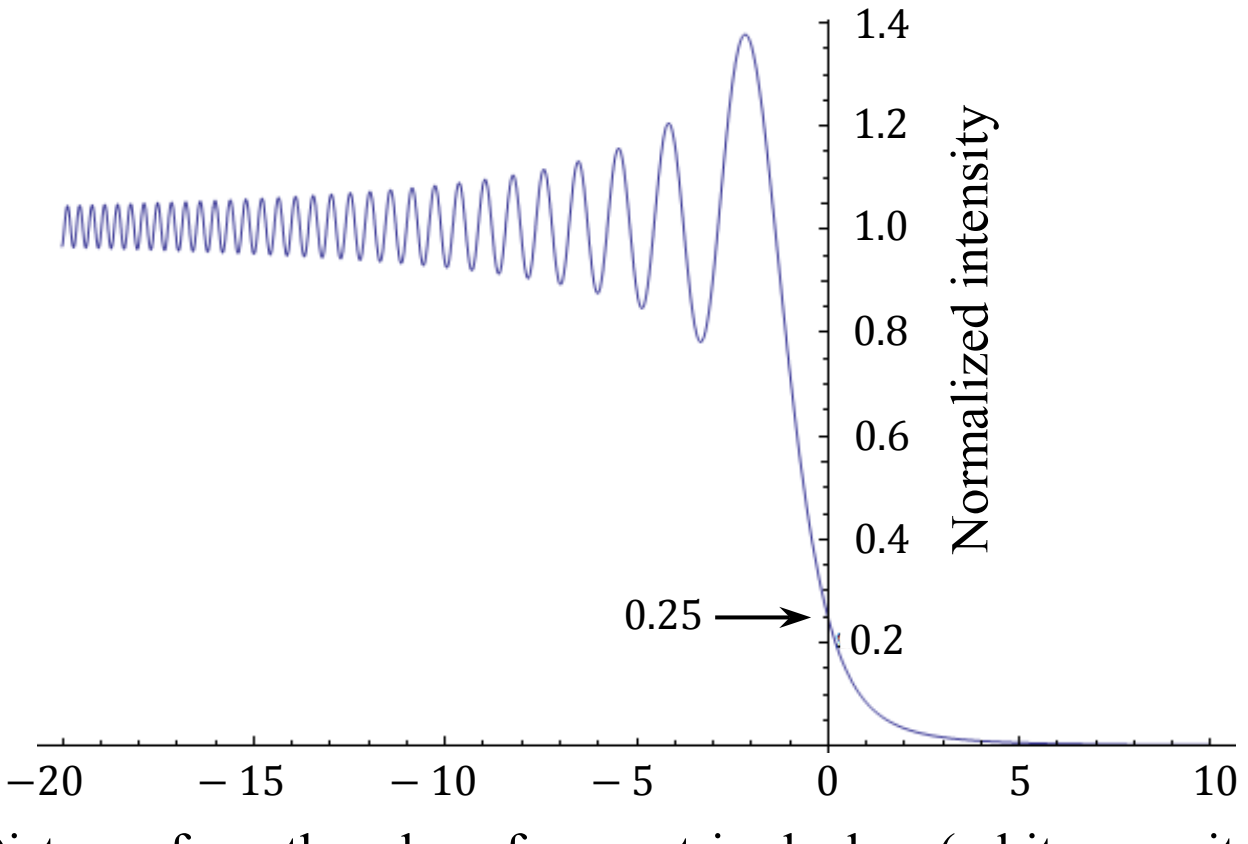

**Fig.3**. Normalized intensity in the far field of a sharp edge as a function of distance along the $x$-axis (at fixed $z_0$) from the edge of the geometric shadow, where $x_0 = z_0\sin\theta_{\text{inc}}$.

**Example 4**. Shown in Fig.4 is a circular aperture of radius $R$ within an otherwise opaque screen located in the $xy$-plane at $z = 0$. A plane-wave $\boldsymbol{E}(\boldsymbol{r},t) = \boldsymbol{E}_{\text{inc}}\exp[\mathrm{i}(k_0\boldsymbol{\sigma}_{\text{inc}}\cdot\boldsymbol{r} - \omega t)]$, where the unit-vector $\boldsymbol{\sigma}_{\text{inc}}$ specifies the direction of incidence, arrives at the aperture from the left-hand side. Maxwell's 3rd equation identifies the incident $B$-field as $\boldsymbol{B} = \boldsymbol{\sigma}_{\text{inc}}\times\boldsymbol{E}/c$. The observation point $\boldsymbol{r}_0 = r_0\boldsymbol{\sigma}_0$ is sufficiently far from the aperture for the following approximation to apply:

$$G(\boldsymbol{r},\boldsymbol{r}_0)=\frac{\exp(\mathrm{i}k_0|\boldsymbol{r}-\boldsymbol{r}_0|)}{|\boldsymbol{r}-\boldsymbol{r}_0|}=\exp\left(\mathrm{i}k_0\sqrt{r^2+r_0^2-2\boldsymbol{r}\cdot\boldsymbol{r}_0}\right)/|\boldsymbol{r}-\boldsymbol{r}_0|\cong\frac{\exp[\mathrm{i}k_0(r_0-\boldsymbol{\sigma}_0\cdot\boldsymbol{r})]}{r_0}. \quad (50)$$

Let us assume that the screen is a thin sheet of a perfect conductor on whose surface the tangential $E$-field necessarily vanishes. The appropriate diffraction equation will then be Eq.(32), with the tangential $E$-field outside the aperture allowed to vanish (i.e., $\hat{\boldsymbol{n}}\times\boldsymbol{E}=0$ on the opaque parts of the screen). Approximating the $E$-field within the aperture with that of the incident plane-wave, we will have

$$\begin{aligned}\int_{S_1}[\hat{\boldsymbol{n}}\times\boldsymbol{E}(\boldsymbol{r})]G(\boldsymbol{r},\boldsymbol{r}_0)\mathrm{d}s &\cong(\hat{\boldsymbol{z}}\times\boldsymbol{E}_{\text{inc}})\int_{\text{aperture}}\exp(\mathrm{i}k_0\boldsymbol{\sigma}_{\text{inc}}\cdot\boldsymbol{r})\times\frac{\exp[\mathrm{i}k_0(r_0-\boldsymbol{\sigma}_0\cdot\boldsymbol{r})]}{r_0}\mathrm{d}s\\ &=\frac{\exp(\mathrm{i}k_0r_0)}{r_0}(\hat{\boldsymbol{z}}\times\boldsymbol{E}_{\text{inc}})\int_{\text{aperture}}\exp[\mathrm{i}k_0(\boldsymbol{\sigma}_{\text{inc}}-\boldsymbol{\sigma}_0)\cdot\boldsymbol{r}]\,\mathrm{d}s\\ &=\frac{\exp(\mathrm{i}k_0r_0)}{r_0}(\hat{\boldsymbol{z}}\times\boldsymbol{E}_{\text{inc}})\int_{r=0}^{R}\int_{\varphi=0}^{2\pi}\exp(\mathrm{i}k_0\zeta r\cos\varphi)\,r\mathrm{d}\varphi\mathrm{d}r. \quad (51)\end{aligned}$$

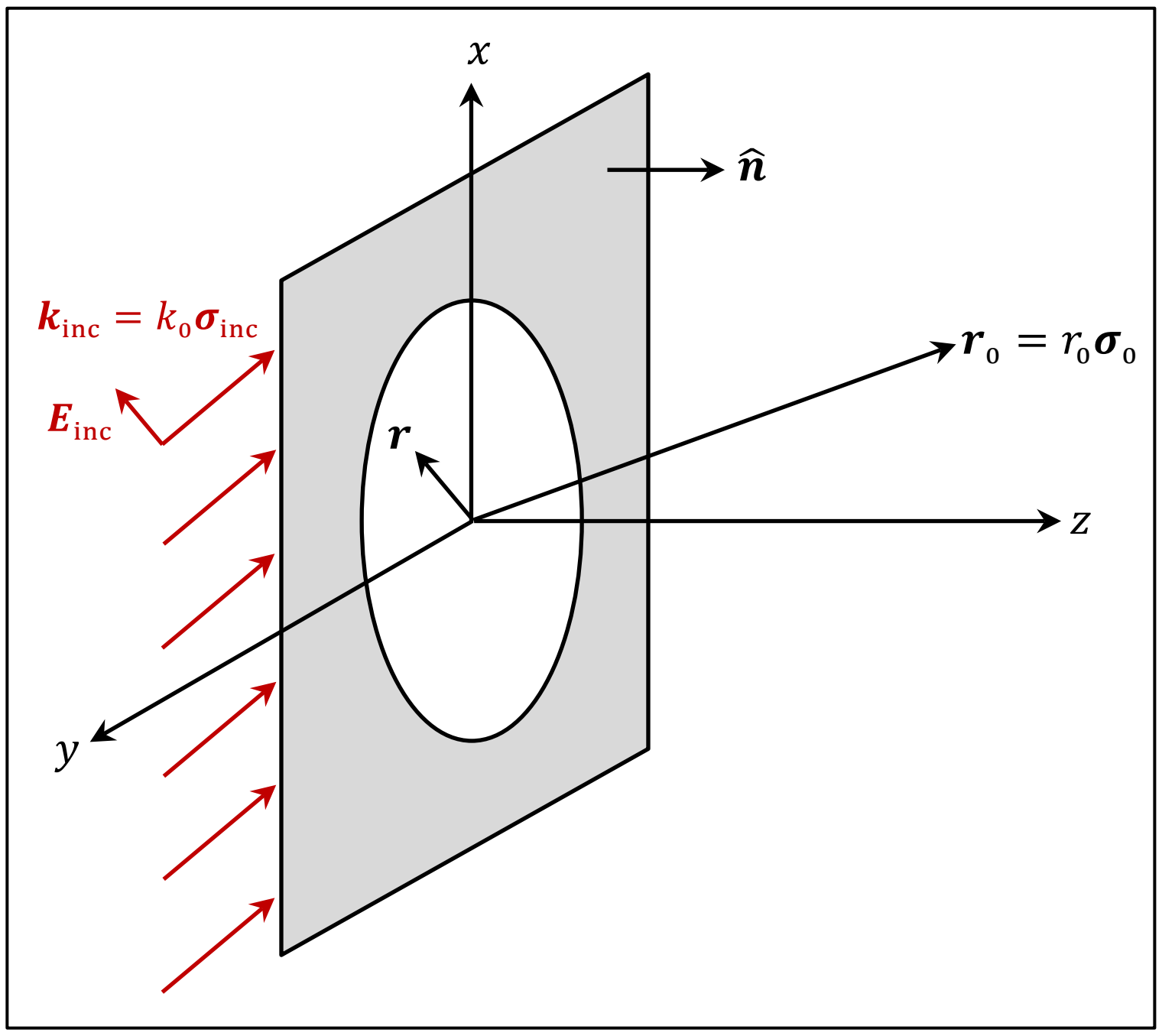

**Fig.4**. A circular aperture of radius $R$ inside an otherwise opaque screen located at $z=0$ is illuminated by a plane-wave whose $E$-field amplitude and propagation direction are specified as $\boldsymbol{E}_{\text{inc}}$ and $\boldsymbol{\sigma}_{\text{inc}}$, respectively. The observation point $\boldsymbol{r}_0=r_0\boldsymbol{\sigma}_0$ is in the far field; that is, $r_0\gg R$. The unit-vectors $\boldsymbol{\sigma}_{\text{inc}}$ and $\boldsymbol{\sigma}_0$ have respective polar coordinates $(\theta_{\text{inc}},\varphi_{\text{inc}})$ and $(\theta_0,\varphi_0)$. The projection of the vector $\boldsymbol{\sigma}_{\text{inc}}-\boldsymbol{\sigma}_0$ onto the $xy$-plane of the aperture is a vector of length $\zeta$ that makes an angle $\varphi$ with the position vector $\boldsymbol{r}=x\hat{\boldsymbol{x}}+y\hat{\boldsymbol{y}}$.

Here, $\zeta$ is the magnitude of the projection of $\boldsymbol{\sigma}_{\text{inc}}-\boldsymbol{\sigma}_0$ onto the $xy$-plane of the aperture, while $\varphi$ is the angle between that projection and the position vector $\boldsymbol{r}=x\hat{\boldsymbol{x}}+y\hat{\boldsymbol{y}}$; that is,

$$\boldsymbol{\sigma}_{\text{inc}}-\boldsymbol{\sigma}_0=(\sin\theta_{\text{inc}}\cos\varphi_{\text{inc}}-\sin\theta_0\cos\varphi_0)\hat{\boldsymbol{x}}+(\sin\theta_{\text{inc}}\sin\varphi_{\text{inc}}-\sin\theta_0\sin\varphi_0)\hat{\boldsymbol{y}}+(\cos\theta_{\text{inc}}-\cos\theta_0)\hat{\boldsymbol{z}}. \quad (52)$$

$$\zeta=\sqrt{\sin^2\theta_{\text{inc}}+\sin^2\theta_0-2\sin\theta_{\text{inc}}\sin\theta_0\cos(\varphi_{\text{inc}}-\varphi_0)}. \quad (53)$$

Consequently,

$$\int_{S_1}[\hat{\boldsymbol{n}}\times\boldsymbol{E}(\boldsymbol{r})]G(\boldsymbol{r},\boldsymbol{r}_0)\mathrm{d}s \cong \frac{2\pi\exp(\mathrm{i}k_0r_0)}{r_0}(\hat{\boldsymbol{z}}\times\boldsymbol{E}_{\mathrm{inc}})\int_0^R rJ_0(k_0\zeta r)\mathrm{d}r$$

$J_0(\cdot)$ and $J_1(\cdot)$ are Bessel functions of the first kind, orders 0 and 1.

$$= \frac{2\pi R\exp(\mathrm{i}k_0r_0)}{k_0r_0\zeta}J_1(k_0\zeta R)\,\hat{\boldsymbol{z}}\times\boldsymbol{E}_{\mathrm{inc}}. \quad (54)$$

The $E$-field at the observation point $\boldsymbol{r}_0 = r_0\boldsymbol{\sigma}_0$ (which is in the far field of the aperture, i.e., $r_0 \gg R$) is now found by computing the curl of the expression on the right-hand side of Eq.(54). Considering that $\hat{\boldsymbol{z}}\times\boldsymbol{E}_{\mathrm{inc}}$ is independent of $\boldsymbol{r}_0$, we use the vector identity $\boldsymbol{\nabla}\times(\psi\boldsymbol{V}_0) = \boldsymbol{\nabla}\psi\times\boldsymbol{V}_0$, then ignore the small (far field) contributions to $\boldsymbol{\nabla}\psi$ due to the dependence of $\zeta$ on $(\theta_0,\varphi_0)$, to arrive at

$$\begin{aligned}
\boldsymbol{E}(\boldsymbol{r}_0) &= (2\pi)^{-1}\boldsymbol{\nabla}_0\times\int_{S_1}[\hat{\boldsymbol{n}}\times\boldsymbol{E}(\boldsymbol{r})]G(\boldsymbol{r},\boldsymbol{r}_0)\mathrm{d}s\\
&\cong (R/k_0\zeta)J_1(k_0\zeta R)\boldsymbol{\nabla}_0[\exp(\mathrm{i}k_0r_0)/r_0]\times(\hat{\boldsymbol{z}}\times\boldsymbol{E}_{\mathrm{inc}})\\
&= (R/k_0\zeta)J_1(k_0\zeta R)(\mathrm{i}k_0 - r_0^{-1})[\exp(\mathrm{i}k_0r_0)/r_0]\boldsymbol{\sigma}_0\times(\hat{\boldsymbol{z}}\times\boldsymbol{E}_{\mathrm{inc}})\\
&\cong \frac{\mathrm{i}RJ_1(k_0\zeta R)e^{\mathrm{i}k_0r_0}}{r_0\zeta}[(\boldsymbol{E}_{\mathrm{inc}}\cdot\boldsymbol{\sigma}_0)\hat{\boldsymbol{z}} - \cos(\theta_0)\,\boldsymbol{E}_{\mathrm{inc}}]. \quad (55)
\end{aligned}$$

The approximate nature of these calculations should be borne in mind when comparing the various estimates of an observed field obtained via different routes.[9] For instance, had we started with Eq.(33) and proceeded by setting $\hat{\boldsymbol{n}}\times\boldsymbol{B} = 0$ on the opaque areas of the screen, we would have arrived at the following estimate of the observed $E$-field:

$$\boldsymbol{E}(\boldsymbol{r}_0) \cong \frac{\mathrm{i}RJ_1(k_0\zeta R)e^{\mathrm{i}k_0r_0}}{r_0\zeta}\{(\boldsymbol{E}_{\mathrm{inc}}\cdot\hat{\boldsymbol{z}})[\boldsymbol{\sigma}_{\mathrm{inc}} - (\boldsymbol{\sigma}_{\mathrm{inc}}\cdot\boldsymbol{\sigma}_0)\boldsymbol{\sigma}_0] - \cos\theta_{\mathrm{inc}}[\boldsymbol{E}_{\mathrm{inc}} - (\boldsymbol{E}_{\mathrm{inc}}\cdot\boldsymbol{\sigma}_0)\boldsymbol{\sigma}_0]\}. \quad (56)$$

The differences between Eqs.(55) and (56), which, in general, are *not* insignificant, can be traced to the assumptions regarding the nature of the opaque screen and the approximations involved in equating the $\boldsymbol{E}$ and $\boldsymbol{B}$ fields within the aperture to those of the incident plane-wave.

**7. Sommerfeld's analysis of diffraction from a perfectly conducting half-plane**. A rare example of an exact solution of Maxwell's equations as applied to EM diffraction was published by A. Sommerfeld in 1896.[25] The simplified version of Sommerfeld's original analysis presented in this section closely parallels that of Ref. [1], Chapter 11. Consider an EM plane-wave propagating in free space and illuminating a thin, perfectly conducting screen that sits in the upper half of the $xy$-plane at $z = 0$. The geometry of the system is shown in Fig.5, where the incident $k$-vector is denoted by $\boldsymbol{k}_{\mathrm{inc}}$, the oscillation frequency of the monochromatic wave is $\omega$, the speed of light in vacuum is $c$, the wave-number is $k_0 = \omega/c$, and the unit-vector along the direction of incidence is $\boldsymbol{\sigma}_{\mathrm{inc}}$. The plane-wave is linearly polarized with its $E$-field amplitude $E_0$ along the $y$-axis and $H$-field amplitude $H_0 = E_0/Z_0$ in the $xz$-plane of incidence. ($Z_0 = \sqrt{\mu_0/\varepsilon_0}$ is the impedance of free space.) The electric current density induced in the semi-infinite screen is denoted by $\boldsymbol{J}(\boldsymbol{r},t)$. In the chosen geometry, all the fields (incident as well as scattered) are uniform along the $y$-axis; consequently, it suffices to specify the position of an arbitrary point in the Cartesian $xyz$ space by its $x$ and $z$ coordinates only; that is,

$$\boldsymbol{r} = x\hat{\boldsymbol{x}} + z\hat{\boldsymbol{z}} = r(\cos\theta\,\hat{\boldsymbol{x}} + \sin\theta\,\hat{\boldsymbol{z}}). \quad (57)$$

Here, $r = \sqrt{x^2+z^2}$, and the angle $\theta$ is measured clockwise from the positive $x$-axis. The range of $\theta$ is $(0,\pi)$ for points $\boldsymbol{r}$ on the right-hand side of the screen, and $(-\pi,0)$ for the points on the left-hand side. Figure 5 shows that $\boldsymbol{k}_{\text{inc}}$ makes an angle $\theta_0 \in (0, \pi)$ with the positive $x$-axis. The incident plane-wave is, thus, fully specified by the following equations:

$$\boldsymbol{k}_{\text{inc}} = k_x\hat{\boldsymbol{x}} + k_z\hat{\boldsymbol{z}} = k_0\boldsymbol{\sigma}_{\text{inc}} = k_0(\cos\theta_0\,\hat{\boldsymbol{x}} + \sin\theta_0\,\hat{\boldsymbol{z}}). \tag{58}$$

$$\begin{aligned}\boldsymbol{E}_{\text{inc}}(\boldsymbol{r},t) &= E_0\hat{\boldsymbol{y}}\exp[\mathrm{i}(\boldsymbol{k}_{\text{inc}}\cdot\boldsymbol{r} - \omega t)] = E_0\hat{\boldsymbol{y}}\exp[\mathrm{i}k_0(\boldsymbol{\sigma}_{\text{inc}}\cdot\boldsymbol{r} - ct)]\\ &= E_0\hat{\boldsymbol{y}}e^{\mathrm{i}k_0 r\cos(\theta-\theta_0)}e^{-\mathrm{i}\omega t}.\end{aligned} \tag{59}$$

$$\boldsymbol{H}_{\text{inc}}(\boldsymbol{r},t) = -(E_0/Z_0)(\sin\theta_0\,\hat{\boldsymbol{x}} - \cos\theta_0\,\hat{\boldsymbol{z}})e^{\mathrm{i}k_0 r\cos(\theta-\theta_0)}e^{-\mathrm{i}\omega t}. \tag{60}$$

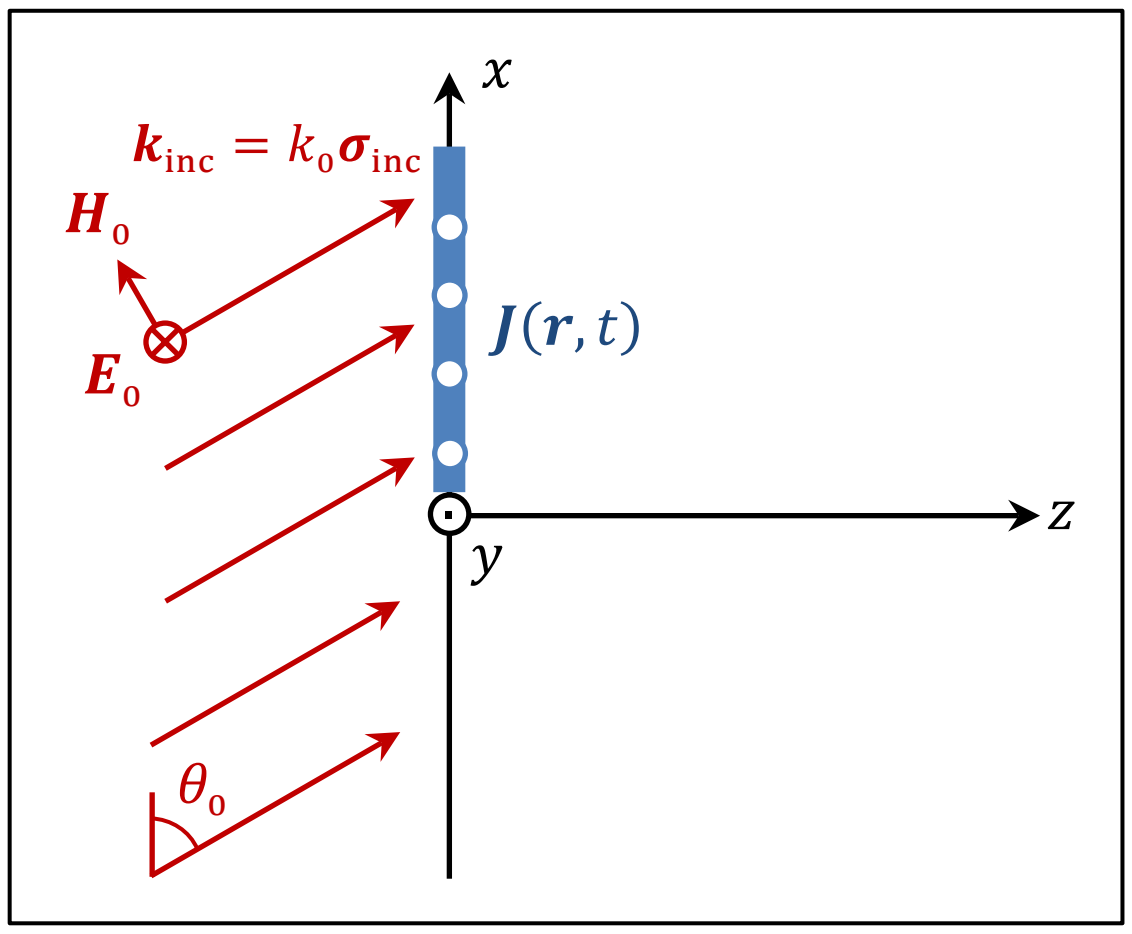

**Fig.5**. A plane, monochromatic EM wave propagating along the unit-vector $\boldsymbol{\sigma}_{\text{inc}} = \cos\theta_0\,\hat{\boldsymbol{x}} + \sin\theta_0\,\hat{\boldsymbol{z}}$ arrives at a thin, semi-infinite, perfectly electrically conducting screen located in the upper half of the $xy$-plane. The plane-wave is linearly polarized, with its $E$-field aligned with the $y$-axis, while its $H$-field has components along the $x$ and $z$ directions. The induced current sheet, denoted by $\boldsymbol{J}(\boldsymbol{r},t)$, oscillates parallel to the incident $E$-field along the direction of the $y$-axis. (The system depicted here is essentially the same as that in Fig.2, with the exception of the angle of incidence $\theta_0$ being the complement of $\theta_{\text{inc}}$ of Fig.2.)

With the aid of the step-function step$(x)$ and Dirac's delta-function $\delta(z)$, we now express the electric current density $\boldsymbol{J}(\boldsymbol{r},t)$ excited on the thin-sheet conductor as a one-dimensional Fourier transform, namely,

$$\boldsymbol{J}(\boldsymbol{r},t) = \text{step}(x)J_s(x)\delta(z)e^{-\mathrm{i}\omega t}\hat{\boldsymbol{y}} = \left[\int_{-\infty}^{\infty}\mathcal{J}(\sigma_x)e^{\mathrm{i}k_0\sigma_x x}\mathrm{d}\sigma_x\right]\delta(z)e^{-\mathrm{i}\omega t}\hat{\boldsymbol{y}}. \tag{61}$$

In the above equation, $\sigma_x$ represents spatial frequency along the $x$-axis, and $\mathcal{J}(\sigma_x)$ is the (complex) amplitude of the induced surface-current-density along $\hat{\boldsymbol{y}}$, having spatial and temporal frequencies $\sigma_x$ and $\omega$, respectively. The first goal of our analysis is to find the function $\mathcal{J}(\sigma_x)$ such that its Fourier transform vanishes along the negative $x$-axis — as demanded by the step-function in Eq.(61) — while its radiated $E$-field cancels the incident plane-wave's $E$-field at the surface of the screen along the positive $x$-axis. Now, a current sheet $\mathcal{J}(\sigma_x)e^{\mathrm{i}(k_0\sigma_x x-\omega t)}\delta(z)\hat{\boldsymbol{y}}$ that fills the entire $xy$-plane at $z = 0$ radiates EM fields into the surrounding free space that can easily be shown to have the following structure:

$$\boldsymbol{E}(\boldsymbol{r},t) = -½(Z_0k_0\hat{\boldsymbol{y}}/k_z)\mathcal{J}(\sigma_x)\exp[\mathrm{i}(k_x x + k_z|z| - \omega t)]. \tag{62}$$

$$\boldsymbol{H}(\boldsymbol{r},t) = ½\mathcal{J}(\sigma_x)[\pm\hat{\boldsymbol{x}} - (k_x/k_z)\hat{\boldsymbol{z}}]\exp[\mathrm{i}(k_x x + k_z|z| - \omega t)]. \tag{63}$$

Here, $k_0 = \omega/c$ is the wave-number in free space, while $k_x = k_0\sigma_x$, and $k_z = \sqrt{k_0^2 - k_x^2}$ are the $k$-vector components along the $x$ and $z$ axes. In general, $k_z$ must be real and positive when $|k_x| \le k_0$, and imaginary and positive otherwise. The $\pm$ signs associated with $H_x$ indicate that the plus sign must be used for the half-space on the right-hand side of the sheet, where $z > 0$, while the minus sign is reserved for the left-hand side, where $z < 0$. The discontinuity of $H_x$ at $z = 0$ thus equals the surface-current-density $\mathcal{J}(\sigma_x)$ along the $y$-direction, in compliance with Maxwell's requisite boundary conditions. Needless to say, Eqs.(62) and (63) represent a single EM plane-wave on either side of the screen's $xy$-plane, which satisfy the symmetry requirement of radiation from the current sheet, and with the tangential component $H_x$ of the radiated $H$-field chosen to satisfy the requisite boundary condition at the plane of the surface current. The plane-waves emanating to the right and left of the $xy$-plane of the current sheet are homogeneous when $k_z$ is real, and inhomogeneous (or evanescent) when $k_z$ is imaginary.

When working in the complex $\sigma_x$-plane, we must ensure that $k_z$ has the correct sign for all values of the real parameter $k_x = k_0\sigma_x$ from $-\infty$ to $\infty$. Considering that $k_z/k_0$ is the square root of the product of $(1-\sigma_x)$ and $(1+\sigma_x)$, we choose for both of these complex numbers the range of phase angles $(-\pi, \pi]$, as depicted in Fig.6(a). The corresponding branch-cuts thus appear as the semi-infinite line segments $(-\infty, -1)$ and $(1, \infty)$, and the integration path along the real axis within the $\sigma_x$-plane, shown in Fig.6(b), will consist of semi-infinite line segments slightly above and slightly below the real axis, as well as a short segment of the real-axis connecting $-1$ to 1. This choice of the integration path ensures that $k_z = k_0\sqrt{1-\sigma_x^2}$ acquires the correct sign for all values of $\sigma_x$.

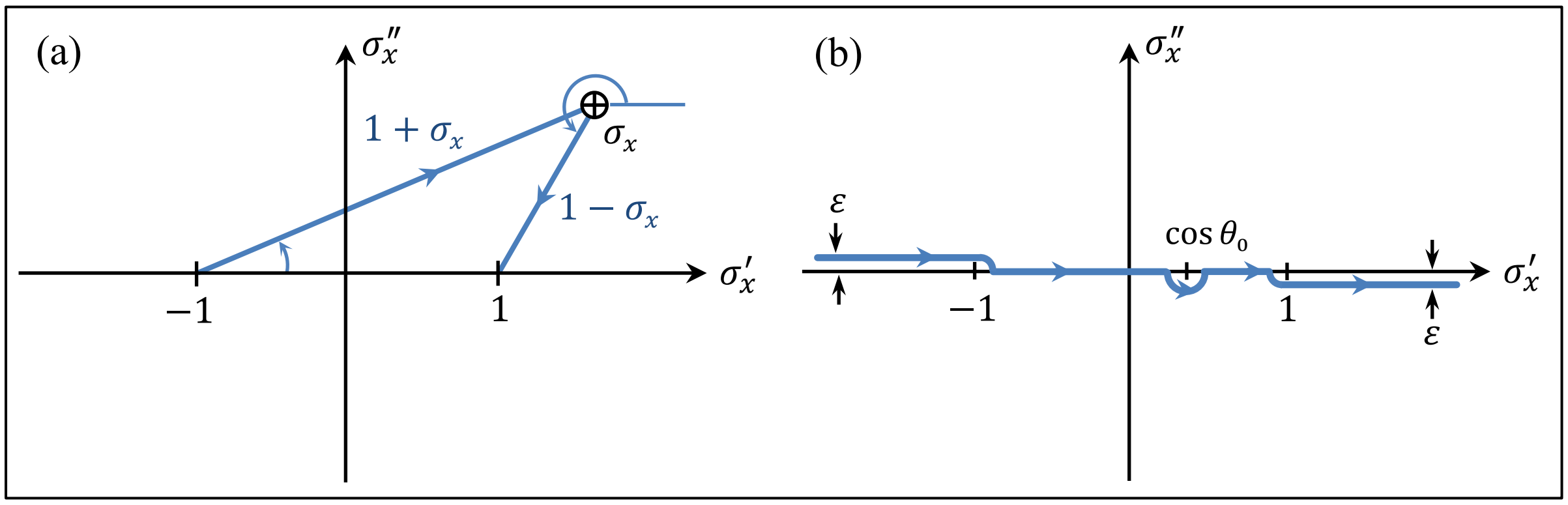

**Fig.6**. (a) In the complex $\sigma_x$-plane, where the phase of the complex numbers $1 \pm \sigma_x$ is measured counterclockwise from the positive $\sigma_x'$ axis, the range of both angles is confined to the $(-\pi, \pi]$ interval. (b) The integration path along the $\sigma_x'$ axis is adjusted by shifting the part from $-\infty$ to $-1$ slightly upward, and the part from 1 to $\infty$ slightly downward, so that $k_z/k_0 = \sqrt{(1-\sigma_x)(1+\sigma_x)}$ has the correct sign everywhere. The pole at $\sigma_x = \cos\theta_0$ is handled by locally deforming the contour into a semi-circular path below the real axis.

For reasons that will become clear as we proceed, Sommerfeld suggested the following mathematical form for the surface current-density $\mathcal{J}$ as a function of $\sigma_x$:

$$\mathcal{J}(\sigma_x) = \mathcal{J}_0\sqrt{1-\sigma_x}/(\sigma_x - \cos\theta_0). \tag{64}$$

The domain of this function is the slightly deformed real axis of the $\sigma_x$-plane depicted in Fig.6(b). The proposed function has a simple pole at $\sigma_x = \cos\theta_0$, where $\theta_0$ is the orientation angle of $\boldsymbol{k}_{\text{inc}}$ shown in Fig.5, and a (complex) constant coefficient $\mathcal{J}_0$ that will be determined shortly. In addition, $\mathcal{J}(\sigma_x)$ contains the term $\sqrt{1-\sigma_x}$, whose branch-cut in the system of Fig.6(a) is the semi-infinite line-segment $(1,\infty)$ along the real axis of the $\sigma_x$-plane. It is now possible to demonstrate that the proposed $\mathcal{J}(\sigma_x)$ satisfies its first required property, namely,

$$\int_{-\infty}^{\infty} \mathcal{J}(\sigma_x) e^{\mathrm{i}k_0\sigma_x x} \mathrm{d}\sigma_x = \mathcal{J}_0 \int_{-\infty}^{\infty} \frac{\sqrt{1-\sigma_x}}{\sigma_x - \cos\theta_0} e^{\mathrm{i}k_0\sigma_x x} \mathrm{d}\sigma_x = 0, \qquad (x < 0). \tag{65}$$

For $x < 0$, the integration contour of Fig.6(b) can be closed with a large semi-circular path in the lower half of the $\sigma_x$-plane. The only part of the integrand in Eq.(65) that requires a branch-cut is $\sqrt{1-\sigma_x}$, whose branch-cut is the line segment $(1,\infty)$. The integration contour of Fig.6(b), when closed in the lower-half of the $\sigma_x$-plane, does not contain this branch-cut. Moreover, the pole at $\sigma_x = \cos\theta_0$ is outside the closed loop of integration, and the contributions to the integral by the singular points $\sigma_x = \pm 1$ are zero. Consequently, the current-density $\boldsymbol{J}(\boldsymbol{r},t)$ in the lower half of the $xy$-plane turns out to be zero, exactly as required.

The radiated $E$-field, a superposition of contributions from the various $\mathcal{J}(\sigma_x)$ in accordance with Eqs.(62) and (64), must cancel out the incident $E$-field at the surface of the screen; that is,

$$E_y(x,y,z=0) = \int_{-\infty}^{\infty} E_y(\sigma_x) e^{\mathrm{i}k_0\sigma_x x} \mathrm{d}\sigma_x = -\tfrac{1}{2} Z_0 k_0 \int_{-\infty}^{\infty} [\mathcal{J}(\sigma_x)/k_z] e^{\mathrm{i}k_0\sigma_x x} \mathrm{d}\sigma_x$$

$$= -\tfrac{1}{2} Z_0 \mathcal{J}_0 \int_{-\infty}^{\infty} \frac{e^{\mathrm{i}k_0\sigma_x x}}{\sqrt{1+\sigma_x}\,(\sigma_x - \cos\theta_0)} \mathrm{d}\sigma_x = -E_0 e^{\mathrm{i}k_0\cos\theta_0 x}, \qquad (x > 0). \tag{66}$$

For $x > 0$, the contour of integration in Eq.(66) can be closed with a large semi-circle in the upper-half of the $\sigma_x$-plane. The branch-cut for $\sqrt{1+\sigma_x}$ in the denominator of the integrand is the line-segment $(-\infty,-1)$, which is below the integration path and, therefore, irrelevant for a contour that closes in the upper-half-plane. For the singularities at $\sigma_x = \pm 1$, the residues are zero, whereas for the pole at $\sigma_x = \cos\theta_0$, the residue is $e^{ik_0\cos\theta_0 x}/\sqrt{1+\cos\theta_0}$. The requisite boundary condition at the screen's surface is thus seen to be satisfied if $\mathcal{J}_0$ is specified as

$$\mathcal{J}_0 = -\mathrm{i}(E_0/\pi Z_0)\sqrt{1+\cos\theta_0}. \tag{67}$$

Having found the functional form of $\mathcal{J}(\sigma_x)$, we now turn to the problem of computing the scattered $E$-field $\boldsymbol{E}_s(\boldsymbol{r})$ at the arbitrary observation point $\boldsymbol{r} = x\hat{\boldsymbol{x}} + z\hat{\boldsymbol{z}} = r(\cos\theta\,\hat{\boldsymbol{x}} + \sin\theta\,\hat{\boldsymbol{z}})$ in accordance with Eqs.(62), (64), and (67); that is,

$$\boldsymbol{E}_s(\boldsymbol{r}) = \mathrm{i}(E_0\hat{\boldsymbol{y}}/2\pi) \int_{-\infty}^{\infty} \frac{\sqrt{1+\cos\theta_0}}{\sqrt{1+\sigma_x}\,(\sigma_x - \cos\theta_0)} \exp[\mathrm{i}k_0(\sigma_x x + \sigma_z|z|)]\, \mathrm{d}\sigma_x. \tag{68}$$

The integral in Eq.(68) must be evaluated at positive as well as negative values of $x$, and also for values of $z$ on both sides of the screen. The presence of $k_z$ in the exponent of the integrand requires that the branch-cuts on both line-segments $(-\infty,-1)$ and $(1,\infty)$ be taken into account. For these reasons, the integration path of Fig.6(b) ceases to be convenient and we need to change the variable $\sigma_x$ to something that avoids the need for branch-cuts. We now switch the variable from $\sigma_x$ to $\varphi$, where $\sigma_x = \cos\varphi$, with the integration path in the complex $\varphi$-plane shown in Fig.7(a). Considering that

$$\cos\varphi = \cos(\varphi' + \mathrm{i}\varphi'') = \cos\varphi' \cosh\varphi'' - \mathrm{i}\sin\varphi' \sinh\varphi'', \tag{69}$$

the depicted integration path represents the continuous variation of $\sigma_x$ from $-\infty$ to $\infty$. Similarly,

$$\sigma_z = \sqrt{1-\cos^2\varphi} = \sin\varphi = \sin\varphi'\cosh\varphi'' + \mathrm{i}\cos\varphi'\sinh\varphi'' \qquad (70)$$

is positive real on the horizontal branch, and positive imaginary on both vertical branches of the depicted contour. It is also easy to verify that $\sqrt{1-\sigma_x} = \sqrt{2}\sin(\varphi/2)$ has identical values at corresponding points on the contours of Figs.(6b) and (7a), as does $\sqrt{1+\sigma_x} = \sqrt{2}\cos(\varphi/2)$. Recalling that $\mathrm{d}\sigma_x = -\sin\varphi\,\mathrm{d}\varphi = -\sqrt{1-\cos^2\varphi}\,\mathrm{d}\varphi$, we can rewrite Eq.(68) as

$$\boldsymbol{E}_s(\boldsymbol{r}) = -\mathrm{i}(E_0\hat{\boldsymbol{y}}/2\pi)\int_{\pi-\mathrm{i}\infty}^{0+\mathrm{i}\infty} \frac{\sqrt{1+\cos\theta_0}\,\sqrt{1-\cos\varphi}}{\cos\varphi-\cos\theta_0}\exp[\mathrm{i}k_0 r\cos(\varphi\mp\theta)]\,\mathrm{d}\varphi. \qquad (71)$$

The above equation yields the scattered $E$-field on both sides of the screen, with the minus sign in the exponent corresponding to $z > 0$, and the plus sign to $z < 0$. In what follows, noting the natural symmetry of the scattered field on the opposite sides of the $xy$-plane, we confine our attention to Eq.(71) with only the minus sign in the exponent; the scattered $E$-field on the left hand side of the screen is subsequently obtained by a simple change of the sign of $z$.

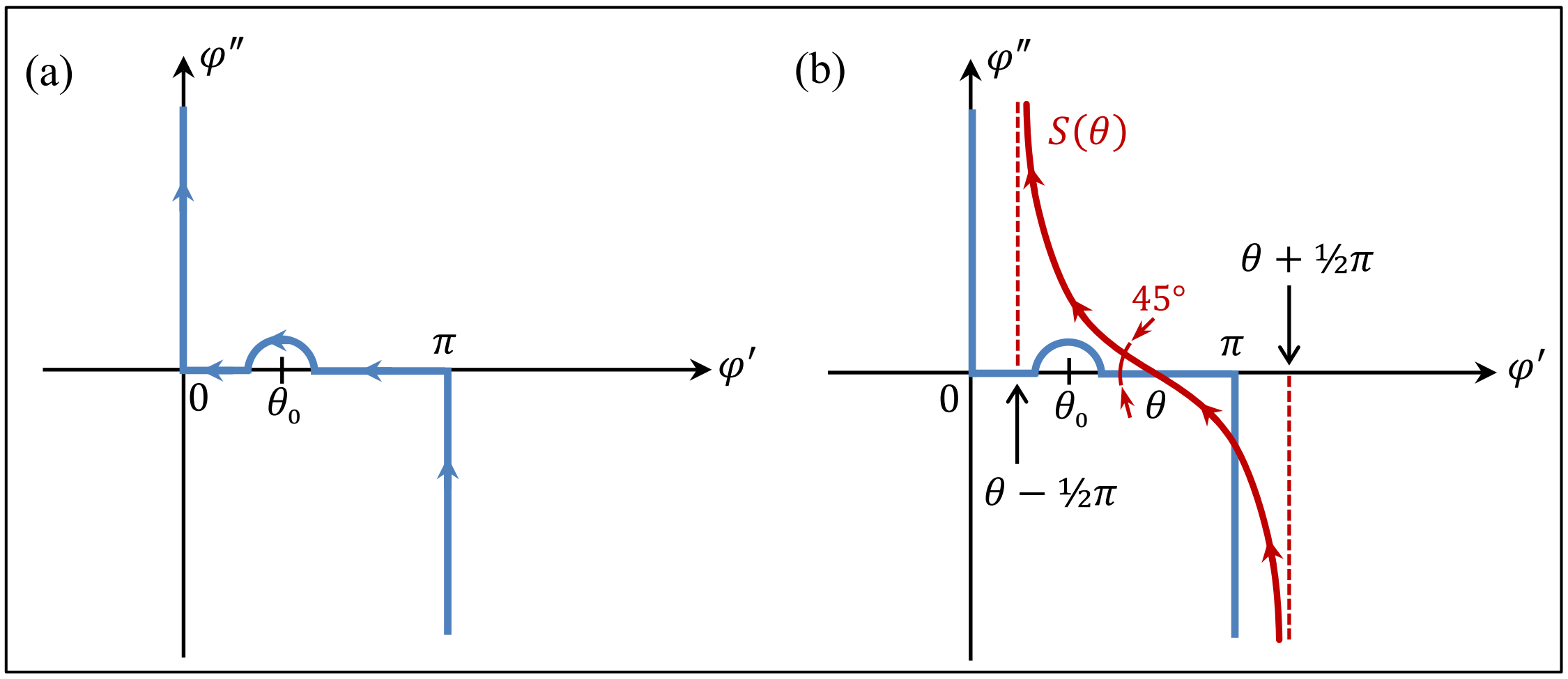

**Fig.7**. (a) Contour of integration in the complex $\varphi$-plane corresponding to the integration path in the $\sigma_x$-plane shown in Fig.6(b). By definition, $\sigma_x = \cos\varphi$, which results in $k_z = k_0\sigma_z = k_0\sin\varphi$, a well-defined function everywhere in the $\varphi$-plane that does not need branch-cuts. The small bump in the integration path around $\varphi = \theta_0$ corresponds to the small semi-circular part of the contour in Fig.6(b). The negative values of $\sigma_x''$ on the semi-circle translate, in accordance with Eq.(69), into positive value of $\varphi''$ on the corresponding bump. (b) Along the steepest-descent contour $S(\theta)$ that passes through the saddle-point $\varphi = \theta$, $\mathrm{Re}[\cos(\varphi-\theta)]$ is constant. At the saddle-point, $S(\theta)$ makes a 45° angle with the horizontal and vertical lines, which represent contours along which $\mathrm{Im}[\cos(\varphi-\theta)]$ is constant. $S(\theta)$ also has the property that $\mathrm{Im}[\cos(\varphi-\theta)]$, which is zero at the saddle-point, rises toward $\infty$ (continuously and symmetrically on both sides of the saddle) as $\varphi$ moves away from the saddle-point along the contour.

The integration path of Fig.7(a) is now replaced with the steepest-descent contour $S(\theta)$ depicted in Fig.7(b). Passing through the saddle-point of $\exp[\mathrm{i}kr\cos(\varphi-\theta)]$ at $\varphi = \theta$, this contour has the property that $\mathrm{Re}[\cos(\varphi-\theta)]$ everywhere on the contour equals 1. In contrast, $\mathrm{Im}[\cos(\varphi-\theta)]$ starts at zero at the saddle, then rises toward infinity (continuously and symmetrically on opposite sides of the saddle) as $\varphi$ moves away from the saddle-point at $\varphi = \theta$. It is easy to show that the original integration path of Fig.7(a) can be joined to $S(\theta)$ to form a closed loop with negligible contributions to the overall loop integral by the segments that connect the two contours at infinity.[**]

The only time when the single pole of the integrand at $\varphi = \theta_0$ needs to be accounted for is when $\theta < \theta_0$, at which point the pole is inside the closed contour, as Fig.7(b) clearly indicates. In the vicinity of the pole, the denominator of the integrand in Eq.(71) can be approximated by the first two terms of the Taylor series expansion of $\cos\varphi$, as follows:

$$\cos\varphi - \cos\theta_0 = [\cos\theta_0 - \sin\theta_0\,(\varphi - \theta_0) + \cdots] - \cos\theta_0 \cong -\sin\theta_0\,(\varphi - \theta_0). \qquad (72)$$

The residue at the pole is thus seen to be $-\exp[\mathrm{i}k_0 r\cos(\theta_0 - \theta)]$, with a corresponding contribution of $-E_0\hat{\boldsymbol{y}}\exp[\mathrm{i}k_0 r\cos(\theta - \theta_0)]$ to the scattered $E$-field. It must be emphasized that the scattered field contributed by the pole at $\varphi = \theta_0$ is relevant only when $\theta < \theta_0$, in which case it cancels the contribution of the incident plane-wave of Eq.(59) to the overall EM field on the right-hand side of the screen, where the screen casts its geometric shadow. Outside this geometric shadow, where $\theta_0 < \theta \le \pi$, the incident beam spills into the $z \ge 0$ region — without the counter-balancing effect of the scattered field produced by the pole at $\varphi = \theta_0$.

Returning to the scattered $E$-field produced by the integral in Eq.(71) over the steepest-descent contour $S(\theta)$, the first term in the integrand can be further streamlined, as follows:

$$\frac{\sqrt{1+\cos\theta_0}\,\sqrt{1-\cos\varphi}}{\cos\varphi - \cos\theta_0} = \frac{2\cos(\theta_0/2)\sin(\varphi/2)}{-2\sin[½(\varphi+\theta_0)]\sin[½(\varphi-\theta_0)]} = -\frac{½}{\sin[½(\varphi+\theta_0)]} - \frac{½}{\sin[½(\varphi-\theta_0)]}. \qquad (73)$$

The scattered $E$-field on the right-hand side of the screen may thus be written as

$$\boldsymbol{E}_s(\boldsymbol{r}) = \mathrm{i}(E_0\hat{\boldsymbol{y}}/4\pi)e^{\mathrm{i}k_0 r}\int_{S(\theta)}\left(\frac{1}{\sin[½(\varphi+\theta_0)]} + \frac{1}{\sin[½(\varphi-\theta_0)]}\right)e^{\mathrm{i}k_0 r[\cos(\varphi-\theta)-1]}\mathrm{d}\varphi\,. \qquad (74)$$

Note that we have factored out the imaginary part of the exponent and taken it outside the integral as $e^{\mathrm{i}k_0 r}$; what remains of the exponent, namely, $\mathrm{i}k_0 r[\cos(\varphi - \theta) - 1]$, is purely real on $S(\theta)$. We proceed to compute the integral in Eq.(74) only for the first term of the integrand and refer to it as $\boldsymbol{E}_{s,1}$; the contribution of the second term, $\boldsymbol{E}_{s,2}$, will then be found by switching the sign of $\theta_0$.

A change of the variable from $\varphi$ to $\varphi - \theta$ would cause the steepest-descent contour to go through the origin of the $\varphi$-plane; this shifted contour will now be denoted by $S_0$. Taking advantage of the symmetry of $S_0$ with respect to the origin, we express the final result of integration in terms of the integral on the upper half of $S_0$, which is denoted by $S_0^+$. We will have

$$\begin{aligned}\boldsymbol{E}_{s,1}(\boldsymbol{r}) &= \mathrm{i}(E_0\hat{\boldsymbol{y}}/4\pi)e^{\mathrm{i}k_0 r}\int_{S_0^+}\left\{\frac{1}{\sin[½(\varphi+\theta+\theta_0)]} + \frac{1}{\sin[½(-\varphi+\theta+\theta_0)]}\right\}e^{\mathrm{i}k_0 r(\cos\varphi-1)}\mathrm{d}\varphi \\ &= \mathrm{i}(E_0\hat{\boldsymbol{y}}/\pi)e^{\mathrm{i}k_0 r}\int_{S_0^+}\frac{\sin[(\theta+\theta_0)/2]\cos(\varphi/2)}{\cos(\varphi)-\cos(\theta+\theta_0)}e^{-\mathrm{i}2k_0 r\sin^2(\varphi/2)}\mathrm{d}\varphi \\ &= -\mathrm{i}(E_0\hat{\boldsymbol{y}}/2\pi)\sin[(\theta+\theta_0)/2]\,e^{\mathrm{i}k_0 r}\int_{S_0^+}\frac{\cos(\varphi/2)\exp[-\mathrm{i}2k_0 r\sin^2(\varphi/2)]}{\sin^2(\varphi/2)-\sin^2[(\theta+\theta_0)/2]}\mathrm{d}\varphi. \qquad (75)\end{aligned}$$

---

[**] On the short line-segments that connect the two contours at $\varphi'' = \pm\infty$, the magnitude of the exponential factor in the integrand in Eq.(71) is $\exp[k_0 r\sin(\varphi' - \theta)\sinh(\varphi'')]$. In the upper half-plane, $-1 \le \sin(\varphi' - \theta) \le 0$ and $\sinh(\varphi'') \to \infty$, whereas in the lower half-plane, $0 \le \sin(\varphi' - \theta) \le 1$ and $\sinh(\varphi'') \to -\infty$. Thus, in both cases, the integrand vanishes.

Another change of variable, this time from $\varphi$ to the real-valued $\zeta = -\exp(\mathrm{i}\pi/4)\sin(\varphi/2)$, where $\zeta$ ranges from 0 to $\infty$ along the steepest-descent contour $S_0^+$, now yields[††]

$$\boldsymbol{E}_{s,1}(\boldsymbol{r}) = -(E_0\hat{\boldsymbol{y}}/\pi)\sin[(\theta+\theta_0)/2]\,e^{\mathrm{i}k_0 r}e^{-\mathrm{i}\pi/4}\int_0^\infty \frac{\exp(-2k_0 r\zeta^2)}{\zeta^2 - \mathrm{i}\sin^2[(\theta+\theta_0)/2]}\mathrm{d}\zeta. \tag{76}$$

The integral appearing in the above equation is evaluated in Appendix G, where it is shown that

$$\int_0^\infty \frac{\exp(-\lambda\zeta^2)}{\zeta^2 - \mathrm{i}\eta^2}\mathrm{d}\zeta = \sqrt{\pi}|\eta|^{-1}e^{-\mathrm{i}\lambda\eta^2}F(|\eta|\sqrt{\lambda}). \tag{77}$$

Here, $F(\alpha) = \int_\alpha^\infty \exp(\mathrm{i}x^2)\,\mathrm{d}x$ is the complex Fresnel integral defined in Eq.(46). We thus find

$$\boldsymbol{E}_{s,1}(\boldsymbol{r}) = -\big(E_0\hat{\boldsymbol{y}}/\sqrt{\pi}\big)e^{-\mathrm{i}\pi/4}e^{\mathrm{i}k_0 r}e^{-\mathrm{i}2k_0 r\sin^2[(\theta+\theta_0)/2]}F\big(\sqrt{2k_0 r}\sin[(\theta+\theta_0)/2]\big). \tag{78}$$

The expression for $\boldsymbol{E}_{s,2}(\boldsymbol{r})$ is derived from Eq.(78) by switching the sign of $\theta_0$. One has to be careful in this case, since $\sin[(\theta-\theta_0)/2]$ may be negative. Given that both $\theta$ and $\theta_0$ are in the $(0,\pi)$ interval, the sign of $\sin[(\theta-\theta_0)/2]$ will be positive if $\theta > \theta_0$, and negative if $\theta < \theta_0$. The scattered $E$-field of Eq.(74) is thus given by

$$\begin{aligned}\boldsymbol{E}_s(\boldsymbol{r}) = \boldsymbol{E}_{s,1} + \boldsymbol{E}_{s,2} = -\big(E_0\hat{\boldsymbol{y}}/\sqrt{\pi}\big)e^{-\mathrm{i}\pi/4}\big\{&e^{\mathrm{i}k_0 r\cos(\theta+\theta_0)}F\big(\sqrt{2k_0 r}\sin[(\theta+\theta_0)/2]\big)\\ &\pm e^{\mathrm{i}k_0 r\cos(\theta-\theta_0)}F\big(\pm\sqrt{2k_0 r}\sin[(\theta-\theta_0)/2]\big)\big\}.\end{aligned} \tag{79}$$

To this result we must add the contribution of the pole, namely, $-E_0\hat{\boldsymbol{y}}\exp[\mathrm{i}k_0 r\cos(\theta-\theta_0)]$ when $\theta < \theta_0$, and the incident beam $E_0\hat{\boldsymbol{y}}\exp[\mathrm{i}k_0 r\cos(\theta-\theta_0)]$ for the entire range $0 \le \theta \le \pi$. Recalling that

$$F(\alpha) + F(-\alpha) = \int_{-\infty}^\infty \exp(\mathrm{i}x^2)\,\mathrm{d}x = \sqrt{\pi}e^{\mathrm{i}\pi/4}, \tag{80}$$

the total $E$-field on the right-hand side of the $xy$-plane, where $0 \le \theta \le \pi$, becomes

$$\begin{aligned}\boldsymbol{E}_{\text{total}}(\boldsymbol{r}) = \big(e^{-\mathrm{i}\pi/4}E_0\hat{\boldsymbol{y}}/\sqrt{\pi}\big)\big\{&e^{\mathrm{i}k_0 r\cos(\theta-\theta_0)}F\big(-\sqrt{2k_0 r}\sin[(\theta-\theta_0)/2]\big)\\ &-e^{\mathrm{i}k_0 r\cos(\theta+\theta_0)}F\big(\sqrt{2k_0 r}\sin[(\theta+\theta_0)/2]\big)\big\}.\end{aligned} \tag{81}$$

On the left-hand side of the screen, where $z < 0$ and, therefore, $-\pi \le \theta \le 0$, the scattered $E$-field is obtained by replacing $\theta$ with $|\theta|$ in Eq.(79) as well as in the contribution by the pole at $\varphi = \theta_0$. (Appendix H shows that the scattered $E$-field in the $z < 0$ region can also be evaluated by direct integration over a modified contour in the complex $\varphi$-plane.) Once again, adding the incident $E$-field and invoking Eq.(80), we find

$$\begin{aligned}\boldsymbol{E}_{\text{total}}(\boldsymbol{r}) = \big(e^{-\mathrm{i}\pi/4}E_0\hat{\boldsymbol{y}}/\sqrt{\pi}\big)\big\{&e^{\mathrm{i}k_0 r\cos(\theta-\theta_0)}F\big(\sqrt{2k_0 r}\sin[(\theta-\theta_0)/2]\big)\\ &-e^{\mathrm{i}k_0 r\cos(\theta+\theta_0)}F\big(-\sqrt{2k_0 r}\sin[(\theta+\theta_0)/2]\big)\big\}.\end{aligned} \tag{82}$$

One obtains Eq.(82) by replacing $\theta$ in Eq.(81) with $2\pi + \theta$, the latter $\theta$ being in the $(-\pi, 0)$ interval. Thus, Eq.(81) with $0 \le \theta \le 2\pi$ covers the entire range of observation points $\boldsymbol{r}$. A clear understanding of Eq.(81) requires familiarity with the general behavior of the Fresnel integral $F(\alpha)$; Appendix F contains a detailed explanation in terms of the Cornu spiral representation of $F(\alpha)$.‡‡ The general expression for the total (i.e., incident plus scattered) $E$-field given in Eq.(81) is somewhat simplified in terms of the new function $\Phi(\alpha) = e^{-\mathrm{i}\alpha^2}F(\alpha)$, as follows:

$$\boldsymbol{E}_{\text{total}}(\boldsymbol{r}) = \frac{E_0\hat{\boldsymbol{y}}e^{\mathrm{i}k_0 r}}{\sqrt{\pi}e^{\mathrm{i}\pi/4}}\Phi\big(-\sqrt{2k_0 r}\sin[(\theta-\theta_0)/2]\big) - \Phi\big(\sqrt{2k_0 r}\sin[(\theta+\theta_0)/2]\big); \quad (0 \le \theta \le 2\pi). \quad (83)$$

[††] Upon setting $\zeta^2 = \mathrm{i}\sin^2(\varphi/2)$, we find two possible choices for the new variable, namely, $\zeta = \pm e^{\mathrm{i}\pi/4}\sin(\varphi/2)$. Of these, the one with the minus sign represents the upper-half $S_0^+$ of the steepest-descent trajectory. To see this, note that as $S_0^+$ approaches its extremity when $\varphi \to -½\pi + \mathrm{i}\infty$, we have $\sin(\varphi/2) \to (-1+\mathrm{i})\exp(\varphi''/2)/2\sqrt{2}$, which must be multiplied by $-e^{\mathrm{i}\pi/4}$ for $\zeta$ to approach $+\infty$.

It is worth mentioning that, the contribution to the scattered $E$-field by the pole at $\varphi = \theta_0$, namely, $\boldsymbol{E}_{s,3}(\boldsymbol{r}) = -E_0\hat{\boldsymbol{y}}\exp[\mathrm{i}k_0 r\cos(\theta-\theta_0)]$, which exists only in the intervals $0 \le \theta < \theta_0$ and $2\pi - \theta_0 < \theta \le 2\pi$, cancels the incident $E$-field in the shadow region behind the screen, while acting as the reflected field in front of the perfectly conducting half-mirror.

**7.1. The magnetic field**. The total $H$-field is computed from the $E$-field of Eq.(83) with the aid of Maxwell's equation $\mathrm{i}k_0 Z_0\boldsymbol{H}(\boldsymbol{r}) = \boldsymbol{\nabla}\times\boldsymbol{E}(\boldsymbol{r}) = -(r^{-1}\partial_\theta E_y)\hat{\boldsymbol{r}} + (\partial_r E_y)\hat{\boldsymbol{\theta}}$. The identity $\Phi'(\alpha) = -1 - 2\mathrm{i}\alpha\Phi(\alpha)$ will be used in this calculation. To simplify the notation, we introduce the new variables $\xi_1 = -\sqrt{2k_0 r}\sin[(\theta-\theta_0)/2]$ and $\xi_2 = \sqrt{2k_0 r}\sin[(\theta+\theta_0)/2]$. We find

$$H_r(\boldsymbol{r}) = \frac{E_0 e^{\mathrm{i}k_0 r}}{\sqrt{\pi}\,e^{\mathrm{i}\pi/4}Z_0}\left[\Phi(\xi_1)\sin(\theta-\theta_0) - \Phi(\xi_2)\sin(\theta+\theta_0) + \frac{\mathrm{i}\sqrt{2}\cos(\theta/2)\cos(\theta_0/2)}{\sqrt{k_0 r}}\right]. \quad (84)$$

$$H_\theta(\boldsymbol{r}) = \frac{E_0 e^{\mathrm{i}k_0 r}}{\sqrt{\pi}\,e^{\mathrm{i}\pi/4}Z_0}\left[\Phi(\xi_1)\cos(\theta-\theta_0) - \Phi(\xi_2)\cos(\theta+\theta_0) - \frac{\mathrm{i}\sqrt{2}\sin(\theta/2)\cos(\theta_0/2)}{\sqrt{k_0 r}}\right]. \quad (85)$$

The Cartesian components $H_x = H_r\cos\theta - H_\theta\sin\theta$ and $H_z = H_r\sin\theta + H_\theta\cos\theta$ of the magnetic field may now be obtained from the polar components $H_r$ and $H_\theta$, as follows:

$$H_x(\boldsymbol{r}) = -\frac{E_0 e^{\mathrm{i}k_0 r}}{\sqrt{\pi}\,e^{\mathrm{i}\pi/4}Z_0}\left\{[\Phi(\xi_1)+\Phi(\xi_2)]\sin\theta_0 - \frac{\mathrm{i}\sqrt{2}\cos(\theta/2)\cos(\theta_0/2)}{\sqrt{k_0 r}}\right\}. \quad (86)$$

$$H_z(\boldsymbol{r}) = +\frac{E_0 e^{\mathrm{i}k_0 r}}{\sqrt{\pi}\,e^{\mathrm{i}\pi/4}Z_0}\left\{[\Phi(\xi_1)-\Phi(\xi_2)]\cos\theta_0 + \frac{\mathrm{i}\sqrt{2}\sin(\theta/2)\cos(\theta_0/2)}{\sqrt{k_0 r}}\right\}. \quad (87)$$

It is readily verified that $H_z = 0$ at the surface of the conductor, where $\theta = 0$, and that $H_x$ equals the $x$ component of the incident $H$-field in the open half of the $xy$-plane, where $\theta = \pi$.

**8. Far field scattering and the optical theorem**. In the system of Fig.1(b), let the object inside the closed surface $S_1$ be illuminated from the outside, and let the observation point $\boldsymbol{r}_0 = r_0\boldsymbol{\sigma}_0$ be far away from the object, so that the approximate form of $G(\boldsymbol{r},\boldsymbol{r}_0)$ given in Eq.(50) along with its corresponding gradient $\boldsymbol{\nabla}G \cong -\mathrm{i}k_0\boldsymbol{\sigma}_0 G(\boldsymbol{r},\boldsymbol{r}_0)$ would be applicable. Denoting by $\boldsymbol{E}_s(\boldsymbol{r})$ and $\boldsymbol{B}_s(\boldsymbol{r})$ the scattered fields appearing on $S_1$, Eq.(28) yields the $E$-field at the observation point as

$$\boldsymbol{E}_s(\boldsymbol{r}_0) \cong \frac{\mathrm{i}k_0\exp(\mathrm{i}k_0 r_0)}{4\pi r_0}\int_{S_1}[c(\hat{\boldsymbol{n}}\times\boldsymbol{B}_s) - (\hat{\boldsymbol{n}}\times\boldsymbol{E}_s)\times\boldsymbol{\sigma}_0 - (\hat{\boldsymbol{n}}\cdot\boldsymbol{E}_s)\boldsymbol{\sigma}_0]\mathrm{e}^{-\mathrm{i}k_0\boldsymbol{\sigma}_0\cdot\boldsymbol{r}}\mathrm{d}s. \quad (88)$$

Considering that the local field in the vicinity of $\boldsymbol{r}_0$ has the character of a plane-wave, the last term in the above integrand, which represents a contribution to $\boldsymbol{E}_s(\boldsymbol{r}_0)$ that is aligned with the local $k$-vector, is expected to be cancelled out by an equal but opposite contribution from the first term.[9] We thus arrive at the following simplified version of Eq.(88):

‡‡ Our Eq.(81) agrees with the corresponding result in Born & Wolf's *Principles of Optics*,[1] provided that the angles $\theta$ and $\alpha_0$ in their Eq.(22) of Chapter 11, Section 5, are recognized as $2\pi - \theta$ and $\pi - \theta_0$ in our notation.

$$\boldsymbol{E}_s(\boldsymbol{r}_0) \cong -\frac{\mathrm{i}k_0 \exp(\mathrm{i}k_0 r_0)}{4\pi r_0}\boldsymbol{\sigma}_0 \times \int_{S_1}[c\boldsymbol{\sigma}_0 \times (\hat{\boldsymbol{n}} \times \boldsymbol{B}_s) - \hat{\boldsymbol{n}} \times \boldsymbol{E}_s]\mathrm{e}^{-\mathrm{i}k_0\boldsymbol{\sigma}_0\cdot\boldsymbol{r}}\mathrm{d}s. \tag{89}$$

With reference to Fig.8, suppose now that the object is illuminated (and thus excited) by a plane-wave arriving along the unit-vector $\boldsymbol{\sigma}_{\text{inc}}$, whose $\boldsymbol{E}$ and $\boldsymbol{B}$ fields are

$$\boldsymbol{E}_{\text{inc}}(\boldsymbol{r},t) = \boldsymbol{E}_{\text{i}} \exp[\mathrm{i}k_0(\boldsymbol{\sigma}_{\text{inc}} \cdot \boldsymbol{r} - ct)]. \tag{90}$$

$$\boldsymbol{B}_{\text{inc}}(\boldsymbol{r},t) = c^{-1}\boldsymbol{\sigma}_{\text{inc}} \times \boldsymbol{E}_{\text{i}} \exp[\mathrm{i}k_0(\boldsymbol{\sigma}_{\text{inc}} \cdot \boldsymbol{r} - ct)]. \tag{91}$$

The time-averaged total Poynting vector on the surface $S_1$ is readily evaluated, as follows:

$$\begin{aligned}
\langle \boldsymbol{S}_{\text{total}}(\boldsymbol{r})\rangle &= \tfrac{1}{2}\mathrm{Re}\big[(\boldsymbol{E}_{\text{i}}e^{\mathrm{i}k_0\boldsymbol{\sigma}_{\text{inc}}\cdot\boldsymbol{r}} + \boldsymbol{E}_s) \times \mu_0^{-1}(\boldsymbol{B}_{\text{i}}e^{\mathrm{i}k_0\boldsymbol{\sigma}_{\text{inc}}\cdot\boldsymbol{r}} + \boldsymbol{B}_s)^*\big] \\
&= \tfrac{1}{2}\mu_0^{-1}\mathrm{Re}\big(\boldsymbol{E}_{\text{i}} \times \boldsymbol{B}_{\text{i}}^* + \boldsymbol{E}_{\text{s}} \times \boldsymbol{B}_{\text{s}}^* + \boldsymbol{E}_{\text{i}}e^{\mathrm{i}k_0\boldsymbol{\sigma}_{\text{inc}}\cdot\boldsymbol{r}} \times \boldsymbol{B}_{\text{s}}^* + \boldsymbol{E}_{\text{s}} \times \boldsymbol{B}_{\text{i}}^* e^{-\mathrm{i}k_0\boldsymbol{\sigma}_{\text{inc}}\cdot\boldsymbol{r}}\big) \\
&= \tfrac{1}{2}Z_0^{-1}(\boldsymbol{E}_{\text{i}} \cdot \boldsymbol{E}_{\text{i}}^*)\boldsymbol{\sigma}_{\text{inc}} + \tfrac{1}{2}\mu_0^{-1}\mathrm{Re}(\boldsymbol{E}_{\text{s}} \times \boldsymbol{B}_{\text{s}}^*) \quad \leftarrow \boxed{Z_0 = \sqrt{\mu_0/\varepsilon_0} \cong 377\,\Omega} \\
&\quad + \tfrac{1}{2}Z_0^{-1}\mathrm{Re}\big\{[c\boldsymbol{E}_{\text{i}}^* \times \boldsymbol{B}_{\text{s}} + \boldsymbol{E}_{\text{s}} \times (\boldsymbol{\sigma}_{\text{inc}} \times \boldsymbol{E}_{\text{i}}^*)]e^{-\mathrm{i}k_0\boldsymbol{\sigma}_{\text{inc}}\cdot\boldsymbol{r}}\big\}.
\end{aligned} \tag{92}$$

If we dot-multiply both sides of Eq.(92) by $-\hat{\boldsymbol{n}}$, then integrate over the closed surface $S_1$, we obtain, on the left-hand side, the total rate of the inward flow of EM energy, which is the absorbed EM power by the object. On the right-hand side, the first term integrates to zero, because $(\boldsymbol{E}_{\text{i}} \cdot \boldsymbol{E}_{\text{i}}^*)\boldsymbol{\sigma}_{\text{inc}}$ is a constant and $\oint_{S_1} \hat{\boldsymbol{n}}\mathrm{d}s = 0$. The integral of the second term will be the negative time-rate of the energy departure from the object via scattering, which can be moved to the left-hand side of the equation. The combination of the two terms on the left-hand side now yields the total EM power that is taken away from the incident beam — either by absorption or due to scattering. We will have

$$\begin{aligned}
\textit{Absorbed + Scattered Power} &= -\tfrac{1}{2}Z_0^{-1}\mathrm{Re}\oint_{S_1} \hat{\boldsymbol{n}} \cdot [c\boldsymbol{E}_{\text{i}}^* \times \boldsymbol{B}_{\text{s}} + \boldsymbol{E}_{\text{s}} \times (\boldsymbol{\sigma}_{\text{inc}} \times \boldsymbol{E}_{\text{i}}^*)]e^{-\mathrm{i}k_0\boldsymbol{\sigma}_{\text{inc}}\cdot\boldsymbol{r}}\mathrm{d}s \\
\boxed{\boldsymbol{a}\cdot(\boldsymbol{b}\times\boldsymbol{c}) = (\boldsymbol{a}\times\boldsymbol{b})\cdot\boldsymbol{c}} \rightarrow &= \tfrac{1}{2}Z_0^{-1}\mathrm{Re}\oint_{S_1}[c(\hat{\boldsymbol{n}} \times \boldsymbol{B}_{\text{s}}) \cdot \boldsymbol{E}_{\text{i}}^* - (\hat{\boldsymbol{n}} \times \boldsymbol{E}_{\text{s}}) \cdot (\boldsymbol{\sigma}_{\text{inc}} \times \boldsymbol{E}_{\text{i}}^*)]e^{-\mathrm{i}k_0\boldsymbol{\sigma}_{\text{inc}}\cdot\boldsymbol{r}}\mathrm{d}s \\
&= \tfrac{1}{2}Z_0^{-1}\mathrm{Re}\left\{\boldsymbol{E}_{\text{i}}^* \cdot \oint_{S_1}[c(\hat{\boldsymbol{n}} \times \boldsymbol{B}_{\text{s}}) - (\hat{\boldsymbol{n}} \times \boldsymbol{E}_{\text{s}}) \times \boldsymbol{\sigma}_{\text{inc}}]e^{-\mathrm{i}k_0\boldsymbol{\sigma}_{\text{inc}}\cdot\boldsymbol{r}}\mathrm{d}s\right\}.
\end{aligned} \tag{93}$$

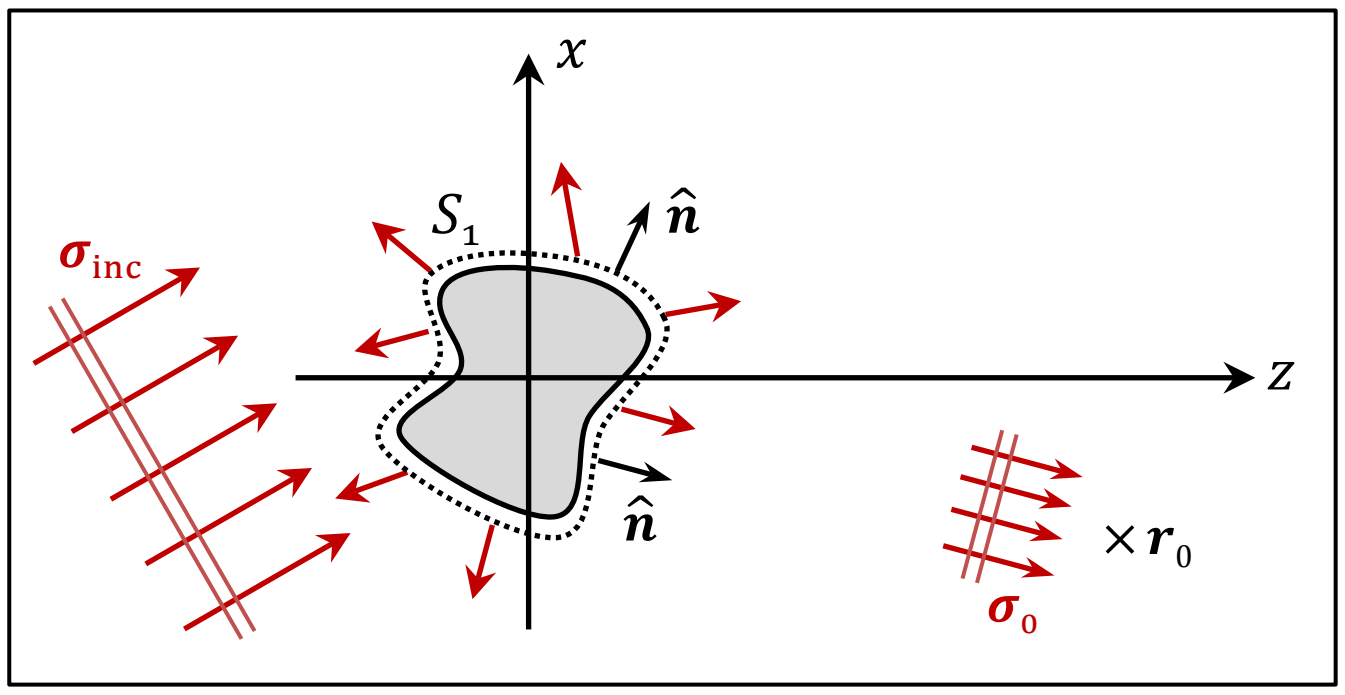

**Fig.8**. A monochromatic plane-wave propagating along the unit-vector $\boldsymbol{\sigma}_{\text{inc}}$ is scattered from a small object in the vicinity of the origin of the coordinate system. The scattered electric and magnetic fields on the closed surface $S_1$ surrounding the object are denoted by $\boldsymbol{E}_s(\boldsymbol{r})$ and $\boldsymbol{B}_s(\boldsymbol{r})$. The surface normals $\hat{\boldsymbol{n}}$ at every point on $S_1$ are outward directed. The scattered light reaching the far away observation point $\boldsymbol{r}_0 = r_0\boldsymbol{\sigma}_0$ has the $k$-vector $k_0\boldsymbol{\sigma}_0$ and the EM fields $\boldsymbol{E}_s(\boldsymbol{r}_0)$ and $\boldsymbol{B}_s(\boldsymbol{r}_0)$.

Comparison with Eq.(88) reveals that the integral in Eq.(93) is proportional to the scattered $E$-field in the direction of $\boldsymbol{\sigma}_0 = \boldsymbol{\sigma}_{\text{inc}}$ as observed in the far field. (Recall that the term $(\hat{\boldsymbol{n}} \cdot \boldsymbol{E}_s)\boldsymbol{\sigma}_0$ in the integrand of Eq.(88) has been deemed inconsequential.) If Eq.(93) is normalized by the incident EM power per unit area, namely, $P_{\text{inc}} = \tfrac{1}{2}Z_0^{-1}\text{Re}(\boldsymbol{E}_{\text{i}} \cdot \boldsymbol{E}_{\text{i}}^*)$, the left-hand side will become the scattering cross-section of the object, while the right-hand side, aside from the coefficient $\mathrm{i}k_0 e^{\mathrm{i}k_0 r_0}/(4\pi r_0)$, will be the projection of the forward-scattered $E$-field (i.e., $\boldsymbol{\sigma}_0 = \boldsymbol{\sigma}_{\text{inc}}$) on the incident $E$-field. This important result in the theory of scattering has come to be known as the *optical theorem* (or the optical cross-section theorem).[1,9,24]

**9. Scattering from weak inhomogeneities.**[9] Figure 9 shows a monochromatic plane-wave of frequency $\omega$ passing through a transparent, linear, isotropic medium that has a region of weak inhomogeneities in the vicinity of the origin of coordinates. The host medium is described by its relative permittivity $\varepsilon(\boldsymbol{r},\omega)$ and permeability $\mu(\boldsymbol{r},\omega)$, which consist of a spatially homogeneous background plus slight variations (localized in the vicinity of $\boldsymbol{r} = 0$) on this background; that is,

$$\varepsilon(\boldsymbol{r},\omega) = \varepsilon_h(\omega) + \delta\varepsilon(\boldsymbol{r},\omega). \tag{94}$$

$$\mu(\boldsymbol{r},\omega) = \mu_h(\omega) + \delta\mu(\boldsymbol{r},\omega). \tag{95}$$

The displacement field is thus written as $\boldsymbol{D}(\boldsymbol{r},\omega) = \varepsilon_0\varepsilon(\boldsymbol{r},\omega)\boldsymbol{E}(\boldsymbol{r},\omega)$ and the magnetic induction is given by $\boldsymbol{B}(\boldsymbol{r},\omega) = \mu_0\mu(\boldsymbol{r},\omega)\boldsymbol{H}(\boldsymbol{r},\omega)$. In the absence of free charges and currents, i.e., when $\rho_{\text{free}}(\boldsymbol{r},\omega) = 0$ and $\boldsymbol{J}_{\text{free}}(\boldsymbol{r},\omega) = 0$, Maxwell's macroscopic equations will be[4-11]

$$\boldsymbol{\nabla} \cdot \boldsymbol{D} = 0; \qquad \boldsymbol{\nabla} \times \boldsymbol{H} = -\mathrm{i}\omega\boldsymbol{D}; \qquad \boldsymbol{\nabla} \times \boldsymbol{E} = \mathrm{i}\omega\boldsymbol{B}; \qquad \boldsymbol{\nabla} \cdot \boldsymbol{B} = 0. \tag{96}$$

Noting that the $\boldsymbol{D}$ and $\boldsymbol{B}$ fields depart only slightly from the respective values $\varepsilon_0\varepsilon_h\boldsymbol{E}$ and $\mu_0\mu_h\boldsymbol{H}$ that they would have had in the absence of the $\delta\varepsilon$ and $\delta\mu$ perturbations, we write

$$\begin{aligned}\boldsymbol{\nabla} \times \boldsymbol{\nabla} \times (\boldsymbol{D} - \varepsilon_0\varepsilon_h\boldsymbol{E}) &= \boldsymbol{\nabla}(\overset{\to 0}{\boldsymbol{\nabla} \cdot \boldsymbol{D}}) - \boldsymbol{\nabla}^2\boldsymbol{D} - \mathrm{i}\omega\varepsilon_0\varepsilon_h\boldsymbol{\nabla} \times \boldsymbol{B} \\ &= -\boldsymbol{\nabla}^2\boldsymbol{D} - \mathrm{i}\omega\varepsilon_0\varepsilon_h[\boldsymbol{\nabla} \times (\boldsymbol{B} - \mu_0\mu_h\boldsymbol{H}) + \mu_0\mu_h\boldsymbol{\nabla} \times \boldsymbol{H}] \\ &= -\boldsymbol{\nabla}^2\boldsymbol{D} - \mathrm{i}\omega\varepsilon_0\varepsilon_h\boldsymbol{\nabla} \times (\boldsymbol{B} - \mu_0\mu_h\boldsymbol{H}) - (\omega/c)^2\mu_h\varepsilon_h\boldsymbol{D}. \end{aligned} \tag{97}$$

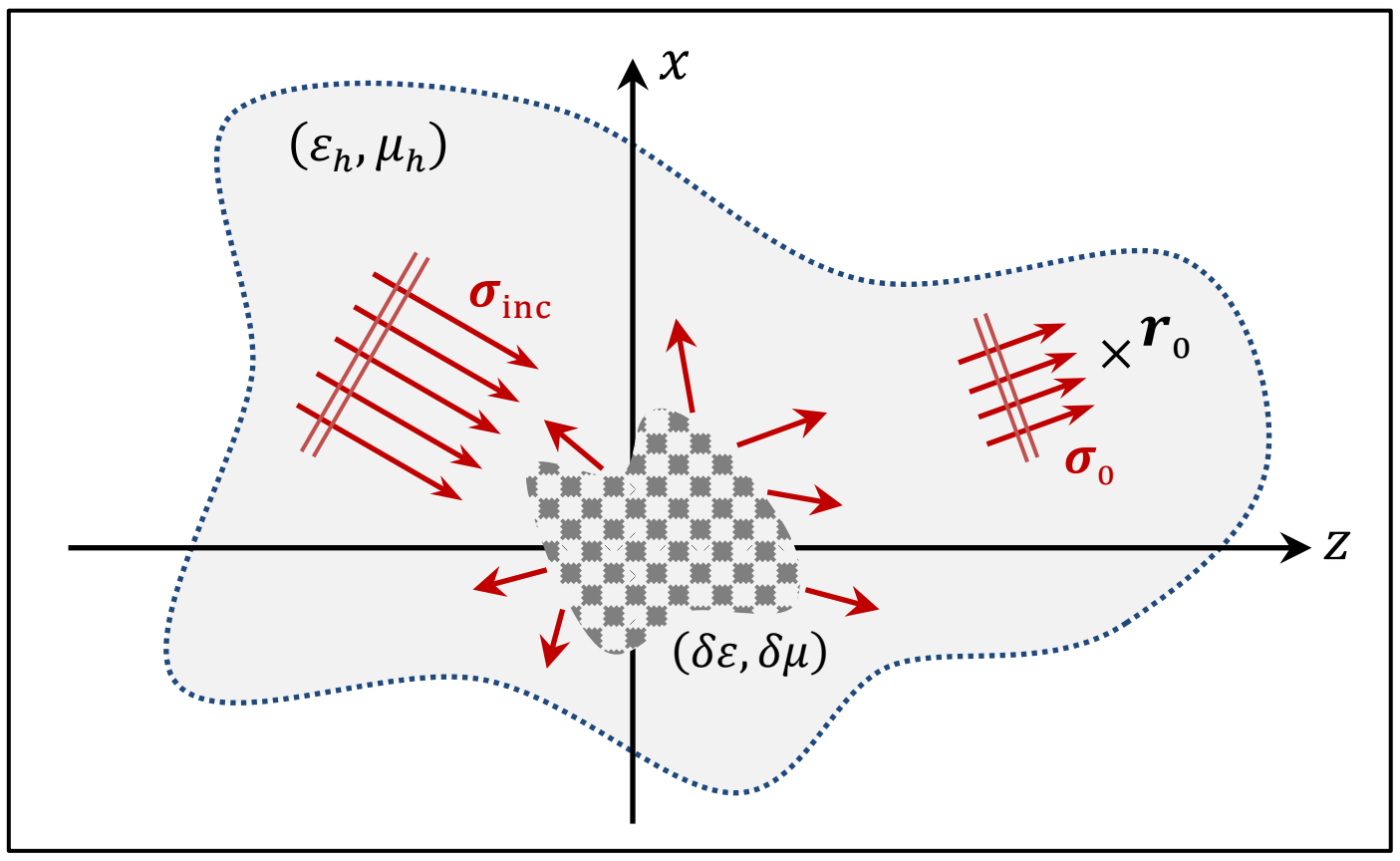

**Fig.9**. A monochromatic plane-wave of frequency $\omega$ and $E$-field amplitude $\boldsymbol{E}_{\text{inc}}$ propagates along the unit-vector $\boldsymbol{\sigma}_{\text{inc}}$ within a mildly inhomogeneous host medium of refractive index $n_h(\omega) = (\mu_h\varepsilon_h)^{1/2}$. The inhomogeneous region of the host, a small patch in the vicinity of the origin of coordinates, is specified by its relative permittivity $\varepsilon_h(\omega) + \delta\varepsilon(\boldsymbol{r},\omega)$ and relative permeability $\mu_h(\omega) + \delta\mu(\boldsymbol{r},\omega)$. The scattered field is observed at the faraway point $\boldsymbol{r}_0 = r_0\boldsymbol{\sigma}_0$.

Recalling that the refractive index of the homogeneous (background) material is defined as $n_h = (\mu_h \varepsilon_h)^{½}$, and that the free-space wave-number is $k_0 = \omega/c$, Eq.(97) may be rewritten as

$$\boldsymbol{\nabla}^2 \boldsymbol{D} + (k_0 n_h)^2 \boldsymbol{D} = -\mathrm{i}(k_0 \varepsilon_h/c)\boldsymbol{\nabla} \times (\delta\mu \boldsymbol{H}) - \varepsilon_0 \boldsymbol{\nabla} \times \boldsymbol{\nabla} \times (\delta\varepsilon \boldsymbol{E}). \tag{98}$$

This Helmholtz equation has a homogeneous solution, which we denote by $\boldsymbol{D}_h(\boldsymbol{r}, \omega)$, and an inhomogeneous solution, which arises from the local deviations $\delta\varepsilon$ and $\delta\mu$ of the host medium. Recalling that the Green function $G(\boldsymbol{r}, \boldsymbol{r}_0) = \exp(\mathrm{i}k_0 n_h |\boldsymbol{r} - \boldsymbol{r}_0|)/|\boldsymbol{r} - \boldsymbol{r}_0|$ is a solution of the Helmholtz equation $\boldsymbol{\nabla}^2 G + (k_0 n_h)^2 G = -4\pi\delta(\boldsymbol{r} - \boldsymbol{r}_0)$, an integral relation for the scattered field solution $\boldsymbol{D}_s(\boldsymbol{r}, \omega)$ of Eq.(98) in terms of the total fields $\boldsymbol{E}(\boldsymbol{r}, \omega)$ and $\boldsymbol{H}(\boldsymbol{r}, \omega)$ will be

d$\boldsymbol{r}$ stands for d$x$d$y$d$z$

$$\boldsymbol{D}_s(\boldsymbol{r}_0, \omega) = (4\pi)^{-1} \int_{\text{volume}} [\mathrm{i}(k_0 \varepsilon_h/c)\boldsymbol{\nabla} \times (\delta\mu \boldsymbol{H}) + \varepsilon_0 \boldsymbol{\nabla} \times \boldsymbol{\nabla} \times (\delta\varepsilon \boldsymbol{E})]\, G(\boldsymbol{r}, \boldsymbol{r}_0)\mathrm{d}\boldsymbol{r}. \tag{99}$$

Using a far-field approximation to $G(\boldsymbol{r}, \boldsymbol{r}_0)$ similar to that in Eq.(50), we will have

$$\boldsymbol{D}_s(\boldsymbol{r}_0, \omega) \cong \frac{\exp(\mathrm{i}k_0 n_h r_0)}{4\pi r_0} \int_{\text{volume}} [\mathrm{i}(k_0 \varepsilon_h/c)\boldsymbol{\nabla} \times (\delta\mu \boldsymbol{H}) + \varepsilon_0 \boldsymbol{\nabla} \times \boldsymbol{\nabla} \times (\delta\varepsilon \boldsymbol{E})] e^{-\mathrm{i}k_0 n_h \boldsymbol{\sigma}_0 \cdot \boldsymbol{r}} \mathrm{d}\boldsymbol{r}. \tag{100}$$

The vector identity $(\boldsymbol{\nabla} \times \boldsymbol{V})e^{-\mathrm{i}\boldsymbol{k}\cdot\boldsymbol{r}} = \mathrm{i}\boldsymbol{k} \times \boldsymbol{V}e^{-\mathrm{i}\boldsymbol{k}\cdot\boldsymbol{r}} + \boldsymbol{\nabla} \times (\boldsymbol{V}e^{-\mathrm{i}\boldsymbol{k}\cdot\boldsymbol{r}})$ can be used to replace the first term in the integrand of Eq.(100) with $-(k_0^2 \varepsilon_h n_h/c)e^{-\mathrm{i}k_0 n_h \boldsymbol{\sigma}_0\cdot\boldsymbol{r}} \boldsymbol{\sigma}_0 \times \delta\mu \boldsymbol{H}$. The volume integral of $\boldsymbol{\nabla} \times (\boldsymbol{V}e^{-\mathrm{i}\boldsymbol{k}\cdot\boldsymbol{r}})$ becomes the surface integral of $e^{-\mathrm{i}\boldsymbol{k}\cdot\boldsymbol{r}}\boldsymbol{V} \times \mathrm{d}\boldsymbol{s}$, which subsequently vanishes because, for away from the inhomogeneous region, $\delta\mu \to 0$. Similarly, the second term of the integrand is replaced by $\mathrm{i}\varepsilon_0 k_0 n_h \boldsymbol{\sigma}_0 \times [\boldsymbol{\nabla} \times (\delta\varepsilon \boldsymbol{E})]e^{-\mathrm{i}k_0 n_h \boldsymbol{\sigma}_0\cdot\boldsymbol{r}}$. Another application of the vector identity then replaces the remaining term with $-\varepsilon_0 k_0^2 n_h^2 \boldsymbol{\sigma}_0 \times (\boldsymbol{\sigma}_0 \times \delta\varepsilon \boldsymbol{E})e^{-\mathrm{i}k_0 n_h \boldsymbol{\sigma}_0\cdot\boldsymbol{r}}$. We thus arrive at

$$\boldsymbol{D}_s(\boldsymbol{r}_0, \omega) \cong \frac{(k_0 n_h)^2 \exp(\mathrm{i}k_0 n_h r_0)}{4\pi r_0} \int_{\text{volume}} [(\varepsilon_h/n_h c)\delta\mu \boldsymbol{H} + \varepsilon_0 \boldsymbol{\sigma}_0 \times \delta\varepsilon \boldsymbol{E}] \times \boldsymbol{\sigma}_0 e^{-\mathrm{i}k_0 n_h \boldsymbol{\sigma}_0\cdot\boldsymbol{r}} \mathrm{d}\boldsymbol{r}. \tag{101}$$

In the first Born approximation, the $\boldsymbol{E}(\boldsymbol{r})$ and $\boldsymbol{H}(\boldsymbol{r})$ fields in the integrand of Eq.(101) are replaced with the solutions $\boldsymbol{E}_h(\boldsymbol{r})$ and $\boldsymbol{H}_h(\boldsymbol{r})$ of the homogeneous Helmholtz equation. When the homogeneous background wave is a plane-wave, we will have

$$\boldsymbol{E}_h(\boldsymbol{r}) = \boldsymbol{E}_{\text{inc}} \exp(\mathrm{i}k_0 n_h \boldsymbol{\sigma}_{\text{inc}} \cdot \boldsymbol{r}). \tag{102}$$

$$\boldsymbol{H}_h(\boldsymbol{r}) = \sqrt{\frac{\varepsilon_0 \varepsilon_h}{\mu_0 \mu_h}} \boldsymbol{\sigma}_{\text{inc}} \times \boldsymbol{E}_{\text{inc}} \exp(\mathrm{i}k_0 n_h \boldsymbol{\sigma}_{\text{inc}} \cdot \boldsymbol{r}). \tag{103}$$

A final substitution from Eqs.(102) and (103) into Eq.(101) yields

$$\boldsymbol{D}_s(\boldsymbol{r}_0, \omega) \cong \frac{\varepsilon_0 \varepsilon_h k_0^2 \exp(\mathrm{i}k_0 n_h r_0)}{4\pi r_0} \int_{\text{volume}} [(\varepsilon_h \delta\mu \boldsymbol{\sigma}_{\text{inc}} + \mu_h \delta\varepsilon \boldsymbol{\sigma}_0) \times \boldsymbol{E}_{\text{inc}}] \times \boldsymbol{\sigma}_0 e^{\mathrm{i}k_0 n_h (\boldsymbol{\sigma}_{\text{inc}} - \boldsymbol{\sigma}_0)\cdot\boldsymbol{r}} \mathrm{d}\boldsymbol{r}. \tag{104}$$

Thus, in the first Born approximation, the scattered field $\boldsymbol{D}_s(\boldsymbol{r}_0, \omega) = \varepsilon_0 \varepsilon_h \boldsymbol{E}_s(\boldsymbol{r}_0, \omega)$ is directly related to the host medium perturbations $\delta\varepsilon(\boldsymbol{r}, \omega)$ and $\delta\mu(\boldsymbol{r}, \omega)$ via the volume integral in Eq.(104). Here, $\boldsymbol{E}_{\text{inc}}$ embodies not only the strength but also the polarization state of the incident plane-wave, the unit-vector $\boldsymbol{\sigma}_{\text{inc}}$ is the direction of incidence, $\boldsymbol{\sigma}_0 = \boldsymbol{r}_0/r_0$ is a unit-vector pointing from the origin of coordinates to the observation point $\boldsymbol{r}_0$, and $\boldsymbol{q} = k_0 n_h (\boldsymbol{\sigma}_{\text{inc}} - \boldsymbol{\sigma}_0)$ is the difference between the incident and scattered $k$-vectors.

**10. Neutron scattering from magnetic electrons in Born's first approximation**. The scattering of slow neutrons from ferromagnetic materials can be treated in ways that are similar to our analysis of EM scattering from mild inhomogeneities discussed in the preceding section. The wave function $\psi(\boldsymbol{r},t)$ of a particle of mass $m$ in the scalar potential field $V(\boldsymbol{r},t)$ satisfies the following Schrödinger equation:[6-8, 20]

$$\mathrm{i}\hbar\partial_t\psi(\boldsymbol{r},t) = -(\hbar^2/2m)\nabla^2\psi(\boldsymbol{r},t) + V(\boldsymbol{r},t)\psi(\boldsymbol{r},t). \tag{105}$$

When the potential is time-independent and the particle is in an eigenstate of energy $\mathcal{E}_n$, we will have the time-independent Schrödinger equation, namely,

$$[(\hbar^2/2m)\nabla^2 + \mathcal{E}_n]\psi(\boldsymbol{r}) = V(\boldsymbol{r})\psi(\boldsymbol{r}). \tag{106}$$

The Green function for Eq.(106) is $G(\boldsymbol{r},\boldsymbol{r}') = e^{\mathrm{i}k|\boldsymbol{r}-\boldsymbol{r}'|}/|\boldsymbol{r}-\boldsymbol{r}'|$, where $\hbar k = (2m\mathcal{E}_n)^{½}$. If, in the absence of the potential $V(\boldsymbol{r})$, the solution of the homogeneous equation is found to be $\psi_0(\boldsymbol{r})$, then, when the potential is introduced, we will have

d$\boldsymbol{r}'$ stands for d$x'$d$y'$d$z'$

$$\psi(\boldsymbol{r}) = \psi_0(\boldsymbol{r}) - (m/2\pi\hbar^2)\iiint_{-\infty}^{\infty} G(\boldsymbol{r},\boldsymbol{r}')V(\boldsymbol{r}')\psi(\boldsymbol{r}')\mathrm{d}\boldsymbol{r}'. \tag{107}$$

Note that Eq.(107) is *not* an actual solution of Eq.(106); rather, considering that the desired wave-function $\psi(\boldsymbol{r})$ appears in the integrand on the right hand-side, Eq.(107) is an integral form of the differential equation (106). In the case of Born's first approximation, one assumes that $V(\boldsymbol{r})$ is a fairly weak potential, in which case $\psi_0(\boldsymbol{r}')$ can be substituted for $\psi(\boldsymbol{r}')$, yielding

$$\psi(\boldsymbol{r}) \cong \psi_0(\boldsymbol{r}) - (m/2\pi\hbar^2)\iiint_{-\infty}^{\infty} G(\boldsymbol{r},\boldsymbol{r}')V(\boldsymbol{r}')\psi_0(\boldsymbol{r}')\mathrm{d}\boldsymbol{r}'. \tag{108}$$

In a typical scattering problem, an incoming particle of mass $m$ and well-defined momentum $\boldsymbol{p} = \hbar\boldsymbol{k}$ has the initial wave-function $\psi_0(\boldsymbol{r}) = e^{\mathrm{i}\boldsymbol{k}\cdot\boldsymbol{r}}$. Upon interacting with a weak scattering potential $V(\boldsymbol{r})$, the wave-function will change in accordance with Eq.(108). Let $V(\boldsymbol{r})$ have significant values only in the vicinity of the origin, $\boldsymbol{r} = 0$, and assume that the scattering process is elastic, so that the momentum $\tilde{\boldsymbol{p}}$ of the particle after scattering will have the same magnitude $\hbar k$ as before, but the direction of propagation changes from that of $\boldsymbol{k}$ to that of $\tilde{\boldsymbol{k}}$. At an observation point $\boldsymbol{r}$ far from the origin, that is, $|\boldsymbol{r}| \gg |\boldsymbol{r}'|$, the scattered particle's momentum is expected to be $\tilde{\boldsymbol{p}} = \hbar\tilde{\boldsymbol{k}} = \hbar k\hat{\boldsymbol{r}}$, and the Green function may be approximated as

$$\frac{\exp(\mathrm{i}k|\boldsymbol{r}-\boldsymbol{r}'|)}{|\boldsymbol{r}-\boldsymbol{r}'|} \cong \frac{\exp[\mathrm{i}k\sqrt{(\boldsymbol{r}-\boldsymbol{r}')\cdot(\boldsymbol{r}-\boldsymbol{r}')}]}{r} \cong \frac{\exp[\mathrm{i}k(r-\boldsymbol{r}'\cdot\hat{\boldsymbol{r}})]}{r} = \frac{\exp(\mathrm{i}kr)}{r}e^{-\mathrm{i}\tilde{\boldsymbol{k}}\cdot\boldsymbol{r}'}. \tag{109}$$

Substitution into Eq.(108) now yields

$$\psi(\boldsymbol{r}) \cong e^{\mathrm{i}\boldsymbol{k}\cdot\boldsymbol{r}} - \left(\frac{m}{2\pi\hbar^2}\right)\frac{e^{\mathrm{i}kr}}{r}\int_{-\infty}^{\infty} V(\boldsymbol{r}')e^{\mathrm{i}(\boldsymbol{k}-\tilde{\boldsymbol{k}})\cdot\boldsymbol{r}'}\mathrm{d}\boldsymbol{r}'. \tag{110}$$

Denoting the change in the direction of the particle's momentum by $\boldsymbol{q} = \boldsymbol{p} - \tilde{\boldsymbol{p}}$, and noting that $e^{\mathrm{i}kr}/r$ is simply a spherical wave emanating from the origin, the scattering amplitude $f(\boldsymbol{q})$ is readily seen to be

$$f(\boldsymbol{q}) \cong -\frac{m}{2\pi\hbar^2}\int_{-\infty}^{\infty} V(\boldsymbol{r}')e^{\mathrm{i}\boldsymbol{q}\cdot\boldsymbol{r}'/\hbar}\,\mathrm{d}\boldsymbol{r}'. \tag{111}$$

Here, $f(\boldsymbol{q})$ has the dimensions of length (meter in *SI*). Note that the presence of $e^{\mathrm{i}kr}/r$ in Eq.(110) makes the wave-function $\psi(\boldsymbol{r})$ dimensionless. Let $\mathrm{d}\Omega = \sin\theta\,\mathrm{d}\theta\mathrm{d}\varphi$ be the differential element of the solid angle viewed from the origin of the coordinates. The differential scattering cross-section will then be $\mathrm{d}\sigma/\mathrm{d}\Omega = |f(\boldsymbol{q})|^2$, with the total cross-section being $\sigma = \int|f(\boldsymbol{q})|^2\mathrm{d}\Omega$.

**Example 1**. A particle of mass $m$ is scattered by the spherically symmetric potential $V(r)$ corresponding to a fixed particle located at $\boldsymbol{r}=0$. The scattering amplitude, computed from Eq.(111), will be

$$f(\boldsymbol{q})=-\frac{m}{2\pi\hbar^2}\int_{r=0}^{\infty}\int_{\theta=0}^{\pi}V(r)\exp(\mathrm{i}qr\cos\theta/\hbar)\,(2\pi r^2\sin\theta)\mathrm{d}\theta\mathrm{d}r=-\frac{2m}{\hbar q}\int_0^{\infty}rV(r)\sin(qr/\hbar)\,\mathrm{d}r. \quad (112)$$

Considering that $\boldsymbol{q}=\boldsymbol{p}-\tilde{\boldsymbol{p}}$, and that $\boldsymbol{p}$ and $\tilde{\boldsymbol{p}}$ have the same magnitude $p$, we denote by $\theta$ the angle between $\boldsymbol{p}$ and $\tilde{\boldsymbol{p}}$, and proceed to write $q=2p\sin(\theta/2)$. The scattering amplitude thus has circular symmetry around the direction of the incident momentum $\boldsymbol{p}$. The ambiguity of Eq.(112) with regard to forward scattering at $\theta=0$ is resolved if the forward amplitude $f(0)$ is directly computed from Eq.(110) — with the destructive interference between the incident and scattered amplitudes properly taken into account.

For the Yukawa potential $V(\boldsymbol{r})=v_0e^{-\alpha r}/r$, with $\alpha>0$ being the range parameter, the scattering amplitude is obtained upon integrating Eq.(112), as follows:

$$f(\boldsymbol{q})=-\frac{2mv_0}{\hbar q}\int_0^{\infty}e^{-\alpha r}\sin(qr/\hbar)\,\mathrm{d}r=-\frac{2mv_0}{q^2+(\alpha\hbar)^2}. \quad (113)$$

In the limit when $\alpha\to 0$, the Yukawa potential approaches the Coulomb potential $V(\boldsymbol{r})=v_0/r$. When a particle having electric charge $\pm e$ and energy $\mathcal{E}=p^2/2m$ is scattered from another particle of charge $\pm e$, the scattering cross-section will be

$$\frac{\mathrm{d}\sigma}{\mathrm{d}\Omega}=|f(\boldsymbol{q})|^2=\left(\frac{e^2}{16\pi\varepsilon_0\mathcal{E}}\right)^2\frac{1}{\sin^4(\theta/2)}. \quad (114)$$

This is the famous Rutherford scattering cross-section.[9]

---

**Example 2**. In low-energy scattering, $k\cong 0$ and the scattering amplitude in all directions becomes $f(\theta,\varphi)\cong-(m/2\pi\hbar^2)\int_{-\infty}^{\infty}V(\boldsymbol{r})\mathrm{d}\boldsymbol{r}$. In the case of low-energy, soft-sphere scattering, where $V(\boldsymbol{r})=V_0$ when $r\le r_0$ and zero otherwise, we find

$$f(\theta,\varphi)=-(m/2\pi\hbar^2)(4\pi r_0^3/3)V_0. \quad (115a)$$

$$\mathrm{d}\sigma/\mathrm{d}\Omega=|f(\theta,\varphi)|^2=(2mr_0^3V_0/3\hbar^2)^2. \quad (115b)$$

$$\sigma=4\pi(\mathrm{d}\sigma/\mathrm{d}\Omega)=16\pi m^2r_0^6V_0^2/9\hbar^4. \quad (115c)$$

---

Let us now consider the case of a polarized neutron entering a ferromagnetic medium and getting scattered from the host's magnetic electrons.[29] To obtain an estimate of the corresponding scattering potential $V(\boldsymbol{r})$, we begin by noting that the magnetic field surrounding a point-dipole $m\delta(\boldsymbol{r})\hat{\boldsymbol{z}}$ in free space is

The magnetic dipole moment $m$ should not be confused with the particle's mass $m$.

$$\boldsymbol{H}(\boldsymbol{r})=m(2\cos\theta\,\hat{\boldsymbol{r}}+\sin\theta\,\hat{\boldsymbol{\theta}})/(4\pi\mu_0r^3). \quad (116)$$

A second magnetic point-dipole $\boldsymbol{m}'$, located at $\boldsymbol{r}\neq 0$, will have the energy $\mathcal{E}=-\boldsymbol{m}'\cdot\boldsymbol{H}(\boldsymbol{r})$, which may be written as follows:

$$\mathcal{E}=-\boldsymbol{m}'\cdot m[3\cos\theta\,\hat{\boldsymbol{r}}-(\cos\theta\,\hat{\boldsymbol{r}}-\sin\theta\,\hat{\boldsymbol{\theta}})]/(4\pi\mu_0r^3)$$
$$=\boldsymbol{m}'\cdot m(\hat{\boldsymbol{z}}-3\cos\theta\,\hat{\boldsymbol{r}})/(4\pi\mu_0r^3)=[\boldsymbol{m}\cdot\boldsymbol{m}'-3(\boldsymbol{m}\cdot\hat{\boldsymbol{r}})(\boldsymbol{m}'\cdot\hat{\boldsymbol{r}})]/(4\pi\mu_0r^3). \quad (117)$$

These results are consistent with the Einstein-Laub formula $\boldsymbol{F}=(\boldsymbol{m}\cdot\boldsymbol{\nabla})\boldsymbol{H}$ for the force as well as $\boldsymbol{T}=\boldsymbol{m}\times\boldsymbol{H}$ for the torque experienced by a point-dipole in a magnetic field.[30,31] Recall that, in contrast to the standard formula for the dipole moment, our definition of $\boldsymbol{B}$ as $\mu_0\boldsymbol{H}+\boldsymbol{M}$ maintains that the magnitude of $\boldsymbol{m}$ equals $\mu_0$ times the electrical current times the area of a small current loop. Consequently, the aforementioned expression for $\mathcal{E}$ coincides with the well-known expression $\mathcal{E}=-\boldsymbol{m}\cdot\boldsymbol{B}(\boldsymbol{r})$ of the energy of the dipole $\boldsymbol{m}$ in the external field $\boldsymbol{B}(\boldsymbol{r})=\mu_0\boldsymbol{H}(\boldsymbol{r})$.[6,9]

Equation (117) must be augmented by the contact term $-2\boldsymbol{m}\cdot\boldsymbol{m}'\delta(\boldsymbol{r})/3\mu_0$ to account for the energy of the dipole pair when $\boldsymbol{m}$ and $\boldsymbol{m}'$ overlap in space.[29] We will have

$$\mathcal{E}(\boldsymbol{r})=\frac{\boldsymbol{m}\cdot\boldsymbol{m}'-3(\boldsymbol{m}\cdot\hat{\boldsymbol{r}})(\boldsymbol{m}'\cdot\hat{\boldsymbol{r}})}{4\pi\mu_0 r^3}-\frac{2\boldsymbol{m}\cdot\boldsymbol{m}'\delta(\boldsymbol{r})}{3\mu_0}. \qquad (118)$$

Suppose the electron has wave-function $\psi(\boldsymbol{r}_\mathrm{e})$ and magnetic dipole moment $\boldsymbol{\mu}_\mathrm{e}$, while the incoming neutron has wave-function $\exp(\mathrm{i}\boldsymbol{k}\cdot\boldsymbol{r}_\mathrm{n})$, magnetic dipole moment $\boldsymbol{\mu}_\mathrm{n}$, and mass $m_\mathrm{n}$. We assume the scattering process does not involve a spin flip, so that both $\boldsymbol{\mu}_\mathrm{e}$ and $\boldsymbol{\mu}_\mathrm{n}$ retain their orientations after the collision. Moreover, we assume the electron — being bound to its host lattice — does not get dislocated or otherwise distorted, so that $\psi(\boldsymbol{r}_\mathrm{e})$ is the same before and after the collision. Thus, the potential energy distribution across the landscape of the incoming neutron is the integral over $\boldsymbol{r}_\mathrm{e}$ of the product of the electron's probability-density function $|\psi(\boldsymbol{r}_\mathrm{e})|^2$ and the dipole-dipole interaction energy $\mathcal{E}(\boldsymbol{r}_\mathrm{e}-\boldsymbol{r}_\mathrm{n})$ between the neutron and the electron. Consequently, in the first Born approximation, the scattering amplitude from an initial neutron momentum $\boldsymbol{p}=\hbar\boldsymbol{k}$ to a final momentum $\boldsymbol{p}'=\hbar\boldsymbol{k}'$, is given by

$$f(\boldsymbol{p}',\boldsymbol{p})=-\left(\frac{m_\mathrm{n}}{2\pi\hbar^2}\right)\int_{-\infty}^{\infty}|\psi(\boldsymbol{r}_\mathrm{e})|^2\mathcal{E}(\boldsymbol{r}_\mathrm{n}-\boldsymbol{r}_\mathrm{e})\exp[\mathrm{i}(\boldsymbol{k}-\boldsymbol{k}')\cdot\boldsymbol{r}_\mathrm{n}]\,\mathrm{d}\boldsymbol{r}_\mathrm{e}\mathrm{d}\boldsymbol{r}_\mathrm{n}. \qquad (119)$$

Defining the electronic magnetization (i.e., magnetic moment density of the electron) by $\boldsymbol{M}(\boldsymbol{r}_\mathrm{e})=|\psi(\boldsymbol{r}_\mathrm{e})|^2\boldsymbol{\mu}_\mathrm{e}$ and its Fourier transform by $\widetilde{\boldsymbol{M}}(\boldsymbol{k})=\int_{-\infty}^{\infty}\boldsymbol{M}(\boldsymbol{r}_\mathrm{e})\exp(\mathrm{i}\boldsymbol{k}\cdot\boldsymbol{r}_\mathrm{e})\,\mathrm{d}\boldsymbol{r}_\mathrm{e}$, upon substitution from Eq.(118) into Eq.(119) and setting $\boldsymbol{r}=\boldsymbol{r}_\mathrm{n}-\boldsymbol{r}_\mathrm{e}$, we find

$$f(\boldsymbol{p}',\boldsymbol{p})=-\frac{m_\mathrm{n}}{8\pi^2\mu_0\hbar^2}\int_{-\infty}^{\infty}\left[\frac{\boldsymbol{\mu}_\mathrm{n}-3(\boldsymbol{\mu}_\mathrm{n}\cdot\hat{\boldsymbol{r}})\hat{\boldsymbol{r}}}{r^3}-\frac{8\pi\delta(\boldsymbol{r})\boldsymbol{\mu}_\mathrm{n}}{3}\right]\cdot\boldsymbol{M}(\boldsymbol{r}_\mathrm{e})\exp[\mathrm{i}(\boldsymbol{p}-\boldsymbol{p}')\cdot\boldsymbol{r}_\mathrm{n}/\hbar]\,\mathrm{d}\boldsymbol{r}_\mathrm{e}\mathrm{d}\boldsymbol{r}_\mathrm{n}. \qquad (120)$$

Defining $\boldsymbol{q}=\boldsymbol{p}-\boldsymbol{p}'$ and changing the variables from $(\boldsymbol{r}_\mathrm{e},\boldsymbol{r}_\mathrm{n})$ to $(\boldsymbol{r}_\mathrm{e},\boldsymbol{r}=\boldsymbol{r}_\mathrm{n}-\boldsymbol{r}_\mathrm{e})$ — whose transformation Jacobian is 1.0 — substantially simplifies the above integral, yielding

$$f(\boldsymbol{q})=-\frac{m_\mathrm{n}}{8\pi^2\mu_0\hbar^2}\int_{-\infty}^{\infty}\left[\frac{\boldsymbol{\mu}_\mathrm{n}-3(\boldsymbol{\mu}_\mathrm{n}\cdot\hat{\boldsymbol{r}})\hat{\boldsymbol{r}}}{r^3}-\frac{8\pi\delta(\boldsymbol{r})\boldsymbol{\mu}_\mathrm{n}}{3}\right]\cdot\widetilde{\boldsymbol{M}}(\boldsymbol{q}/\hbar)\exp(\mathrm{i}\boldsymbol{q}\cdot\boldsymbol{r}/\hbar)\,\mathrm{d}\boldsymbol{r}. \qquad (121)$$

Appendix I shows that the exact evaluation of the integral in Eq.(121) leads to

$$f(\boldsymbol{q})=\left(\frac{m_\mathrm{n}}{2\pi\mu_0\hbar^2}\right)\boldsymbol{\mu}_\mathrm{n}\cdot\{\widetilde{\boldsymbol{M}}(\boldsymbol{q}/\hbar)-[\widetilde{\boldsymbol{M}}(\boldsymbol{q}/\hbar)\cdot\hat{\boldsymbol{q}}]\hat{\boldsymbol{q}}\}. \qquad (122)$$

This is the same result as given in Ref. [29], Eq.(23), in the case of $\lambda=1$. The coefficient $4\pi\mu_0$ appears here because we have worked in the *SI* system of units with $\boldsymbol{B}=\mu_0\boldsymbol{H}+\boldsymbol{M}$.

**11. Concluding remarks**. In this paper, we described some of the fundamental theories of EM scattering and diffraction using the electrodynamics of Maxwell and Lorentz in conjunction with standard mathematical methods of the vector calculus, complex analysis, differential equations, and Fourier transform theory. The scalar Huygens-Fresnel-Kirchhoff and Rayleigh-Sommerfeld theories were presented at first, followed by their extensions that cover the case of vector diffraction of EM waves. Examples were provided to showcase the application of these vector diffraction and scattering formulas to certain problems of practical interest. We did not discuss the alternate method of diffraction calculations by means of Fourier transformation, which involves an expansion of the initial field profile in the $xy$-plane at $z = 0$ into its plane-wave constituents. In fact, with the aid of the two-dimensional Fourier transform of $G(\boldsymbol{r}, 0)$ given in Eq.(37), it is rather easy to establish the equivalence of the Fourier expansion method with the Rayleigh-Sommerfeld formula in Eq.(25), and also with the related vector formulas in Eqs.(32) and (33). Appendix J outlines the mathematical steps needed to establish these equivalencies.

The Sommerfeld solution to the problem of diffraction from a thin, perfectly conducting half-plane described in Sec.7 is one of the few problems in the EM theory of diffraction for which an exact analytical solution has been found; for a discussion of related problems of this type, see Ref.[1], Chapter 11. Scattering of plane-waves from spherical particles of known relative permittivity $\varepsilon(\omega)$ and permeability $\mu(\omega)$, the so-called Mie scattering, is another problem for which an exact solution (albeit in the form of an infinite series) exists; for a discussion of this and related problems the reader is referred to the vast literature of Mie scattering.[1,4,9-11,22]

In our analysis of neutron scattering from ferromagnets in Sec.10, we used the contact term $-2\boldsymbol{m} \cdot \boldsymbol{m}'\delta(\boldsymbol{r})/3\mu_0$ to account for the interaction energy of the dipole pair $\boldsymbol{m}, \boldsymbol{m}'$ when they happen to overlap at the same location in space. This is tantamount to assuming that the dipolar magnetic moments are produced by circulating electrical currents. The contact term would have been $\boldsymbol{m} \cdot \boldsymbol{m}'\delta(\boldsymbol{r})/3\mu_0$ had each magnetic moment been produced by a pair of equal and opposite magnetic monopoles residing within the corresponding particle. Since the Amperian current loop model has been found to agree most closely with experimental findings, we used the former expression for the contact term in Eq.(118); see Ref. [9], Sec.5.7, and Ref. [29] for a pedagogical discussion of the experimental evidence — from neutron scattering as well as the existence of the famous 21 cm astrophysical spectral line of atomic hydrogen — in favor of the Amperian current loop model of the intrinsic magnetic dipole moments of subatomic particles.

Although we did not discuss the Babinet principle of complementary screens that is well known in classical optics, it is worth mentioning here that a rigorous version of this principle has been proven in Maxwellian electrodynamics.[1,9] The original version of Babinet's principle is based on the Kirchhoff diffraction integral of Eq.(23), and the notion that, if $S_1$ consists of apertures in an opaque screen, then the complement of $S_1$ would be opaque where $S_1$ is transmissive, and transmissive where $S_1$ is opaque. Considering that, in Kirchhoff's approximation, $\psi(\boldsymbol{r})$ and $\partial_n\psi(\boldsymbol{r})$ in Eq.(23) retain the values of the incident beam in the open aperture(s) but vanish in the opaque regions, it is a reasonable conjecture that the observed field in the presence of $S_1$ and that in the presence of $S_1$'s complement would add up to the observed field when all screens are removed — i.e., when the unobstructed beam reaches the observation point. Similar arguments can be based on either of the Rayleigh-Sommerfeld diffraction integrals in Eqs.(24) and (25), provided, of course, that the Kirchhoff approximation remains applicable. Appendix K describes the rigorous version of the Babinet principle and provides a simple proof that relies on symmetry arguments similar to those used in Example 2 of Sec.6.

Finally, to keep the size and scope of this tutorial within reasonable boundaries, we did not broach the important problem of EM scattering from small dielectric spheres, nor that of EM scattering from small perfectly conducting spheres. The interested reader can find a detailed discussion of these problems in Appendices L and M, respectively.

## Appendix A

The integrals in Eqs.(17) and (18) are evaluated with the aid of Euler's beta and gamma functions, $B(x,y)$ and $\Gamma(x)$, as follows:

$$\int_0^\infty 4\pi r^2(r^2+\varepsilon)^{-5/2}\mathrm{d}r = 4\pi\varepsilon^{-5/2}\int_0^\infty r^2(1+\varepsilon^{-1}r^2)^{-5/2}\mathrm{d}r = 2\pi\varepsilon^{-5/2}\varepsilon^{3/2}B\left(\tfrac{3}{2},1\right)$$

$$= 2\pi\varepsilon^{-1}\frac{\Gamma(3/2)\Gamma(1)}{\Gamma(5/2)} = 4\pi/3\varepsilon. \tag{A1}$$

$$\int_0^\infty 4\pi r(r^2+\varepsilon)^{-3/2}\mathrm{d}r = 4\pi\varepsilon^{-3/2}\int_0^\infty r(1+\varepsilon^{-1}r^2)^{-3/2}\mathrm{d}r = 2\pi\varepsilon^{-3/2}\varepsilon B\left(1,\tfrac{1}{2}\right)$$

$$= 2\pi\varepsilon^{-½}\frac{\Gamma(1)\Gamma(½)}{\Gamma(3/2)} = 4\pi/\sqrt{\varepsilon}\cdot \tag{A2}$$

In these derivations, the following identities from Gradshteyn and Ryzhik's *Table of Integrals, Series, and Products*[27] have been used:

$$\int_0^\infty x^{\mu-1}(1+\beta x^p)^{-\nu}\mathrm{d}x = p^{-1}\beta^{-\mu/p}B\left(\frac{\mu}{p},\nu-\frac{\mu}{p}\right);\quad |\arg(\beta)|<\pi,\quad p>0,\quad 0<\mathrm{Re}(\mu)<p\mathrm{Re}(\nu).$$

(G&R[27] **3.251**-11)

$B(x,y) = \Gamma(x)\Gamma(y)/\Gamma(x+y).$ (G&R[27] **8.384**-1)

$\Gamma(x+1) = x\Gamma(x).$ (G&R[27] **8.331**-1)

$\Gamma(1) = \Gamma(2) = 1;\quad \Gamma(½) = \sqrt{\pi}.$ (G&R[27] **8.338**-1,2)

---

## Appendix B

The Green function $G(\boldsymbol{r},\boldsymbol{r}_0)$ is the solution of the inhomogeneous Helmholtz equation:

$$(\nabla^2+k_0^2)G(\boldsymbol{r},\boldsymbol{r}_0) = -4\pi\delta(\boldsymbol{r}-\boldsymbol{r}_0). \tag{B1}$$

Since the function $G$ is expected to be translation invariant, we solve Eq.(B1) for $\boldsymbol{r}_0 = 0$. The function of interest here, namely, the Fourier transform $\tilde{G}(\boldsymbol{k})$ of $G(\boldsymbol{r},0)$, is thus seen to be

$$\tilde{G}(\boldsymbol{k}) = 4\pi/(k^2-k_0^2). \tag{B2}$$

The inverse Fourier transform of $\tilde{G}(\boldsymbol{k})$ can now be found by direct integration, as follows:

$$G(\boldsymbol{r},0) = (2\pi)^{-3}\iiint_{-\infty}^{\infty}\tilde{G}(\boldsymbol{k})e^{\mathrm{i}\boldsymbol{k}\cdot\boldsymbol{r}}\mathrm{d}\boldsymbol{k} = (2\pi)^{-3}\int_{k=0}^{\infty}\int_{\theta=0}^{\pi}\frac{4\pi}{k^2-k_0^2}e^{\mathrm{i}kr\cos\theta}2\pi k^2\sin\theta\,\mathrm{d}k\mathrm{d}\theta$$

$\mathrm{d}\boldsymbol{k}$ stands for $\mathrm{d}k_x\mathrm{d}k_y\mathrm{d}k_z$

$$= \frac{\mathrm{i}}{\pi r}\int_{k=0}^{\infty}\frac{k}{k^2-k_0^2}\left(e^{\mathrm{i}kr\cos\theta}\Big|_{\theta=0}^{\pi}\right)\mathrm{d}k = \frac{\mathrm{i}}{\pi r}\int_0^\infty\frac{k}{k^2-k_0^2}\left(e^{-\mathrm{i}kr}-e^{\mathrm{i}kr}\right)\mathrm{d}k$$

residue calculus

$$= \frac{\mathrm{i}}{2\pi r}\left[\int_{-\infty}^{\infty}\frac{k\exp(-\mathrm{i}kr)}{k^2-k_0^2}\mathrm{d}k - \int_{-\infty}^{\infty}\frac{k\exp(\mathrm{i}kr)}{k^2-k_0^2}\mathrm{d}k\right] = ½\left(\frac{e^{\mathrm{i}k_0r}}{r}+\frac{e^{-\mathrm{i}k_0r}}{r}\right). \tag{B3}$$

It is noteworthy that $G(\boldsymbol{r},0)$ has equal contributions from $\exp(\pm\mathrm{i}k_0r)/r$, even though, in practice, one typically picks only one of these two solutions — either the outgoing spherical wave with the plus sign or the incoming wave with the minus sign. The solution of Eq.(B1), of course, is not unique, since any plane-wave in the form of $A\exp(\mathrm{i}k_0\boldsymbol{\sigma}\cdot\boldsymbol{r})$, where $A$ is an arbitrary constant and $\boldsymbol{\sigma}$ an arbitrary (real-valued) unit-vector, can be added to $G(\boldsymbol{r},0)$ without affecting the solution. A superposition of such plane-waves can shift the balance between $\exp(\pm\mathrm{i}k_0 r)/r$ in Eq.(B3), so that any linear combination of these two functions can act as a perfectly good solution of Eq.(B1). As a matter of fact, $\sin(k_0 r)/r$ is one such homogeneous solution of Eq.(B1), whose linear combination with Eq.(B3) can produce the desired balance.

It is possible to directly compute the Fourier transform of either of the candidate solutions for $G(\boldsymbol{r},0)$, namely, $\exp(\pm\mathrm{i}k_0 r)/r$. We begin by replacing $k_0$ with $k_0-\mathrm{i}\varepsilon$, where $\varepsilon$, a small (positive or negative) number, will eventually be made to approach zero. We find

$$\mathcal{F}\left\{\frac{\exp[\pm\mathrm{i}(k_0-\mathrm{i}\varepsilon)r]}{r}\right\}=\int_{r=0}^{\infty}\int_{\theta=0}^{\pi}\frac{\exp[\pm\mathrm{i}(k_0-\mathrm{i}\varepsilon)r]}{r}e^{-\mathrm{i}kr\cos\theta}2\pi r^2\sin\theta\,\mathrm{d}r\mathrm{d}\theta$$

Fourier operator (arrow pointing to $\mathcal{F}$)

$$=\frac{2\pi}{\mathrm{i}k}\int_{r=0}^{\infty}e^{\pm\mathrm{i}(k_0-\mathrm{i}\varepsilon)r}\left(e^{-\mathrm{i}kr\cos\theta}\Big|_{\theta=0}^{\pi}\right)\mathrm{d}r=\frac{2\pi}{\mathrm{i}k}\int_{r=0}^{\infty}e^{\pm\mathrm{i}(k_0-\mathrm{i}\varepsilon)r}\left(e^{\mathrm{i}kr}-e^{-\mathrm{i}kr}\right)\mathrm{d}r$$

Choose sign of $\varepsilon$ so that $\exp(\pm\varepsilon r)\to 0$ when $r\to\infty$.

$$=\frac{2\pi}{\mathrm{i}k}\int_0^{\infty}e^{\pm\varepsilon r}\{\exp[\mathrm{i}(\pm k_0+k)r]-\exp[\mathrm{i}(\pm k_0-k)r]\}\mathrm{d}r$$

$$=\frac{2\pi}{\mathrm{i}k}\left[-\frac{1}{\pm\varepsilon+\mathrm{i}(\pm k_0+k)}+\frac{1}{\pm\varepsilon+\mathrm{i}(\pm k_0-k)}\right]=\frac{2\pi}{k}\left[\frac{1}{\pm(k_0-\mathrm{i}\varepsilon)+k}-\frac{1}{\pm(k_0-\mathrm{i}\varepsilon)-k}\right]$$

$$=\frac{4\pi}{k^2-(k_0-\mathrm{i}\varepsilon)^2}. \tag{B4}$$

It is thus seen that $\tilde{G}(\boldsymbol{k})=4\pi/(k^2-k_0^2)$ is the Fourier transform of $G(\boldsymbol{r},0)=e^{\pm\mathrm{i}k_0 r}/r$, irrespective of the sign of $k_0$.

---

## Appendix C

Kirchhoff's Eq.(22), applied to the $x$-component of the $E$-field yields Eq.(27), which should similarly be satisfied by the remaining components $E_y, E_z$ of the $E$-field. The vectorial version of the Kirchhoff formula may now be written down for $\boldsymbol{E}(\boldsymbol{r}_0)$, which, upon manipulation by standard vector identities, leads to Eq.(28). A step-by-step procedure is shown below.

$$4\pi\boldsymbol{E}(\boldsymbol{r}_0)=2\int_S(\hat{\boldsymbol{n}}\cdot\boldsymbol{\nabla}G)\boldsymbol{E}\mathrm{d}s+\int_V\boldsymbol{\nabla}^2(G\boldsymbol{E})\mathrm{d}\boldsymbol{r}$$

$\boldsymbol{\nabla}^2\boldsymbol{V}(\boldsymbol{r})=(\nabla^2 V_x)\hat{\boldsymbol{x}}+(\nabla^2 V_y)\hat{\boldsymbol{y}}+(\nabla^2 V_z)\hat{\boldsymbol{z}}$

$\boldsymbol{\nabla}^2\boldsymbol{V}=\boldsymbol{\nabla}(\boldsymbol{\nabla}\cdot\boldsymbol{V})-\boldsymbol{\nabla}\times(\boldsymbol{\nabla}\times\boldsymbol{V})$

$$=2\int_S(\hat{\boldsymbol{n}}\cdot\boldsymbol{\nabla}G)\boldsymbol{E}\mathrm{d}s+\int_V\boldsymbol{\nabla}[\boldsymbol{\nabla}\cdot(G\boldsymbol{E})]\mathrm{d}\boldsymbol{r}-\int_V\boldsymbol{\nabla}\times[\boldsymbol{\nabla}\times(G\boldsymbol{E})]\mathrm{d}\boldsymbol{r}$$

$\hat{\boldsymbol{n}}$ points into the volume $V$ → $\int_V\boldsymbol{\nabla}\phi(\boldsymbol{r})\mathrm{d}\boldsymbol{r}=-\int_S\phi(\boldsymbol{r})\hat{\boldsymbol{n}}\mathrm{d}s$; $\int_V\boldsymbol{\nabla}\times\boldsymbol{V}(\boldsymbol{r})\mathrm{d}\boldsymbol{r}=-\int_S\hat{\boldsymbol{n}}\times\boldsymbol{V}(\boldsymbol{r})\mathrm{d}s$

$$=2\int_S(\hat{\boldsymbol{n}}\cdot\boldsymbol{\nabla}G)\boldsymbol{E}\mathrm{d}s-\int_S[\boldsymbol{\nabla}\cdot(G\boldsymbol{E})]\hat{\boldsymbol{n}}\mathrm{d}s+\int_S\hat{\boldsymbol{n}}\times[\boldsymbol{\nabla}\times(G\boldsymbol{E})]\mathrm{d}s$$

$\boldsymbol{\nabla}\cdot(\phi\boldsymbol{V})=\phi\boldsymbol{\nabla}\cdot\boldsymbol{V}+\boldsymbol{V}\cdot\boldsymbol{\nabla}\phi$; $\boldsymbol{\nabla}\times(\phi\boldsymbol{V})=\boldsymbol{\nabla}\phi\times\boldsymbol{V}+\phi\boldsymbol{\nabla}\times\boldsymbol{V}$

$$=\int_S[2(\hat{\boldsymbol{n}}\cdot\boldsymbol{\nabla}G)\boldsymbol{E}-(G\boldsymbol{\nabla}\cdot\boldsymbol{E}+\boldsymbol{E}\cdot\boldsymbol{\nabla}G)\hat{\boldsymbol{n}}+\hat{\boldsymbol{n}}\times(\boldsymbol{\nabla}G\times\boldsymbol{E}+G\boldsymbol{\nabla}\times\boldsymbol{E})]\mathrm{d}s$$

Maxwell's 1st equation → 0; $\boldsymbol{a}\times(\boldsymbol{b}\times\boldsymbol{c})=(\boldsymbol{a}\cdot\boldsymbol{c})\boldsymbol{b}-(\boldsymbol{a}\cdot\boldsymbol{b})\boldsymbol{c}$; $\mathrm{i}\omega\boldsymbol{B}$

$$=\int_S[(\hat{\boldsymbol{n}}\cdot\boldsymbol{\nabla}G)\boldsymbol{E}-(\boldsymbol{E}\cdot\boldsymbol{\nabla}G)\hat{\boldsymbol{n}}+(\hat{\boldsymbol{n}}\cdot\boldsymbol{E})\boldsymbol{\nabla}G+\mathrm{i}\omega(\hat{\boldsymbol{n}}\times\boldsymbol{B})G]\mathrm{d}s$$

$$=\int_S[(\hat{\boldsymbol{n}}\times\boldsymbol{E})\times\boldsymbol{\nabla}G+(\hat{\boldsymbol{n}}\cdot\boldsymbol{E})\boldsymbol{\nabla}G+\mathrm{i}\omega(\hat{\boldsymbol{n}}\times\boldsymbol{B})G]\mathrm{d}s. \tag{C1}$$

We have thus arrived at a vectorial version of the Kirchhoff formula for the observed $E$-field.

---

## Appendix D

It is possible to derive Eq.(30) directly from Eq.(29), and vice versa. Here, Maxwell's equation $\boldsymbol{\nabla}\times\boldsymbol{B}(\boldsymbol{r}) = -(\mathrm{i}\omega/c^2)\boldsymbol{E}(\boldsymbol{r})$ will be used to show that Eq.(29) is a direct consequence of Eq.(30). Since the integration variable over the closed surface $S$ is $\boldsymbol{r}$, whereas the fields are observed at $\boldsymbol{r}_0$, we denote the differential operator by $\boldsymbol{\nabla}$ when operating on $\boldsymbol{r}$, and by $\boldsymbol{\nabla}_0$ when operating on $\boldsymbol{r}_0$. The symmetry of $G(\boldsymbol{r},\boldsymbol{r}_0)$ then allows us to replace $\boldsymbol{\nabla}_0 G$ with $-\boldsymbol{\nabla}G$. Starting with $\boldsymbol{B}(\boldsymbol{r}_0)$ given in Eq.(30) as an integral over the closed surface $S$, we will have

$$\boldsymbol{E}(\boldsymbol{r}_0) = (\mathrm{i}c^2/\omega)\boldsymbol{\nabla}_0\times\boldsymbol{B}(\boldsymbol{r}_0)$$

$$= (\mathrm{i}c^2/4\pi\omega)\oint_S \boldsymbol{\nabla}_0\times[(\hat{\boldsymbol{n}}\times\boldsymbol{B})\times\boldsymbol{\nabla}G + (\hat{\boldsymbol{n}}\cdot\boldsymbol{B})\boldsymbol{\nabla}G - \mathrm{i}(\omega/c^2)(\hat{\boldsymbol{n}}\times\boldsymbol{E})G]\mathrm{d}s$$

$\boldsymbol{\nabla}\times(\boldsymbol{a}\times\boldsymbol{b}) = \boldsymbol{a}(\boldsymbol{\nabla}\cdot\boldsymbol{b}) - \boldsymbol{b}(\boldsymbol{\nabla}\cdot\boldsymbol{a}) + (\boldsymbol{b}\cdot\boldsymbol{\nabla})\boldsymbol{a} - (\boldsymbol{a}\cdot\boldsymbol{\nabla})\boldsymbol{b}$ $\quad$ $\boldsymbol{\nabla}\times(\psi\boldsymbol{a}) = \boldsymbol{\nabla}\psi\times\boldsymbol{a} + \psi\boldsymbol{\nabla}\times\boldsymbol{a}$

$$= (\mathrm{i}c^2/4\pi\omega)\oint_S\{(\hat{\boldsymbol{n}}\times\boldsymbol{B})(\boldsymbol{\nabla}_0\cdot\boldsymbol{\nabla}G) - [(\hat{\boldsymbol{n}}\times\boldsymbol{B})\cdot\boldsymbol{\nabla}_0]\boldsymbol{\nabla}G + (\hat{\boldsymbol{n}}\cdot\boldsymbol{B})(\boldsymbol{\nabla}_0\times\boldsymbol{\nabla}G)$$

$-\boldsymbol{\nabla}\times\boldsymbol{\nabla}G = 0$

$-\boldsymbol{\nabla}^2 G$

$$-\mathrm{i}(\omega/c^2)(\boldsymbol{\nabla}_0 G)\times(\hat{\boldsymbol{n}}\times\boldsymbol{E})\}\mathrm{d}s$$

$\delta(\boldsymbol{r}-\boldsymbol{r}_0) = 0$ on the surface $S$; $k_0 = \omega/c$

$$= (\mathrm{i}c^2/4\pi\omega)\oint_S\{(\hat{\boldsymbol{n}}\times\boldsymbol{B})[4\pi\delta(\boldsymbol{r}-\boldsymbol{r}_0) + k_0^2 G] + [(\hat{\boldsymbol{n}}\times\boldsymbol{B})\cdot\boldsymbol{\nabla}]\boldsymbol{\nabla}G$$

$$-\mathrm{i}(\omega/c^2)(\hat{\boldsymbol{n}}\times\boldsymbol{E})\times\boldsymbol{\nabla}G\}\mathrm{d}s$$

$$= (4\pi)^{-1}\oint_S\{\mathrm{i}\omega(\hat{\boldsymbol{n}}\times\boldsymbol{B})G + (\mathrm{i}c^2/\omega)[(\hat{\boldsymbol{n}}\times\boldsymbol{B})\cdot\boldsymbol{\nabla}]\boldsymbol{\nabla}G + (\hat{\boldsymbol{n}}\times\boldsymbol{E})\times\boldsymbol{\nabla}G\}\mathrm{d}s. \qquad \text{(D1)}$$

The middle term in the above integrand can be expanded, as follows:

$$[(\hat{\boldsymbol{n}}\times\boldsymbol{B})\cdot\boldsymbol{\nabla}]\boldsymbol{\nabla}G = [(n_yB_z - n_zB_y)\partial_x + (n_zB_x - n_xB_z)\partial_y + (n_xB_y - n_yB_x)\partial_z]\boldsymbol{\nabla}G$$

$$= [n_x(B_y\partial_z - B_z\partial_y) + n_y(B_z\partial_x - B_x\partial_z) + n_z(B_x\partial_y - B_y\partial_x)]\boldsymbol{\nabla}G$$

$$= n_x[\partial_z(B_y\boldsymbol{\nabla}G) - \partial_y(B_z\boldsymbol{\nabla}G) + (\partial_yB_z - \partial_zB_y)\boldsymbol{\nabla}G]$$

$$+n_y[\partial_x(B_z\boldsymbol{\nabla}G) - \partial_z(B_x\boldsymbol{\nabla}G) + (\partial_zB_x - \partial_xB_z)\boldsymbol{\nabla}G]$$

$$+n_z[\partial_y(B_x\boldsymbol{\nabla}G) - \partial_x(B_y\boldsymbol{\nabla}G) + (\partial_xB_y - \partial_yB_x)\boldsymbol{\nabla}G]$$

$$= [\hat{\boldsymbol{n}}\cdot\boldsymbol{\nabla}\times(\boldsymbol{B}\partial_xG)]\hat{\boldsymbol{x}} + [\hat{\boldsymbol{n}}\cdot\boldsymbol{\nabla}\times(\boldsymbol{B}\partial_yG)]\hat{\boldsymbol{y}} + [\hat{\boldsymbol{n}}\cdot\boldsymbol{\nabla}\times(\boldsymbol{B}\partial_zG)]\hat{\boldsymbol{z}}$$

$$+(\hat{\boldsymbol{n}}\cdot\boldsymbol{\nabla}\times\boldsymbol{B})\boldsymbol{\nabla}G. \qquad \text{(D2)}$$

Gauss's theorem now allows us to replace the surface integral of $\hat{\boldsymbol{n}}\cdot\boldsymbol{\nabla}\times(\boldsymbol{B}\partial_xG)$ with the volume integral of $\boldsymbol{\nabla}\cdot[\boldsymbol{\nabla}\times(\boldsymbol{B}\partial_xG)]$, which is zero since the divergence of the curl of any vector field is identically zero. By the same token, the surface integrals of the 2nd and 3rd terms on the right-hand side of Eq.(D2) vanish. Substituting the remaining term into Eq.(D1) and recalling that $(\mathrm{i}c^2/\omega)\boldsymbol{\nabla}\times\boldsymbol{B} = \boldsymbol{E}$, we arrive at

$$\boldsymbol{E}(\boldsymbol{r}_0) = (4\pi)^{-1}\oint_S[\mathrm{i}\omega(\hat{\boldsymbol{n}}\times\boldsymbol{B})G + (\hat{\boldsymbol{n}}\cdot\boldsymbol{E})\boldsymbol{\nabla}G + (\hat{\boldsymbol{n}}\times\boldsymbol{E})\times\boldsymbol{\nabla}G]\mathrm{d}s. \qquad \text{(D3)}$$

This is the same expression for $\boldsymbol{E}(\boldsymbol{r}_0)$ as that in Eq.(28). The proof will be complete when the contribution of the remote spherical (or hemi-spherical) surface $S_2$ to the overall integral is recognized as negligible. Consequently, the integral over the closed surface $S = S_1 + S_2$ in Eq.(D3) is equivalent to the integral over $S_1$ in Eq.(29).

## Appendix E

Evaluating the 2D Fourier transform relation of Eq.(37) requires the following identities:

$$\int_0^\infty \frac{\sin\left(a\sqrt{x^2+b^2}\right)\cos(cx)}{\sqrt{x^2+b^2}}\mathrm{d}x = \begin{cases} \tfrac{1}{2}\pi J_0\left(b\sqrt{a^2-c^2}\right), & 0<c<a \\ 0, & 0<a<c \end{cases} \quad (b>0). \quad \text{(G\&R[27] 3.876-1)}$$

$$\int_0^\infty \frac{\cos\left(a\sqrt{x^2+b^2}\right)\cos(cx)}{\sqrt{x^2+b^2}}\mathrm{d}x = \begin{cases} -\tfrac{1}{2}\pi Y_0\left(b\sqrt{a^2-c^2}\right), & 0<c<a \\ K_0\left(b\sqrt{c^2-a^2}\right), & 0<a<c \end{cases} \quad (b>0). \quad \text{(G\&R[27] 3.876-2)}$$

$$\int_0^\infty J_0\left(\alpha\sqrt{x^2+z^2}\right)\cos(\beta x)\,\mathrm{d}x = \begin{cases} \dfrac{\cos\left(z\sqrt{\alpha^2-\beta^2}\right)}{\sqrt{\alpha^2-\beta^2}}, & 0<\beta<\alpha \\ 0, & 0<\alpha<\beta \end{cases} \quad (z>0). \quad \text{(G\&R[27] 6.677-3)}$$

$$\int_0^\infty Y_0\left(\alpha\sqrt{x^2+z^2}\right)\cos(\beta x)\,\mathrm{d}x = \begin{cases} \dfrac{\sin\left(z\sqrt{\alpha^2-\beta^2}\right)}{\sqrt{\alpha^2-\beta^2}}, & 0<\beta<\alpha \\ -\dfrac{\exp\left(-z\sqrt{\beta^2-\alpha^2}\right)}{\sqrt{\beta^2-\alpha^2}}, & 0<\alpha<\beta \end{cases} \quad (z>0). \quad \text{(G\&R[27] 6.677-4)}$$

$$\int_0^\infty K_0\left(\alpha\sqrt{x^2+z^2}\right)\cos(\beta x)\,\mathrm{d}x = \frac{\pi\exp\left(-z\sqrt{\alpha^2+\beta^2}\right)}{2\sqrt{\alpha^2+\beta^2}}, \; [z>0,\; \mathrm{Re}(\alpha)>0,\; \beta>0]. \quad \text{(G\&R[27] 6.677-5)}$$

Here, $J_0(\cdot)$ is the Bessel function of first kind, order zero, $Y_0(\cdot)$ is the Bessel function of second kind, order zero, and $K_0(\cdot)$ is the modified Bessel function of imaginary argument, order zero, defined as $K_0(z) = \tfrac{1}{2}\mathrm{i}\pi[J_0(\mathrm{i}z) + \mathrm{i}Y_0(\mathrm{i}z)]$. In cases where $k_z = \sqrt{k_0^2 - k_x^2 - k_y^2}$ is real-valued, evaluating the Fourier integral yields

$$\iint_{-\infty}^{\infty} \frac{\exp\left(\mathrm{i}k_0\sqrt{x^2+y^2+z^2}\right)}{\sqrt{x^2+y^2+z^2}} e^{\mathrm{i}(k_x x + k_y y)}\mathrm{d}x\mathrm{d}y$$

(odd function of $y$: $\mathrm{i}\sin(k_y y)$)

$$= \int_{x=-\infty}^{\infty} e^{\mathrm{i}k_x x} \int_{y=-\infty}^{\infty} \frac{\left[\cos\left(k_0\sqrt{x^2+y^2+z^2}\right) + \mathrm{i}\sin\left(k_0\sqrt{x^2+y^2+z^2}\right)\right] \times \left[\cos(k_y y) + \mathrm{i}\sin(k_y y)\right]}{\sqrt{x^2+y^2+z^2}}\mathrm{d}y\mathrm{d}x$$

(odd function of $x$: $\mathrm{i}\sin(k_x x)$)

$$= \mathrm{i}\pi \int_{-\infty}^{\infty} [\cos(k_x x) + \mathrm{i}\sin(k_x x)] \times \left[J_0\left(\sqrt{k_0^2 - k_y^2}\,\sqrt{x^2+z^2}\right) + \mathrm{i}Y_0\left(\sqrt{k_0^2 - k_y^2}\,\sqrt{x^2+z^2}\right)\right]\mathrm{d}x$$

$$= \mathrm{i}2\pi \exp\left(\mathrm{i}z\sqrt{k_0^2 - k_x^2 - k_y^2}\right) \Big/ \sqrt{k_0^2 - k_x^2 - k_y^2} = \mathrm{i}2\pi\exp(\mathrm{i}k_z z)/k_z. \quad \text{(E1)}$$

If $k_z$ happens to be imaginary, either because $k_x > k_0$, or $k_y > k_0$ or $k_x^2 + k_y^2 > k_0^2$, the intermediate steps leading to the final result in Eq.(E1) could be different, but the final result will be the same. The Fourier integral may now be written $2\pi\exp\left(-z\sqrt{k_x^2 + k_y^2 - k_0^2}\right)\Big/\sqrt{k_x^2 + k_y^2 - k_0^2}$.

## Appendix F

The Cornu spiral, a graph of the complex Fresnel integral $\int_0^x e^{i\pi\zeta^2/2}d\zeta = C(x) + iS(x)$ in the complex plane, is shown in Fig.F1. Note that the Fresnel integral as well as its real and imaginary parts, $C(x)$ and $S(x)$, are defined differently here than in Eq.(46). Whereas Eq.(46) complies with the standard definition of these functions in accordance with Gradshteyn and Ryzhik's *Table of Integral, Series, and Products*,[27] the definitions adopted here are more convenient for describing the spiral. The center of the spiral corresponds to $x = 0$, with positive values of $x$ represented on the right arm, and negative values on the left arm of the spiral.

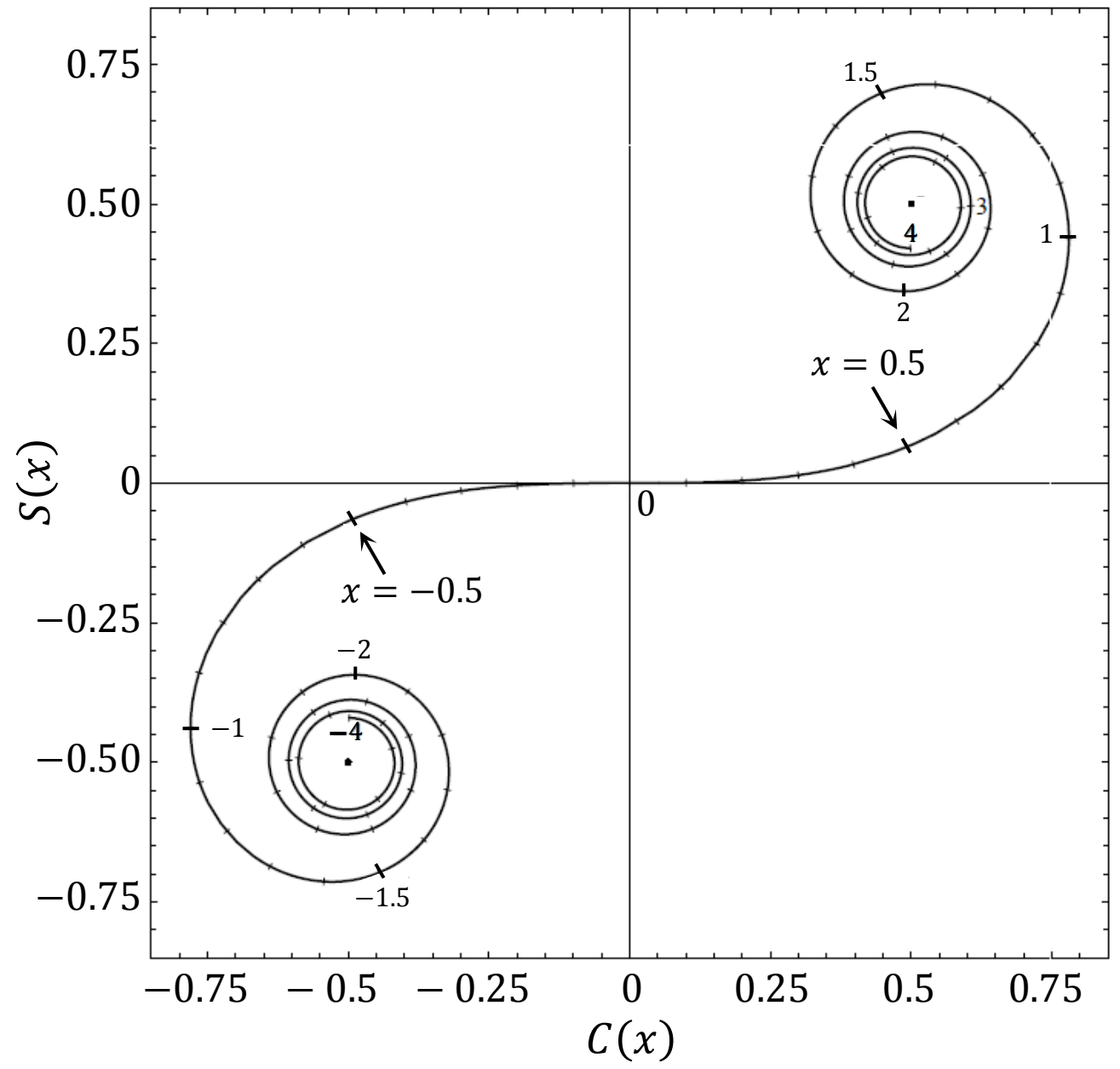

**Fig. F1**. The Cornu spiral is a complex-plane graph of the Fresnel integral $\int_0^x e^{i\pi\zeta^2/2}d\zeta = C(x) + iS(x)$.

The derivative with respect to $x$ of the complex Fresnel integral is $e^{i\pi x^2/2}$, indicating that, along the spiral, the points corresponding to $x$ and $x + dx$ are connected by an infinitesimal arrow of length $dx$ and orientation angle $\pi x^2/2$. (As always, the angle in the complex plane is measured counterclockwise from the positive real axis.) Thus, at $x = 0$, the tangent to the spiral is parallel to the real axis; at $x = 1$, where the angle of the little arrow is $\pi/2$, the tangent is parallel to the imaginary axis; at $x = \sqrt{2}$, the angle is $\pi$ and the tangent is anti-parallel to the real axis; at $x = 2$, the angle is $2\pi$ and the tangent is, once again, parallel to the real axis. Figure F1 shows a few more points corresponding to other values of $x$ along the length of the spiral. As $x \to \infty$, the spiral approaches its limit point of ½ + ½i, which is the value of $\int_0^\infty e^{i\pi\zeta^2/2}d\zeta$ obtained by integrating in the complex $\zeta$-plane along the 45° line. An asymptotic series for the Fresnel integral can be found by a repeated application of integration-by-parts, as follows:

$$\int_0^x e^{i\pi\zeta^2/2}d\zeta = \int_0^\infty e^{i\pi\zeta^2/2}d\zeta - \int_x^\infty e^{i\pi\zeta^2/2}d\zeta = \frac{\exp(i\pi/4)}{\sqrt{2}} - \int_x^\infty (i\pi\zeta)^{-1}(i\pi\zeta e^{i\pi\zeta^2/2})d\zeta$$

$$= \frac{\exp(i\pi/4)}{\sqrt{2}} + \frac{\exp(i\pi x^2/2)}{i\pi x} - i\pi\int_x^\infty (i\pi\zeta)^{-2}e^{i\pi\zeta^2/2}d\zeta$$

$$= \frac{\exp(i\pi/4)}{\sqrt{2}} + \frac{\exp(i\pi x^2/2)}{i\pi x} - \frac{\exp(i\pi x^2/2)}{\pi^2 x^3} + 3\pi^2\int_x^\infty (i\pi\zeta)^{-4}e^{i\pi\zeta^2/2}d\zeta. \quad \text{(F1)}$$

**Interpreting the results of diffraction calculations from a half-plane**. The Fresnel function $F(\zeta)=\int_\zeta^\infty \exp(\mathrm{i}x^2)\,\mathrm{d}x$ is related to the Cornu spiral via Eq.(46). The spiral depicted in Fig.F1 represents the function $\int_0^x \exp(\mathrm{i}\pi\zeta^2/2)\,\mathrm{d}\zeta = \sqrt{2/\pi}\int_0^{\sqrt{\pi/2}x}\exp(\mathrm{i}y^2)\,\mathrm{d}y$; consequently,

$$F(\zeta)=\sqrt{\pi/4}\,e^{\mathrm{i}\pi/4}-\sqrt{\pi/2}\,\mathrm{Cornu}(\sqrt{2/\pi}\,\zeta)=\sqrt{\pi/2}\,[(½+½\mathrm{i})-\mathrm{Cornu}(\sqrt{2/\pi}\,\zeta)]. \quad \text{(F2)}$$

The complex number representing $F(\zeta)$ is thus seen to be $\sqrt{\pi/2}$ times the arrow that goes from the point $x=\sqrt{2/\pi}\,\zeta$ on the Cornu spiral to the spiral's eye at ½ + ½i. Starting at $\zeta=0$, where $F(\zeta)=\sqrt{\pi/4}\,e^{\mathrm{i}\pi/4}$, when $\zeta$ rises from 0 toward $\infty$, the magnitude of $F(\zeta)$ shrinks rapidly, while its phase cycles between 0 and $2\pi$ faster and faster. In contrast, when $\zeta$ goes from 0 toward $-\infty$, the magnitude of $F(\zeta)$ quickly grows toward the limit value of $\sqrt{\pi}e^{\mathrm{i}\pi/4}$, albeit with some oscillations along the way, whereas the phase hovers pretty much around $\pi/4$.

We can now see how the intensity plot in Fig.3 comes about. In accordance with Eq.(49), the observed intensity is the squared magnitude of $F[\sqrt{\pi/(\lambda_0 z_0)}\,(x_0-z_0\sin\theta_{\mathrm{inc}})]$. As $x_0$ rises above $z_0\sin\theta_{\mathrm{inc}}$ into the shadow region, the magnitude of $F(\cdot)$ quickly drops from $½\sqrt{\pi}$ to zero. However, when $x_0$ goes below $z_0\sin\theta_{\mathrm{inc}}$, the magnitude of $F(\cdot)$ rises rapidly — and with rapid oscillations — toward the limiting value of $\sqrt{\pi}$.

One can similarly interpret Sommerfeld's famous result in Eq.(81) pertaining to diffraction from a thin, perfectly conducting half-plane. The plane-wave $e^{ik_0 r\cos(\theta-\theta_0)}$ appearing in Eq.(81) is the incident plane-wave. In the system depicted in Fig.5, consider a circle in the $xz$-plane centered at the origin of coordinates, whose radius $r$ is large enough to make $\sqrt{2k_0 r}$ a fairly large number. If the observation point $\boldsymbol{r}=(r,\theta)$ is located in the shadow region, where $0\le\theta<\theta_0$, the argument of $F(-\sqrt{2k_0 r}\sin[(\theta-\theta_0)/2])$ will be large and positive, making the amplitude of the incident plane-wave in the shadow region exceedingly small. Near the edge of the shadow, where $\theta\cong\theta_0$, the magnitude of $F(\cdot)$ is $\sim ½\sqrt{\pi}$, and the shadow begins to recede. In the remaining part of the circle of radius $r$, where $\theta_0<\theta\le 2\pi$, the argument of $F(\cdot)$ will be large and negative, indicating the presence of the incident plane-wave at essentially its full strength.

The plane-wave $e^{\mathrm{i}k_0 r\cos(\theta+\theta_0)}$, which is also present in Eq.(81), corresponds to the reflected wave at the front facet of the perfectly conducting mirror — because the angle of the reflected $k$-vector with the $x$-axis is $-\theta_0$. Now, on the aforementioned circle of radius $r$, the argument of the coefficient $F(\sqrt{2k_0 r}\sin[(\theta+\theta_0)/2])$ of the reflected wave is large and positive in the region where $0\le\theta<2\pi-\theta_0$. Therefore, in this region, the contribution of the reflected wave to the overall observed $E$-field is negligible. The situation changes drastically as $\theta$ approaches $2\pi-\theta_0$, at which point the magnitude of $F(\cdot)$ climbs to $½\sqrt{\pi}$, and then continues to grow — with characteristic oscillations — as $\theta$ goes from $2\pi-\theta_0$ to $2\pi$.

Note that at $\theta=0$ and $2\pi$, irrespective of the distance $r$ from the edge, the two plane-waves in Eq.(81), being exactly equal and opposite, cancel out, as required by the boundary condition at the front and back facets of the mirror. The incident and reflected waves interfere in the region $2\pi-\theta_0\lesssim\theta\le 2\pi$, where both waves have a strong presence. The oscillations occur near the edge of the shadow, where $\theta\cong\theta_0$, and also near the edge of the reflected beam, $\theta\cong 2\pi-\theta_0$, where the argument of the coefficients $F(\cdot)$ transitions from positive to negative values.

---

## Appendix G

We evaluate the integral that appears in Eq.(76). In what follows, $\lambda$ is real and positive, whereas $\eta$, while real, may be positive or negative. We use the fact that $\int_0^\infty e^{-\lambda\zeta^2}\mathrm{d}\zeta = ½\sqrt{\pi/\lambda}$.

$$\int_0^\infty \frac{\exp(-\lambda\zeta^2)}{\zeta^2 - \mathrm{i}\eta^2}\mathrm{d}\zeta = e^{-\mathrm{i}\lambda\eta^2}\int_0^\infty \frac{\exp[-\lambda(\zeta^2 - \mathrm{i}\eta^2)]}{\zeta^2 - \mathrm{i}\eta^2}\mathrm{d}\zeta = -e^{-\mathrm{i}\lambda\eta^2}\int_\lambda^\infty \left\{\frac{\mathrm{d}}{\mathrm{d}\lambda}\int_{\zeta=0}^\infty \frac{\exp[-\lambda(\zeta^2 - \mathrm{i}\eta^2)]}{\zeta^2 - \mathrm{i}\eta^2}\mathrm{d}\zeta\right\}\mathrm{d}\lambda$$

$$= e^{-\mathrm{i}\lambda\eta^2}\int_\lambda^\infty \left[\int_{\zeta=0}^\infty e^{-\lambda(\zeta^2-\mathrm{i}\eta^2)}\mathrm{d}\zeta\right]\mathrm{d}\lambda = e^{-\mathrm{i}\lambda\eta^2}\int_\lambda^\infty e^{\mathrm{i}\lambda\eta^2}\left(\int_0^\infty e^{-\lambda\zeta^2}\mathrm{d}\zeta\right)\mathrm{d}\lambda$$

$$= ½\sqrt{\pi}e^{-\mathrm{i}\lambda\eta^2}\int_\lambda^\infty \lambda^{-½}e^{\mathrm{i}\lambda\eta^2}\mathrm{d}\lambda = \sqrt{\pi}|\eta|^{-1}e^{-\mathrm{i}\lambda\eta^2}\int_{|\eta|\sqrt{\lambda}}^\infty e^{\mathrm{i}x^2}\mathrm{d}x \leftarrow \boxed{x = |\eta|\sqrt{\lambda}}$$

$$= \sqrt{\pi}|\eta|^{-1}e^{-\mathrm{i}\lambda\eta^2}F(|\eta|\sqrt{\lambda}). \tag{G1}$$

In this equation, $F(\alpha) = \int_\alpha^\infty \exp(\mathrm{i}x^2)\,\mathrm{d}x$ is the complex Fresnel integral as defined in Eq.(46).

---

## Appendix H

In the case of $-\pi \le \theta \le 0$, the scattered $E$-field can be computed directly using the integration path shown in Fig. H1. The blue contour corresponds to the path shown in Fig.6(b), where $\cos\varphi = \sigma_x$ goes from $-\infty$ to $\infty$. Note that $\sigma_z = \sqrt{1-\sigma_x^2} = \sin(\varphi' + \mathrm{i}\varphi'')$ is negative imaginary on the vertical legs of the contour, and negative real on its horizontal leg. This is a necessary condition because $z$ is now negative whereas $\sigma_z z$ must remain positive. The pole at $\sigma_x = \cos\theta_0$ now appears at $\varphi = -\theta_0$, and the small semi-circle around the pole in the $\sigma_x$-plane now becomes a small dip in the vicinity of $\varphi = -\theta_0$. The remaining factor in Eq.(68) is $\sqrt{1+\sigma_x}$, which retains its identity when replaced by $\cos(\varphi/2)$ on the blue trajectory of Fig. H1. The saddle point is at $\varphi = \theta$, and the pole at $\varphi = -\theta_0$ needs to be accounted for only when $\theta > -\theta_0$.

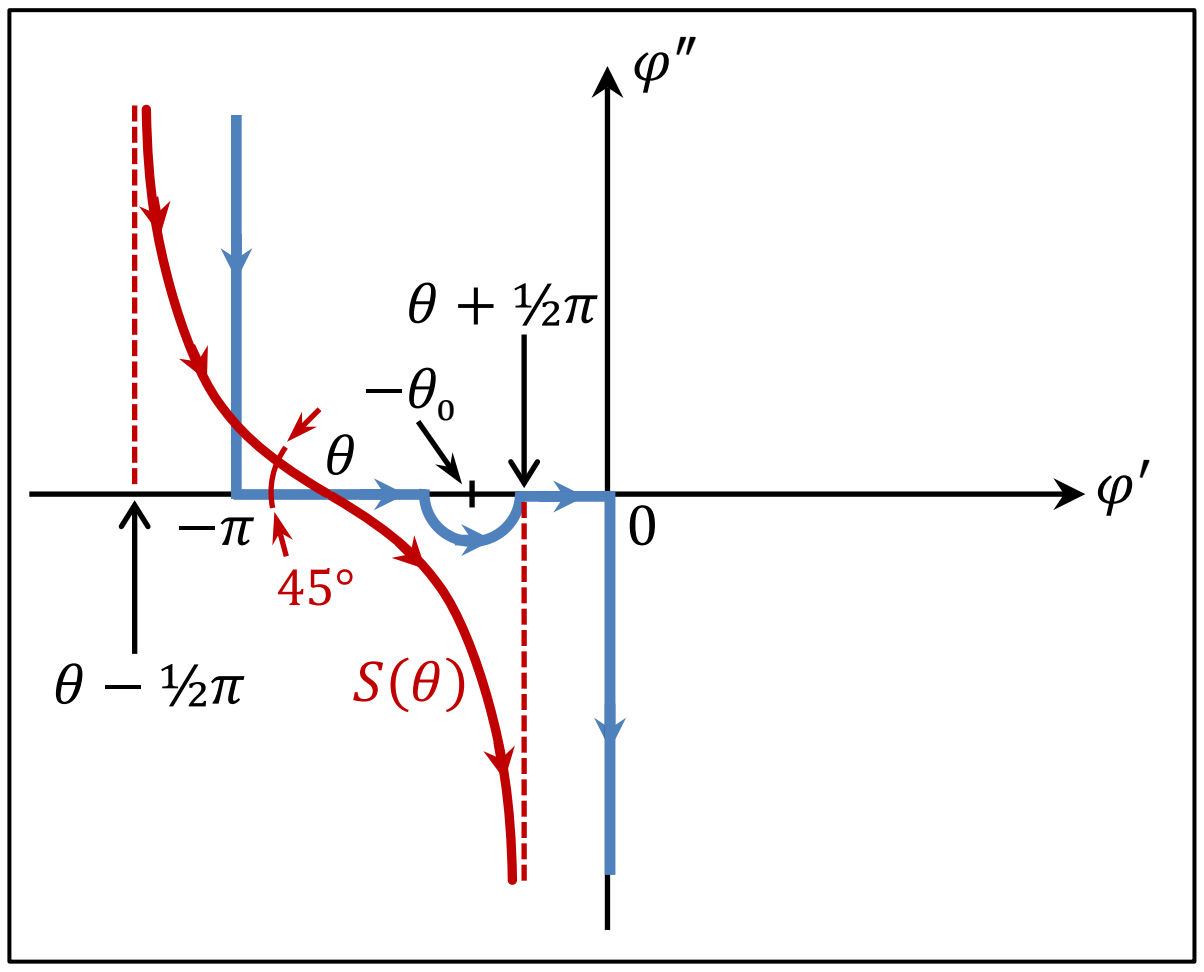

**Fig. H1**. Shown in blue is the integration path in the complex $\varphi$-plane for the range $-\pi \le \theta \le 0$. The corresponding steepest-descent contour $S(\theta)$ is shown in red. The small dip around $\varphi = -\theta_0$ corresponds to the small semi-circle around $\sigma_x = \cos\theta_0$ in the contour depicted in Fig.6(b).

---

## Appendix I

We evaluate the integral in Eq.(121) which, upon straightforward algebraic manipulation, leads to

$$f(\boldsymbol{q}) = \frac{m_\mathrm{n}}{8\pi^2\mu_0\hbar^2}\Big\{\boldsymbol{\mu}_\mathrm{n}\cdot\widetilde{\boldsymbol{M}}(\boldsymbol{q}/\hbar)\Big[(8\pi/3) - \int_{-\infty}^{\infty} r^{-3}\exp(\mathrm{i}\boldsymbol{q}\cdot\boldsymbol{r}/\hbar)\,\mathrm{d}\boldsymbol{r}\Big]$$

$$+3\int_{-\infty}^{\infty}(\boldsymbol{\mu}_\mathrm{n}\cdot\hat{\boldsymbol{r}})\big[\widetilde{\boldsymbol{M}}(\boldsymbol{q}/\hbar)\cdot\hat{\boldsymbol{r}}\big]r^{-3}\exp(\mathrm{i}\boldsymbol{q}\cdot\boldsymbol{r}/\hbar)\,\mathrm{d}\boldsymbol{r}\Big\}. \tag{I1}$$

With reference to Fig.I1, the first integral appearing in Eq.(I1) is readily evaluated as follows:

$$\int_{-\infty}^{\infty} r^{-3}\exp(\mathrm{i}\boldsymbol{q}\cdot\boldsymbol{r}/\hbar)\,\mathrm{d}^3r = \int_{r=0}^{\infty}\int_{\theta=0}^{\pi} r^{-3}\exp(\mathrm{i}qr\cos\theta/\hbar)\,(2\pi r^2\sin\theta)\mathrm{d}\theta\mathrm{d}r$$

$$= 2\pi\int_{r=0}^{\infty}\frac{\exp(\mathrm{i}qr/\hbar)-\exp(-\mathrm{i}qr/\hbar)}{\mathrm{i}qr^2/\hbar}\mathrm{d}r = 4\pi\int_0^{\infty}\frac{\sin(qr/\hbar)}{qr^2/\hbar}\mathrm{d}r$$

$$= 4\pi\int_0^{\infty}(\sin x/x^2)\mathrm{d}x = 4\pi\Big[-(\sin x/x)\big|_0^{\infty} + \int_0^{\infty}(\cos x/x)\mathrm{d}x\Big]$$

$$= 4\pi\Big[1 + \int_0^{\infty}(\cos x/x)\mathrm{d}x\Big]. \tag{I2}$$

The divergent term will eventually cancel out.

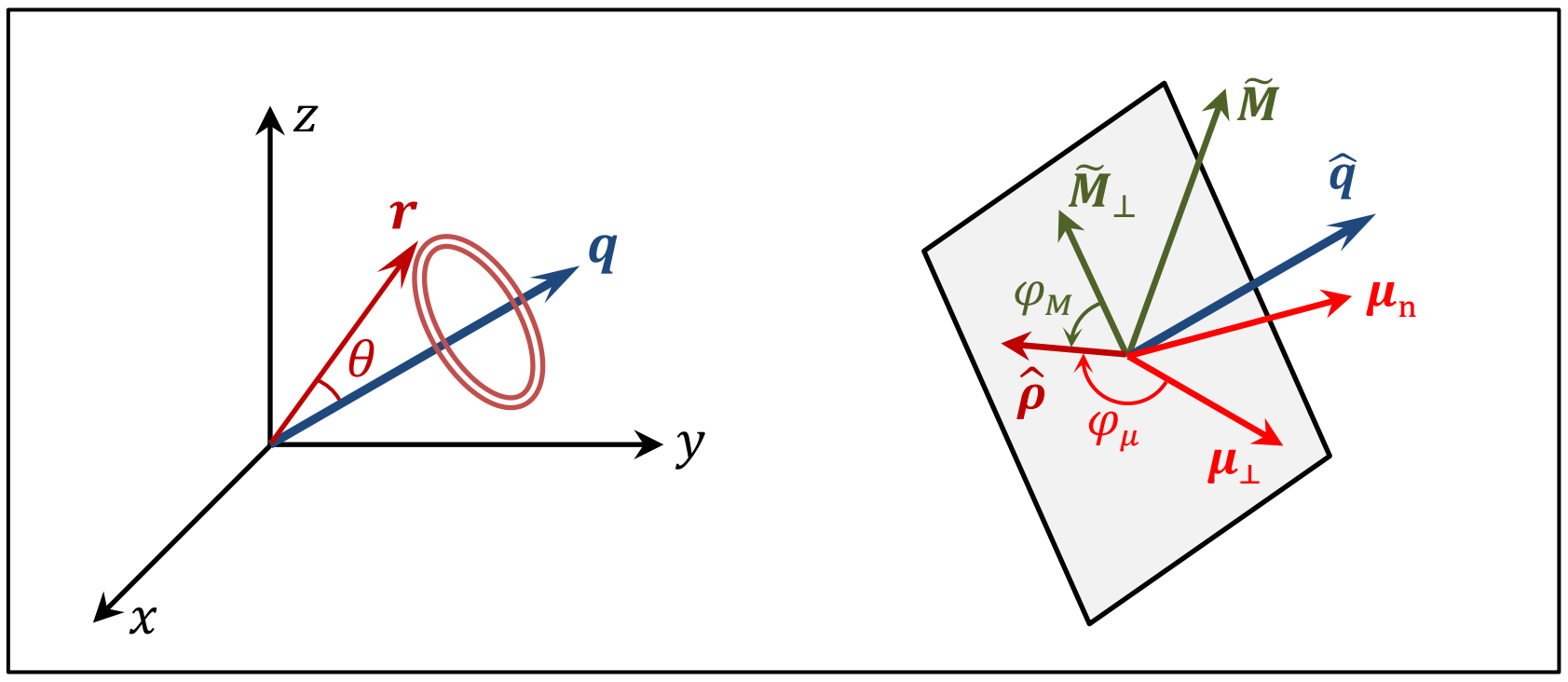

**Fig.I1**. In the 3-dimensional Cartesian space of $\boldsymbol{r} = x\hat{\boldsymbol{x}} + y\hat{\boldsymbol{y}} + z\hat{\boldsymbol{z}}$, the integrands in Eq.(I1) have certain symmetries around the frequency vector $\boldsymbol{q}$. The angle between $\boldsymbol{q}$ and $\boldsymbol{r}$ is denoted by $\theta$, while the azimuthal angle around the $q$-axis is $\varphi$ (not shown). The vectors $\boldsymbol{\mu}_\mathrm{n}$ and $\widetilde{\boldsymbol{M}}$ have projections $\boldsymbol{\mu}_\perp$ and $\widetilde{\boldsymbol{M}}_\perp$ in the plane perpendicular to the unit-vector $\hat{\boldsymbol{q}}$ (associated with the spatial frequency $\boldsymbol{q}$). The projection of $\boldsymbol{r}$ onto the plane perpendicular to $\hat{\boldsymbol{q}}$ is along the unit-vector $\hat{\boldsymbol{\rho}}$ .

As for the remaining integral, we write $\boldsymbol{\mu}_\mathrm{n}$, $\widetilde{\boldsymbol{M}}(\boldsymbol{q}/\hbar)$, $\hat{\boldsymbol{r}}$ as the sum of their projections along the unit-vector $\hat{\boldsymbol{q}}$ and onto the plane orthogonal to $\hat{\boldsymbol{q}}$. Thus, $\boldsymbol{\mu}_\mathrm{n} = (\boldsymbol{\mu}_\mathrm{n}\cdot\hat{\boldsymbol{q}})\hat{\boldsymbol{q}} + \boldsymbol{\mu}_\perp$, and $\widetilde{\boldsymbol{M}} = (\widetilde{\boldsymbol{M}}\cdot\hat{\boldsymbol{q}})\hat{\boldsymbol{q}} + \widetilde{\boldsymbol{M}}_\perp$, while $\hat{\boldsymbol{r}} = (\cos\theta)\hat{\boldsymbol{q}} + (\sin\theta)\hat{\boldsymbol{\rho}}$, where $\hat{\boldsymbol{\rho}}$, the unit-vector along the projection of $\hat{\boldsymbol{r}}$ in the plane orthogonal to $\hat{\boldsymbol{q}}$, makes the angles $\varphi_\mu$ and $\varphi_M$ with $\boldsymbol{\mu}_\perp$ and $\widetilde{\boldsymbol{M}}_\perp$. We will have

$$(\boldsymbol{\mu}_\mathrm{n}\cdot\hat{\boldsymbol{r}})(\widetilde{\boldsymbol{M}}\cdot\hat{\boldsymbol{r}}) = (\boldsymbol{\mu}_\mathrm{n}\cdot\hat{\boldsymbol{q}}\cos\theta + \mu_\perp\sin\theta\cos\varphi_\mu)(\widetilde{\boldsymbol{M}}\cdot\hat{\boldsymbol{q}}\cos\theta + \widetilde{M}_\perp\sin\theta\cos\varphi_M)$$

$$= (\boldsymbol{\mu}_\mathrm{n}\cdot\hat{\boldsymbol{q}})(\widetilde{\boldsymbol{M}}\cdot\hat{\boldsymbol{q}})\cos^2\theta + (\boldsymbol{\mu}_\mathrm{n}\cdot\hat{\boldsymbol{q}})\widetilde{M}_\perp\sin\theta\cos\theta\cos\varphi_M$$

$$+(\widetilde{\boldsymbol{M}}\cdot\hat{\boldsymbol{q}})\mu_\perp\sin\theta\cos\theta\cos\varphi_\mu + ½\mu_\perp\widetilde{M}_\perp\sin^2\theta\cos(\varphi_\mu+\varphi_M)$$

$$+½\mu_\perp\widetilde{M}_\perp\sin^2\theta\cos(\varphi_\mu-\varphi_M). \tag{I3}$$

Integration around the azimuthal direction $\varphi$ causes the 2nd, 3rd and 4th terms (colored blue) on the right-hand side of Eq.(I3) to vanish. The remaining terms now yield

$$\int_{-\infty}^{\infty}(\boldsymbol{\mu}_{\mathrm{n}}\cdot\hat{\boldsymbol{r}})[\widetilde{\boldsymbol{M}}(\boldsymbol{q}/\hbar)\cdot\hat{\boldsymbol{r}}]r^{-3}\exp(\mathrm{i}\boldsymbol{q}\cdot\boldsymbol{r}/\hbar)\,\mathrm{d}\boldsymbol{r}$$

$$=(\boldsymbol{\mu}_{\mathrm{n}}\cdot\hat{\boldsymbol{q}})(\widetilde{\boldsymbol{M}}\cdot\hat{\boldsymbol{q}})\int_{r=0}^{\infty}r^{-3}\int_{\theta=0}^{\pi}\cos^2\theta\exp(\mathrm{i}qr\cos\theta/\hbar)\,(2\pi r^2\sin\theta)\mathrm{d}\theta\mathrm{d}r$$

$$+½(\boldsymbol{\mu}_{\perp}\cdot\widetilde{\boldsymbol{M}}_{\perp})\int_{r=0}^{\infty}r^{-3}\int_{\theta=0}^{\pi}\sin^2\theta\exp(\mathrm{i}qr\cos\theta/\hbar)\,(2\pi r^2\sin\theta)\mathrm{d}\theta\mathrm{d}r$$

$$=2\pi[(\boldsymbol{\mu}_{\mathrm{n}}\cdot\hat{\boldsymbol{q}})(\widetilde{\boldsymbol{M}}\cdot\hat{\boldsymbol{q}})-½(\boldsymbol{\mu}_{\perp}\cdot\widetilde{\boldsymbol{M}}_{\perp})]\int_{r=0}^{\infty}r^{-1}\int_{\theta=0}^{\pi}\sin\theta\cos^2\theta\exp(\mathrm{i}qr\cos\theta/\hbar)\,\mathrm{d}\theta\mathrm{d}r$$

$$+2\pi(\boldsymbol{\mu}_{\perp}\cdot\widetilde{\boldsymbol{M}}_{\perp})\left[1+\int_0^{\infty}(\cos x/x)\mathrm{d}x\right]. \tag{I4}$$

← The divergent term will eventually cancel out.

Next, we evaluate the 2D integral on the right-hand side of Eq.(I4), as follows:

$$\int_{r=0}^{\infty}r^{-1}\int_{\theta=0}^{\pi}\sin\theta\cos^2\theta\exp(\mathrm{i}qr\cos\theta/\hbar)\,\mathrm{d}\theta\mathrm{d}r=\int_{r=0}^{\infty}r^{-1}\left(\frac{\mathrm{i}\hbar}{qr}\right)\Big[\cos^2\theta\exp(\mathrm{i}qr\cos\theta/\hbar)\big|_{\theta=0}^{\pi}$$

$$+2\int_{\theta=0}^{\pi}\sin\theta\cos\theta\exp(\mathrm{i}qr\cos\theta/\hbar)\,\mathrm{d}\theta\Big]\mathrm{d}r$$

$$=\int_{r=0}^{\infty}\left\{\frac{2\hbar\sin(qr/\hbar)}{qr^2}+\left(\frac{2\mathrm{i}\hbar}{qr^2}\right)\left(\frac{\mathrm{i}\hbar}{qr}\right)\left[\cos\theta\exp(\mathrm{i}qr\cos\theta/\hbar)\big|_{\theta=0}^{\pi}+\int_0^{\pi}\sin\theta\exp(\mathrm{i}qr\cos\theta/\hbar)\,\mathrm{d}\theta\right]\right\}\mathrm{d}r$$

$$=\int_0^{\infty}\left\{\frac{2\hbar\sin(qr/\hbar)}{qr^2}+\frac{2\hbar^2}{q^2r^3}\left[2\cos(qr/\hbar)-\frac{2\hbar\sin(qr/\hbar)}{qr}\right]\right\}\mathrm{d}r=\int_0^{\infty}\left(\frac{2\sin x}{x^2}+\frac{4\cos x}{x^3}-\frac{4\sin x}{x^4}\right)\mathrm{d}x$$

$$=2\int_0^{\infty}(\sin x/x^2)\mathrm{d}x+4\int_0^{\infty}(\cos x/x^3)\mathrm{d}x+(4\sin x/3x^3)\big|_0^{\infty}-(4/3)\int_0^{\infty}(\cos x/x^3)\mathrm{d}x$$

$$=(4\sin x/3x^3)\big|_0^{\infty}+2\int_0^{\infty}(\sin x/x^2)\mathrm{d}x-(4\cos x/3x^2)\big|_0^{\infty}-\int_0^{\infty}(4\sin x/3x^2)\mathrm{d}x$$

$$=(4\sin x/3x^3)\big|_0^{\infty}-(4\cos x/3x^2)\big|_0^{\infty}-⅔(\sin x/x)\big|_0^{\infty}+⅔\int_0^{\infty}(\cos x/x)\mathrm{d}x$$

$$=-\frac{4}{3}\left(\cancel{\frac{1}{x^2}}-\frac{1}{3!}+\cdots\right)_{x\to0}+\frac{4}{3}\left(\cancel{\frac{1}{x^2}}-\frac{1}{2!}+\cdots\right)_{x\to0}+\frac{2}{3}+\frac{2}{3}\int_0^{\infty}(\cos x/x)\mathrm{d}x$$

$$=\frac{2}{9}+\frac{2}{3}\int_0^{\infty}(\cos x/x)\mathrm{d}x. \tag{I5}$$

Recognizing that $(\boldsymbol{\mu}_{\mathrm{n}}\cdot\hat{\boldsymbol{q}})(\widetilde{\boldsymbol{M}}\cdot\hat{\boldsymbol{q}})+(\boldsymbol{\mu}_{\perp}\cdot\widetilde{\boldsymbol{M}}_{\perp})=\boldsymbol{\mu}_{\mathrm{n}}\cdot\widetilde{\boldsymbol{M}}$, Eq.(I4) becomes

$$\int_{-\infty}^{\infty}(\boldsymbol{\mu}_{\mathrm{n}}\cdot\hat{\boldsymbol{r}})[\widetilde{\boldsymbol{M}}(\boldsymbol{q}/\hbar)\cdot\hat{\boldsymbol{r}}]r^{-3}\exp(\mathrm{i}\hbar\boldsymbol{q}\cdot\boldsymbol{r})\,\mathrm{d}\boldsymbol{r}=\frac{4\pi}{9}[(\boldsymbol{\mu}_{\mathrm{n}}\cdot\hat{\boldsymbol{q}})(\widetilde{\boldsymbol{M}}\cdot\hat{\boldsymbol{q}})+4(\boldsymbol{\mu}_{\perp}\cdot\widetilde{\boldsymbol{M}}_{\perp})]$$

$$+\frac{4\pi}{3}(\boldsymbol{\mu}_{\mathrm{n}}\cdot\widetilde{\boldsymbol{M}})\int_0^{\infty}(\cos x/x)\mathrm{d}x. \tag{I6}$$

Finally, upon substituting from Eqs.(I2) and (I6) into Eq.(I1), we arrive at

$$f(\boldsymbol{q})=\frac{m_{\mathrm{n}}}{8\pi^2\mu_0\hbar^2}\Big\{\boldsymbol{\mu}_{\mathrm{n}}\cdot\widetilde{\boldsymbol{M}}\Big[(8\pi/3)-4\pi-4\pi\int_0^{\infty}(\cos x/x)\mathrm{d}x\Big]$$

$$+(4\pi/3)[(\boldsymbol{\mu}_{\mathrm{n}}\cdot\hat{\boldsymbol{q}})(\widetilde{\boldsymbol{M}}\cdot\hat{\boldsymbol{q}})+4(\boldsymbol{\mu}_{\perp}\cdot\widetilde{\boldsymbol{M}}_{\perp})]+4\pi\boldsymbol{\mu}_{\mathrm{n}}\cdot\widetilde{\boldsymbol{M}}\int_0^{\infty}(\cos x/x)\mathrm{d}x\Big\}$$

$$=\left(\frac{m_{\mathrm{n}}}{2\pi\mu_0\hbar^2}\right)\boldsymbol{\mu}_{\mathrm{n}}\cdot\{\widetilde{\boldsymbol{M}}(\boldsymbol{q}/\hbar)-[\widetilde{\boldsymbol{M}}(\boldsymbol{q}/\hbar)\cdot\hat{\boldsymbol{q}}]\hat{\boldsymbol{q}}\}. \tag{I7}$$

This is the same result as given in Ref.[29], Eq.(23), in the case of $\lambda=1$. The coefficient $4\pi\mu_0$ appears here because we have worked in the *SI* system of units with $\boldsymbol{B}=\mu_0\boldsymbol{H}+\boldsymbol{M}$.

## Appendix J

To establish the equivalence of the Rayleigh-Sommerfeld diffraction formulation with the plane-wave expansion method based on Fourier transformation of the initial distribution in the $xy$-plane at $z = 0$, we begin by differentiating both sides of Eq.(37) with respect to $z$, as follows:

$$\iint_{-\infty}^{\infty} \partial_z \left[\frac{\exp(\mathrm{i}k_0\sqrt{x^2+y^2+z^2})}{\sqrt{x^2+y^2+z^2}}\right] e^{\mathrm{i}(k_x x+k_y y)}\mathrm{d}x\mathrm{d}y = -2\pi\exp(\mathrm{i}k_z z). \tag{J1}$$

Here, $k_z = \sqrt{k_0^2 - k_x^2 - k_y^2}$ could be real or imaginary, depending on whether $k_x^2 + k_y^2$ is less than or greater than $k_0^2$. An inverse Fourier transformation now yields

$$(2\pi)^{-2}\iint_{-\infty}^{\infty} e^{\mathrm{i}k_z z} e^{-\mathrm{i}(k_x x+k_y y)}\mathrm{d}k_x\mathrm{d}k_y = -\frac{1}{2\pi}\partial_z\left[\frac{\exp(\mathrm{i}k_0\sqrt{x^2+y^2+z^2})}{\sqrt{x^2+y^2+z^2}}\right]. \tag{J2}$$

The properly defined inverse transform of $\exp(\mathrm{i}k_z z)$ is obtained by reversing the signs of $k_x, k_y$ in Eq.(J2), which is tantamount to flipping the signs of $x$ and $y$. Consequently,

$$\mathcal{F}^{-1}\{e^{\mathrm{i}k_z z}\} = (2\pi)^{-2}\iint_{-\infty}^{\infty} e^{\mathrm{i}k_z z}e^{\mathrm{i}(k_x x+k_y y)}\mathrm{d}k_x\mathrm{d}k_y = -\frac{1}{2\pi}\partial_z\left[\frac{\exp(\mathrm{i}k_0\sqrt{x^2+y^2+z^2})}{\sqrt{x^2+y^2+z^2}}\right]$$

$$= \frac{z}{2\pi(x^2+y^2+z^2)}\left(\frac{1}{\sqrt{x^2+y^2+z^2}} - \mathrm{i}k_0\right)e^{\mathrm{i}k_0\sqrt{x^2+y^2+z^2}}. \tag{J3}$$

The above formula is associated with the name of Hermann Weyl. Suppose now that the scalar field $\psi$ in the $xy$-plane at $z = 0$ is represented by its Fourier spectrum $\tilde{\psi}$, namely,

$$\psi(x, y, z = 0) = (2\pi)^{-2}\iint_{-\infty}^{\infty}\tilde{\psi}(k_x, k_y)e^{\mathrm{i}(k_x x+k_y y)}\mathrm{d}k_x\mathrm{d}k_y. \tag{J4}$$

Propagating this field by a distance $z_0$ along the $z$-axis requires the multiplication of $\tilde{\psi}$ by the propagation factor $\exp(\mathrm{i}k_z z_0)$. Recalling that the Fourier transform of the convolution of two functions is the product of the individual transforms of those functions,† we will have

$$\psi(x_0, y_0, z_0) = \mathcal{F}^{-1}\{\tilde{\psi}(k_x, k_y)e^{\mathrm{i}k_z z_0}\} = \psi(x, y, z = 0) * \mathcal{F}^{-1}\{e^{\mathrm{i}k_z z_0}\}$$

$$= -\frac{1}{2\pi}\iint_{-\infty}^{\infty}\psi(x, y, z = 0)\partial_z\left[\frac{\exp\{\mathrm{i}k_0[(x_0-x)^2+(y_0-y)^2+z_0^2]^{1/2}\}}{[(x_0-x)^2+(y_0-y)^2+z_0^2]^{1/2}}\right]\mathrm{d}x\mathrm{d}y. \tag{J5}$$

This is the same as the Rayleigh-Sommerfeld formula given in Eq.(25), with $S_1$ being the $xy$-plane at $z = 0$, and $(\boldsymbol{r} - \boldsymbol{r}_0)\cdot\hat{\boldsymbol{n}}$ being equal to $-z_0$.

In similar fashion, one can relate Eqs.(32) and (33) to their Fourier method counterparts. For instance, recognizing the integral in Eq.(32) as a convolution integral, we use Eq.(37) to write

---

† Convolution theorem: $\mathcal{F}^{-1}\{\tilde{F}(k_x, k_y)\tilde{G}(k_x, k_y)\} = (2\pi)^{-2}\iint_{-\infty}^{\infty}\tilde{F}(k_x, k_y)\tilde{G}(k_x, k_y)e^{\mathrm{i}(k_x x+k_y y)}\mathrm{d}k_x\mathrm{d}k_y$

$$= (2\pi)^{-2}\iint_{-\infty}^{\infty}\left[\iint_{-\infty}^{\infty}F(x', y')e^{-\mathrm{i}(k_x x'+k_y y')}\mathrm{d}x'\mathrm{d}y'\right]\tilde{G}(k_x, k_y)e^{\mathrm{i}(k_x x+k_y y)}\mathrm{d}k_x\mathrm{d}k_y$$

$$= \iint_{-\infty}^{\infty}F(x', y')\left[(2\pi)^{-2}\iint_{-\infty}^{\infty}\tilde{G}(k_x, k_y)e^{\mathrm{i}[k_x(x-x')+k_y(y-y')]}\mathrm{d}k_x\mathrm{d}k_y\right]\mathrm{d}x'\mathrm{d}y'$$

$$= \iint_{-\infty}^{\infty}F(x', y')G(x - x', y - y')\mathrm{d}x'\mathrm{d}y' = F(x, y) * G(x, y).$$

$$
\begin{aligned}
\boldsymbol{E}(\boldsymbol{r}_0) &= (2\pi)^{-1}\boldsymbol{\nabla}_0 \times \iint_{-\infty}^{\infty} [\hat{\boldsymbol{z}} \times \boldsymbol{E}(x, y, z = 0)] \frac{\exp\{\mathrm{i}k_0[(x_0-x)^2+(y_0-y)^2+z_0^2]^{½}\}}{[(x_0-x)^2+(y_0-y)^2+z_0^2]^{½}} \mathrm{d}x\mathrm{d}y \\
&= (2\pi)^{-1}\boldsymbol{\nabla}_0 \times (2\pi)^{-2} \iint_{-\infty}^{\infty} [\hat{\boldsymbol{z}} \times \widetilde{\boldsymbol{E}}(k_x, k_y)](\mathrm{i}2\pi e^{\mathrm{i}k_z z_0}/k_z) e^{\mathrm{i}(k_x x_0 + k_y y_0)} \mathrm{d}k_x \mathrm{d}k_y \\
&= (2\pi)^{-2} \iint_{-\infty}^{\infty} \mathrm{i}^2 (k_x\hat{\boldsymbol{x}} + k_y\hat{\boldsymbol{y}} + k_z\hat{\boldsymbol{z}}) \times (\hat{\boldsymbol{z}} \times \widetilde{\boldsymbol{E}}) k_z^{-1} e^{\mathrm{i}(k_x x_0 + k_y y_0 + k_z z_0)} \mathrm{d}k_x \mathrm{d}k_y \\
&= (2\pi)^{-2} \iint_{-\infty}^{\infty} [(\boldsymbol{k} \cdot \hat{\boldsymbol{z}})\widetilde{\boldsymbol{E}} - \cancelto{0}{(\boldsymbol{k} \cdot \widetilde{\boldsymbol{E}})}\hat{\boldsymbol{z}}] k_z^{-1} e^{\mathrm{i}\boldsymbol{k} \cdot \boldsymbol{r}_0} \mathrm{d}k_x \mathrm{d}k_y \\
&= (2\pi)^{-2} \iint_{-\infty}^{\infty} \widetilde{\boldsymbol{E}}(k_x, k_y) e^{\mathrm{i}\boldsymbol{k} \cdot \boldsymbol{r}_0} \mathrm{d}k_x \mathrm{d}k_y. \qquad \text{(J6)}
\end{aligned}
$$

This, of course, is the expression of the Fourier method of diffraction calculation, in which $E_x(x, y, z = 0)$ and $E_y(x, y, z = 0)$ are Fourier transformed within the $xy$-plane of the initial distribution to yield $\tilde{E}_x(k_x, k_y)$ and $\tilde{E}_y(k_x, k_y)$. Using $\boldsymbol{k} = k_x\hat{\boldsymbol{x}} + k_y\hat{\boldsymbol{y}} + \sqrt{k_0^2 - k_x^2 - k_y^2}\,\hat{\boldsymbol{z}}$ and invoking Maxwell's first equation $k_x\tilde{E}_x + k_y\tilde{E}_y + k_z\tilde{E}_z = 0$ to determine $\tilde{E}_z$, one arrives at the full expression for $\widetilde{\boldsymbol{E}}(k_x, k_y)$, from which $\boldsymbol{E}(\boldsymbol{r}_0)$ is computed via Eq.(J6).

---

## Appendix K

The rigorous version of Babinet's principle of complementary screens pertains to a system such as that of Fig.K1(a) and its complement depicted in Fig.K1(b). The screen in Fig.K1(a) is a perfectly conducting thin sheet located in the $xy$-plane at $z = 0$. The incident beam passes through the aperture and is observed at $\boldsymbol{r}_0$. The observed $E$-field is computed from Eq.(32), whose domain of integration is now confined to the area of the aperture — since the tangential $E$-field on the metallic screen vanishes. We will have

$$\boldsymbol{E}(\boldsymbol{r}_0) = (2\pi)^{-1}\boldsymbol{\nabla}_0 \times \int_{\text{aperture}} [\hat{\boldsymbol{z}} \times \boldsymbol{E}(\boldsymbol{r})] G(\boldsymbol{r}, \boldsymbol{r}_0) \mathrm{d}s. \qquad \text{(K1)}$$

In the complementary setup of Fig.K1(b), the screen is removed and a thin, perfectly conducting obstacle of the same shape as the aperture of Fig.K1(a) is placed at the aperture's location. Also, the incident $\boldsymbol{E}$ and $\boldsymbol{B}$ fields are changed to $\boldsymbol{E}_{\text{inc}}^{(\text{c})} = c\boldsymbol{B}_{\text{inc}}$ and $\boldsymbol{B}_{\text{inc}}^{(\text{c})} = -\boldsymbol{E}_{\text{inc}}/c$.

We construct the new fields $\boldsymbol{\mathcal{E}}(\boldsymbol{r}) = \boldsymbol{E}(\boldsymbol{r}) - c\boldsymbol{B}^{(\text{c})}(\boldsymbol{r})$ and $\boldsymbol{\mathcal{B}}(\boldsymbol{r}) = \boldsymbol{B}(\boldsymbol{r}) + c^{-1}\boldsymbol{E}^{(\text{c})}(\boldsymbol{r})$ in the half-space $z \geq 0$. It is not difficult to verify that $\boldsymbol{\mathcal{E}}$ and $\boldsymbol{\mathcal{B}}$ satisfy all four of Maxwell's equations in this empty half-space. Now, in the system of Fig.K1(a), the tangential $\boldsymbol{E}$ is zero on the metallic surface, while in the aperture, the tangential $\boldsymbol{B}$ equals $\boldsymbol{B}_{\text{inc}}^{\parallel}(x, y, 0)$.‡ Similarly, in the system of Fig.K1(b), the tangential $\boldsymbol{E}^{(\text{c})}$ on the metallic surface vanishes, while in the open areas, the tangential $\boldsymbol{B}^{(\text{c})}$ equals $\boldsymbol{B}_{\text{inc}}^{\parallel(\text{c})}(x, y, 0)$, which is the same as $-\boldsymbol{E}_{\text{inc}}^{\parallel}(x, y, 0)/c$. It is thus seen that, on the metallic surface of Fig.K1(a), $\boldsymbol{\mathcal{E}}^{\parallel}(x, y, 0) = \boldsymbol{E}_{\text{inc}}^{\parallel}(x, y, 0)$, while in the corresponding aperture, $\boldsymbol{\mathcal{B}}^{\parallel}(x,y,0) = \boldsymbol{B}_{\text{inc}}^{\parallel}(x,y,0)$. With these boundary conditions, the solution of Maxwell's equations for $\boldsymbol{\mathcal{E}}$ and $\boldsymbol{\mathcal{B}}$ in the half-space $z \geq 0$ will be $\boldsymbol{\mathcal{E}}(\boldsymbol{r}_0) = \boldsymbol{E}_{\text{inc}}(\boldsymbol{r}_0)$ and $\boldsymbol{\mathcal{B}}(\boldsymbol{r}_0) = \boldsymbol{B}_{\text{inc}}(\boldsymbol{r}_0)$.[§] We thus have

$$\boldsymbol{E}(\boldsymbol{r}_0) - c\boldsymbol{B}^{(\text{c})}(\boldsymbol{r}_0) = \boldsymbol{E}_{\text{inc}}(\boldsymbol{r}_0). \qquad \text{(K2)}$$

$$\boldsymbol{B}(\boldsymbol{r}_0) + c^{-1}\boldsymbol{E}^{(\text{c})}(\boldsymbol{r}_0) = \boldsymbol{B}_{\text{inc}}(\boldsymbol{r}_0). \qquad \text{(K3)}$$

‡The reason that the tangential $\boldsymbol{B}$ in the aperture equals $\boldsymbol{B}_{\text{inc}}^{\parallel}(x, y, 0)$ is that, within the aperture area, the tangential component $\boldsymbol{B}_s^{\parallel}$ of the scattered $B$-field vanishes. To see this, observe that the scattered $\boldsymbol{E}$ and $\boldsymbol{B}$ fields are produced by the induced surface charge- and current-densities in the perfectly conducting regions of the screen $S_1$. Since the surface current-density does not have a $z$-component, the only components of the scattered vector potential $\boldsymbol{A}_s(\boldsymbol{r})$ will be $A_{sx}$ and $A_{sy}$. The symmetry of the system under consideration then ensures that, within the aperture area, $\partial_z A_{sx} = 0$ and $\partial_z A_{sy} = 0$, which results in $B_{sx} = B_{sy} = 0$. (A similar argument reveals that $E_{sz}$ within the aperture vanishes as well, since $\partial_t A_{sz} = 0$ and the symmetry of the scattered scalar potential $\psi_s(\boldsymbol{r})$ ensures that $\partial_z \psi_s = 0$.)

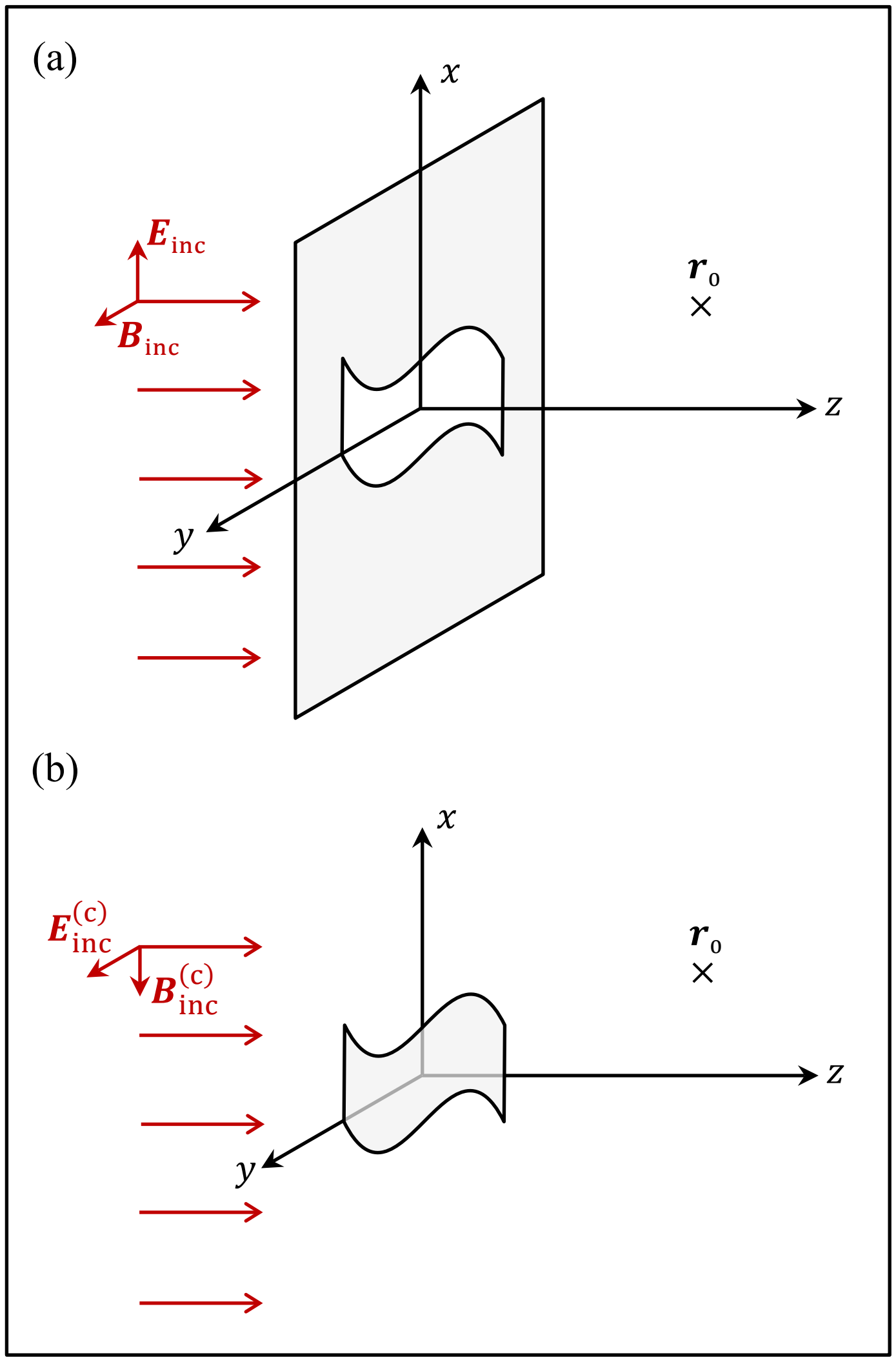

**Fig. K1**. (a) Sitting in the $xy$-plane at $z = 0$ is a thin, perfectly conducting screen containing one or more apertures. The incident beam, having electric field $\boldsymbol{E}_{\text{inc}}(\boldsymbol{r})$ and magnetic field $\boldsymbol{B}_{\text{inc}}(\boldsymbol{r})$, arrives from the left-hand side. The observation point $\boldsymbol{r}_0$ is on the right-hand side of the screen. (b) In the complementary setup, the screen is removed and a thin, perfectly conducting obstacle is placed at the location of the aperture. In addition, the incident $\boldsymbol{E}$ and $\boldsymbol{B}$ fields are changed to $\boldsymbol{E}_{\text{inc}}^{(\text{c})} = c\boldsymbol{B}_{\text{inc}}$ and $\boldsymbol{B}_{\text{inc}}^{(\text{c})} = -\boldsymbol{E}_{\text{inc}}/c$.

[§] The proof of this statement is rather trivial. The incident wave is a superposition of EM plane-waves. For any and all such plane-waves, knowledge of the tangential component of either $\boldsymbol{E}$ or $\boldsymbol{B}$ in the $xy$-plane at $z = 0$ suffices to fully specify the plane-wave in its entirety. It is immaterial whether one specifies $\boldsymbol{E}^{\parallel}$ or $\boldsymbol{B}^{\parallel}$ over the entire plane, or use a patchwork wherein $\boldsymbol{E}^{\parallel}$ is specified in some areas and $\boldsymbol{B}^{\parallel}$ in the remaining areas of the $xy$-plane. Given that $\boldsymbol{\mathcal{E}}^{\parallel}(x,y,0)$ equals $\boldsymbol{E}_{\text{inc}}^{\parallel}$ on the metallic surface in Fig.K1(a), while $\boldsymbol{\mathcal{B}}^{\parallel}(x,y,0)$ equals $\boldsymbol{B}_{\text{inc}}^{\parallel}$ within the aperture, it is inevitable that $(\boldsymbol{\mathcal{E}},\boldsymbol{\mathcal{B}})$ will coincide with $(\boldsymbol{E}_{\text{inc}},\boldsymbol{B}_{\text{inc}})$ everywhere in the $z \geq 0$ half-space.

Equations (K2) and (K3) embody the rigorous version of Babinet's principle of complementary screens for EM waves. Manipulating these equations can yield useful formulas for practical applications. For instance, replacing $\boldsymbol{E}_{\text{inc}}$ in Eq.(K2) with its equal $-c\boldsymbol{B}_{\text{inc}}^{(\text{c})}$ results in

$$\boldsymbol{E}(\boldsymbol{r}_0) = c\boldsymbol{B}^{(\text{c})}(\boldsymbol{r}_0) - c\boldsymbol{B}_{\text{inc}}^{(\text{c})}(\boldsymbol{r}_0) = c\boldsymbol{B}_s^{(\text{c})}(\boldsymbol{r}_0). \tag{K4}$$

Here, $\boldsymbol{B}_s^{(\text{c})}$ is the scattered $B$-field in the complementary system of Fig.K1(b). Considering that the tangential component of the scattered $B$-field vanishes in the open areas of the $xy$-plane at $z = 0$, the scattered $B$-field at the observation point can be computed from Eq.(33), as follows:

$$\boldsymbol{B}_s^{(\text{c})}(\boldsymbol{r}_0) = (2\pi)^{-1}\boldsymbol{\nabla}_0 \times \int_{\text{obstacle}} [\hat{\boldsymbol{z}} \times \boldsymbol{B}_s^{(\text{c})}(\boldsymbol{r})] G(\boldsymbol{r}, \boldsymbol{r}_0) \mathrm{d}s. \tag{K5}$$

Thus, Eqs.(K4) and (K5) provide an alternative to Eq.(K1) for computing the diffracted $E$-field that passes through the aperture in Fig.K1(a).

---

## Appendix L

To compute the scattering cross-section of a small dielectric sphere, let the particle of small radius $R$ and isotropic dielectric constant $\varepsilon_0\varepsilon(\omega) = \varepsilon_0[1 + \chi_{\text{e}}(\omega)]$ be centered at the origin of the coordinate system and excited by the incident plane-wave $\boldsymbol{E}(\boldsymbol{r}, t) = \boldsymbol{E}_{\text{inc}} \exp[\mathrm{i}k_0(\boldsymbol{\sigma}_{\text{inc}} \cdot \boldsymbol{r} - ct)]$. In the static approximation, the induced polarization $\boldsymbol{P}_0 e^{-\mathrm{i}\omega t}$ of the dipole produces the uniform self-field $\boldsymbol{E}_{\text{self}}(t) = -(\boldsymbol{P}_0/3\varepsilon_0)e^{-\mathrm{i}\omega t}$, which, in combination with the incident $E$-field, yields

$$\boldsymbol{P}_0 = \varepsilon_0\chi_{\text{e}}(\omega)\left(\boldsymbol{E}_{\text{inc}} - \frac{\boldsymbol{P}_0}{3\varepsilon_0}\right) \quad \rightarrow \quad \boldsymbol{P}_0 = \varepsilon_0\left[\frac{3\chi_{\text{e}}(\omega)}{\chi_{\text{e}}(\omega) + 3}\right]\boldsymbol{E}_{\text{inc}} = 3\varepsilon_0\left[\frac{\varepsilon(\omega) - 1}{\varepsilon(\omega) + 2}\right]\boldsymbol{E}_{\text{inc}}. \tag{L1}$$

The induced dipole moment of the particle is thus $\boldsymbol{p}(t) = \boldsymbol{p}_0 e^{-\mathrm{i}\omega t}$, where

$$\boldsymbol{p}_0 = (4\pi/3)R^3\boldsymbol{P}_0 = 4\pi\varepsilon_0 R^3\left[\frac{\varepsilon(\omega) - 1}{\varepsilon(\omega) + 2}\right]\boldsymbol{E}_{\text{inc}}. \tag{L2}$$

If the dipole happens to be aligned with the $z$-axis, then at the far away observation point $\boldsymbol{r}_0 = r_0\boldsymbol{\sigma}_0$ the radiated $E$-field in the $(r, \theta, \varphi)$ spherical coordinate system will be[11]

$$\boldsymbol{E}(\boldsymbol{r}_0, t) = -\frac{p_0}{4\pi\varepsilon_0 r_0}(\omega/c)^2 \sin\theta\, e^{-\mathrm{i}\omega(t - r_0/c)}\,\hat{\boldsymbol{\theta}}\,. \tag{L3}$$

In vector notation, we can replace $p_0 \sin\theta\,\hat{\boldsymbol{\theta}}$ with the triple cross product $(\boldsymbol{p}_0 \times \boldsymbol{\sigma}_0) \times \boldsymbol{\sigma}_0$, which no longer restricts $\boldsymbol{p}_0$ to the direction of the $z$-axis. The far-field $\boldsymbol{E}$ and $\boldsymbol{H}$ are now written

$$\boldsymbol{E}(\boldsymbol{r}_0, t) = -\frac{(\omega/c)^2}{4\pi\varepsilon_0 r_0} e^{-\mathrm{i}\omega(t - r_0/c)}\,(\boldsymbol{p}_0 \times \boldsymbol{\sigma}_0) \times \boldsymbol{\sigma}_0. \tag{L4}$$

$$\boldsymbol{H}(\boldsymbol{r}_0, t) = -\frac{(\omega/c)^2}{4\pi\varepsilon_0 Z_0 r_0} e^{-\mathrm{i}\omega(t - r_0/c)}\,\boldsymbol{p}_0 \times \boldsymbol{\sigma}_0. \tag{L5}$$

The time-averaged Poynting vector at the observation point is found to be

$$\langle \boldsymbol{S}(\boldsymbol{r}_0, t)\rangle = ½\mathrm{Re}(\boldsymbol{E} \times \boldsymbol{H}^*) = \frac{\mu_0\omega^4}{32\pi^2 c r_0^2}[(\boldsymbol{p}_0^* \times \boldsymbol{\sigma}_0) \cdot (\boldsymbol{p}_0 \times \boldsymbol{\sigma}_0)]\boldsymbol{\sigma}_0 \leftarrow \boxed{\boldsymbol{a} \times (\boldsymbol{b} \times \boldsymbol{c}) = (\boldsymbol{a} \cdot \boldsymbol{c})\boldsymbol{b} - (\boldsymbol{a} \cdot \boldsymbol{b})\boldsymbol{c}}$$

$$= \frac{\mu_0\omega^4}{32\pi^2 c r_0^2}[\boldsymbol{p}_0 \cdot \boldsymbol{p}_0^* - (\boldsymbol{p}_0 \cdot \boldsymbol{\sigma}_0)(\boldsymbol{p}_0^* \cdot \boldsymbol{\sigma}_0)]\boldsymbol{\sigma}_0 \leftarrow \boxed{(\boldsymbol{a} \times \boldsymbol{b}) \cdot (\boldsymbol{c} \times \boldsymbol{d}) = (\boldsymbol{a} \cdot \boldsymbol{c})(\boldsymbol{b} \cdot \boldsymbol{d}) - (\boldsymbol{a} \cdot \boldsymbol{d})(\boldsymbol{b} \cdot \boldsymbol{c})}$$

$$= \frac{(\omega/c)^4 R^6}{2Z_0 r_0^2}\left|\frac{\varepsilon(\omega) - 1}{\varepsilon(\omega) + 2}\right|^2 [\boldsymbol{E}_{\text{inc}} \cdot \boldsymbol{E}_{\text{inc}}^* - (\boldsymbol{E}_{\text{inc}} \cdot \boldsymbol{\sigma}_0)(\boldsymbol{E}_{\text{inc}}^* \cdot \boldsymbol{\sigma}_0)]\boldsymbol{\sigma}_0. \tag{L6}$$

Since the infinitesimal surface area perpendicular to $\boldsymbol{r}_0$ is $r_0^2 \mathrm{d}\Omega = r_0^2 \sin\theta\, \mathrm{d}\theta \mathrm{d}\varphi$, the EM energy per unit time, $\mathrm{d}\mathcal{E}$, crossing this surface will satisfy the following equation:

$$\frac{\mathrm{d}\mathcal{E}}{\mathrm{d}\Omega} = (\omega/c)^4 R^6 \left|\frac{\varepsilon(\omega)-1}{\varepsilon(\omega)+2}\right|^2 \left[\frac{\boldsymbol{E}_{\text{inc}} \cdot \boldsymbol{E}_{\text{inc}}^* - (\boldsymbol{E}_{\text{inc}} \cdot \boldsymbol{\sigma}_0)(\boldsymbol{E}_{\text{inc}}^* \cdot \boldsymbol{\sigma}_0)}{2Z_0}\right]. \quad \text{(L7)}$$

In general, the time-averaged incident optical energy per unit time per unit cross-sectional area of the incident plane-wave is

$$\mathcal{E}_{\text{inc}} = \tfrac{1}{2}\text{Re}(\boldsymbol{E}_{\text{inc}} \times \boldsymbol{H}_{\text{inc}}^*) = \tfrac{1}{2}\text{Re}[\boldsymbol{E}_{\text{inc}} \times (\boldsymbol{\sigma}_{\text{inc}} \times \boldsymbol{E}_{\text{inc}}^*)/Z_0] = (\boldsymbol{E}_{\text{inc}} \cdot \boldsymbol{E}_{\text{inc}}^*/2Z_0)\boldsymbol{\sigma}_{\text{inc}}. \quad \text{(L8)}$$

Consequently, the differential scattering cross section of the particle is

$$\frac{\mathrm{d}\mathcal{S}}{\mathrm{d}\Omega} = (\omega/c)^4 R^6 \left|\frac{\varepsilon(\omega)-1}{\varepsilon(\omega)+2}\right|^2 \left(1 - \frac{|\boldsymbol{E}_{\text{inc}} \cdot \boldsymbol{\sigma}_0|^2}{|\boldsymbol{E}_{\text{inc}}|^2}\right). \quad \text{(L9)}$$

To find the total scattering cross section, we must integrate the above expression over the surface of a unit sphere centered at the origin of coordinates. The first term integrates to $\int \mathrm{d}\Omega = \int_{\theta=0}^{\pi}\int_{\varphi=0}^{2\pi} \sin\theta\, \mathrm{d}\theta \mathrm{d}\varphi = 4\pi$. As for the second term, writing $\boldsymbol{E}_{\text{inc}} = E_x\hat{\boldsymbol{x}} + E_y\hat{\boldsymbol{y}} + E_z\hat{\boldsymbol{z}}$ and $\boldsymbol{\sigma}_0 = \sin\theta\cos\varphi\,\hat{\boldsymbol{x}} + \sin\theta\sin\varphi\,\hat{\boldsymbol{y}} + \cos\theta\,\hat{\boldsymbol{z}}$, we will have

$$\begin{aligned}
|\boldsymbol{E}_{\text{inc}} \cdot \boldsymbol{\sigma}_0|^2 &= (E_x \sin\theta\cos\varphi + E_y \sin\theta\sin\varphi + E_z\cos\theta) \cdot (E_x^* \sin\theta\cos\varphi + E_y^* \sin\theta\sin\varphi + E_z^* \cos\theta) \\
&= |E_x|^2 \sin^2\theta\cos^2\varphi + |E_y|^2 \sin^2\theta\sin^2\varphi + |E_z|^2\cos^2\theta + \tfrac{1}{2}(E_xE_y^* + E_x^*E_y)\sin^2\theta\sin(2\varphi) \\
&\quad + \tfrac{1}{2}(E_xE_z^* + E_zE_x^*)\sin(2\theta)\cos\varphi + \tfrac{1}{2}(E_yE_z^* + E_zE_y^*)\sin(2\theta)\cos\varphi. \quad \text{(L10)}
\end{aligned}$$

$$\begin{aligned}
\int_{\substack{\text{sphere}\\\text{surface}}} |\boldsymbol{E}_{\text{inc}} \cdot \boldsymbol{\sigma}_0|^2 \mathrm{d}\Omega &= \int_{\theta=0}^{\pi}\int_{\varphi=0}^{2\pi} |\boldsymbol{E}_{\text{inc}} \cdot \boldsymbol{\sigma}_0|^2 \sin\theta\, \mathrm{d}\theta\mathrm{d}\varphi \\
&= \int_{\theta=0}^{\pi} [\pi(|E_x|^2 + |E_y|^2)\sin^3\theta + 2\pi|E_z|^2 \sin\theta\cos^2\theta]\mathrm{d}\theta \\
&= (4\pi/3)(|E_x|^2 + |E_y|^2 + |E_z|^2) = (4\pi/3)|\boldsymbol{E}_{\text{inc}}|^2. \quad \text{(L11)}
\end{aligned}$$

The total scattering cross-section of the isotropic dielectric sphere is thus seen to be independent of the polarization state of the incident beam and given by

$$\mathcal{S} = \left(\frac{8\pi}{3}\right)\left|\frac{\varepsilon(\omega)-1}{\varepsilon(\omega)+2}\right|^2 \left(\frac{\omega}{c}\right)^4 R^6. \quad \text{(L12)}$$

For linearly polarized light, if the incident polarization happens to be perpendicular to the plane of incidence (i.e., the plane defined by $\boldsymbol{\sigma}_{\text{inc}}$ and $\boldsymbol{\sigma}_0$), then, in the differential cross-section formula of Eq.(L9), we will have $\boldsymbol{E}_{\text{inc}} \cdot \boldsymbol{\sigma}_0 = 0$. However, for an incident polarization in the plane of incidence, $\boldsymbol{E}_{\text{inc}} \cdot \boldsymbol{\sigma}_0 = E_{\text{inc}} \sin\vartheta$, where $\vartheta$ is the deviation angle of $\boldsymbol{\sigma}_0$ from $\boldsymbol{\sigma}_{\text{inc}}$. In the case of natural light, which is unpolarized, the incident beam is an equal mixture of linear polarizations along the $\parallel$ and $\perp$ directions. The degree of polarization of the scattered light is then given by

$$\Pi(\vartheta) = \frac{(\mathrm{d}\mathcal{S}_\perp/\mathrm{d}\Omega) - (\mathrm{d}\mathcal{S}_\parallel/\mathrm{d}\Omega)}{(\mathrm{d}\mathcal{S}_\perp/\mathrm{d}\Omega) + (\mathrm{d}\mathcal{S}_\parallel/\mathrm{d}\Omega)} = \frac{1-(1-\sin^2\vartheta)}{1+(1-\sin^2\vartheta)} = \frac{\sin^2\vartheta}{1+\cos^2\vartheta}. \quad \text{(L13)}$$

---

## Appendix M

To compute the scattering cross-section of a small perfectly conducting sphere, we begin by noting that, inside the perfect conductor, both $\boldsymbol{E}$ and $\boldsymbol{B}$ fields must vanish. Assuming the particle radius $R$ is much smaller than the incident wavelength $\lambda = 2\pi c/\omega$, and that the static approximation is reasonably accurate to describe the internal fields, the spherical particle acquires an electric dipole moment $\boldsymbol{p}_0 e^{-\mathrm{i}\omega t}$ along the direction of the incident $E$-field, and a magnetic dipole moment $\boldsymbol{m}_0 e^{-\mathrm{i}\omega t}$ along the direction of the incident $B$-field. Denoting the polarization and magnetization of the particle by $\boldsymbol{P}_0$ and $\boldsymbol{M}_0$, respectively, the vanishing of the internal $E$-field demands that $\boldsymbol{E}_{\text{inc}} = \boldsymbol{P}_0/3\varepsilon_0$, while the vanishing of the internal $B$-field requires that $\boldsymbol{B}_{\text{inc}} = -2\boldsymbol{M}_0/3$.** Consequently,

$$\boldsymbol{p}_0 = 4\pi\varepsilon_0 R^3 \boldsymbol{E}_{\text{inc}}. \tag{M1}$$

$$\boldsymbol{m}_0 = -2\pi\mu_0 R^3 \boldsymbol{H}_{\text{inc}}. \tag{M2}$$

If $\boldsymbol{m}_0$ happens to be aligned with the $z$-axis, then at the far away observation point $\boldsymbol{r}_0 = r_0\boldsymbol{\sigma}_0$ the radiated $E$-field in the $(r, \theta, \varphi)$ spherical coordinate system will be (Ref.[11], Problem 4.17)

$$\boldsymbol{E}(\boldsymbol{r}_0, t) = \frac{m_0\omega^2}{4\pi c r_0} \sin\theta\, e^{-\mathrm{i}\omega(t-r_0/c)}\, \hat{\boldsymbol{\varphi}}\,. \tag{M3}$$

In vector notation, we can replace $m_0 \sin\theta\, \hat{\boldsymbol{\varphi}}$ with the cross-product $\boldsymbol{m}_0 \times \boldsymbol{\sigma}_0$, which no longer restricts $\boldsymbol{m}_0$ to being aligned with the $z$-axis. The far-fields $\boldsymbol{E}$ and $\boldsymbol{H}$ will now be written

$$\boldsymbol{E}(\boldsymbol{r}_0, t) = \frac{\omega^2}{4\pi c r_0} e^{-\mathrm{i}\omega(t-r_0/c)}\, \boldsymbol{m}_0 \times \boldsymbol{\sigma}_0. \tag{M4}$$

$$\boldsymbol{H}(\boldsymbol{r}_0, t) = \frac{\omega^2}{4\pi c Z_0 r_0} e^{-\mathrm{i}\omega(t-r_0/c)}\, \boldsymbol{\sigma}_0 \times (\boldsymbol{m}_0 \times \boldsymbol{\sigma}_0). \tag{M5}$$

The total scattered fields are now found by coherent superposition of the scattered fields of the induced electric and magnetic dipoles given by Eqs.(L4), (L5), (M4), and (M5), as follows:

$$\boldsymbol{E}_{\text{total}}(\boldsymbol{r}_0, t) = \frac{R^3(\omega/c)^2}{r_0}\, \boldsymbol{\sigma}_0 \times [\boldsymbol{E}_{\text{inc}} \times (\boldsymbol{\sigma}_0 - \tfrac{1}{2}\boldsymbol{\sigma}_{\text{inc}})] e^{-\mathrm{i}\omega(t-r_0/c)}. \tag{M6}$$

$$\boldsymbol{H}_{\text{total}}(\boldsymbol{r}_0, t) = \frac{R^3(\omega/c)^2}{Z_0 r_0}\, \boldsymbol{\sigma}_0 \times [\boldsymbol{E}_{\text{inc}} + \tfrac{1}{2}(\boldsymbol{E}_{\text{inc}} \times \boldsymbol{\sigma}_{\text{inc}}) \times \boldsymbol{\sigma}_0] e^{-\mathrm{i}\omega(t-r_0/c)}. \tag{M7}$$

$$\langle \boldsymbol{S}_{\text{total}}(\boldsymbol{r}_0, t)\rangle = \frac{R^6(\omega/c)^4}{2Z_0 r_0^2} \operatorname{Re}\{\boldsymbol{\sigma}_0 \times [\boldsymbol{E}_{\text{inc}} \times (\boldsymbol{\sigma}_0 - \tfrac{1}{2}\boldsymbol{\sigma}_{\text{inc}})]\} \times \{\boldsymbol{\sigma}_0 \times [\boldsymbol{E}^*_{\text{inc}} + \tfrac{1}{2}(\boldsymbol{E}^*_{\text{inc}} \times \boldsymbol{\sigma}_{\text{inc}}) \times \boldsymbol{\sigma}_0]\}$$

$$= \frac{R^6(\omega/c)^4}{2Z_0 r_0^2} \operatorname{Re}\Big\{\{\boldsymbol{\sigma}_0 \times [\boldsymbol{E}_{\text{inc}} \times (\boldsymbol{\sigma}_0 - \tfrac{1}{2}\boldsymbol{\sigma}_{\text{inc}})]\} \cdot [\boldsymbol{E}^*_{\text{inc}} + \tfrac{1}{2}(\boldsymbol{E}^*_{\text{inc}} \times \boldsymbol{\sigma}_{\text{inc}}) \times \boldsymbol{\sigma}_0]\Big\}\boldsymbol{\sigma}_0$$

---

** In the *SI* system of units, where $\boldsymbol{B} = \mu_0\boldsymbol{H} + \boldsymbol{M}$, the magnitude of a magnetic dipole moment $\boldsymbol{m}$ is $\mu_0 \times$ (circulating current around a loop) $\times$ (area of the loop). The magnetization $\boldsymbol{M}$, being the magnetic dipole moment per unit volume, then has the units of $\mu_0$ (i.e., henry/meter) $\times$ ampere/meter. This is the same as the units of $\mu_0\boldsymbol{H}$, namely, henry $\times$ ampere/meter$^2$, which is known as weber/meter$^2$ and often used as the units of $\boldsymbol{B}$.

Our perfectly conducting spherical particle develops surface charges and surface currents in response to the incident plane-wave. The induced surface currents produce the magnetic dipole moment $\boldsymbol{m}_0 e^{-\mathrm{i}\omega t}$, which creates an internal $B$-field in the same way that the circulating current of a solenoid gives rise to a magnetic $B$-field inside the solenoid. If the spherical particle were made of a true magnetic material of uniform magnetization, its internal $\boldsymbol{H}$ and $\boldsymbol{B}$ fields would have been $-\boldsymbol{M}_0/3\mu_0$ and $2\boldsymbol{M}_0/3$, respectively. However, here we only have a spherical solenoid with an internal $B$-field of $2\boldsymbol{M}_0/3$; the corresponding $H$-field is simply $2\boldsymbol{M}_0/3\mu_0$.

$$= \frac{R^6(\omega/c)^4}{2Z_0 r_0^2} \mathrm{Re}\,\{[(1 - ½\boldsymbol{\sigma}_0 \cdot \boldsymbol{\sigma}_{\rm inc})\boldsymbol{E}_{\rm inc} - (\boldsymbol{\sigma}_0 \cdot \boldsymbol{E}_{\rm inc})(\boldsymbol{\sigma}_0 - ½\boldsymbol{\sigma}_{\rm inc})]$$

$$\cdot [(1 - ½\boldsymbol{\sigma}_0 \cdot \boldsymbol{\sigma}_{\rm inc})\boldsymbol{E}_{\rm inc}^* + ½(\boldsymbol{\sigma}_0 \cdot \boldsymbol{E}_{\rm inc}^*)\boldsymbol{\sigma}_{\rm inc}]\}\boldsymbol{\sigma}_0$$

$$= \frac{R^6(\omega/c)^4}{2Z_0 r_0^2} [(1 - ½\boldsymbol{\sigma}_0 \cdot \boldsymbol{\sigma}_{\rm inc})^2 \boldsymbol{E}_{\rm inc} \cdot \boldsymbol{E}_{\rm inc}^* - ¾(\boldsymbol{\sigma}_0 \cdot \boldsymbol{E}_{\rm inc})(\boldsymbol{\sigma}_0 \cdot \boldsymbol{E}_{\rm inc}^*)]\boldsymbol{\sigma}_0. \qquad \text{(M8)}$$

The differential scattering cross section is thus found to be

$$\frac{\mathrm{d}\mathcal{S}}{\mathrm{d}\Omega} = R^6(\omega/c)^4 \left[(1 - ½\boldsymbol{\sigma}_0 \cdot \boldsymbol{\sigma}_{\rm inc})^2 - \frac{3|\boldsymbol{E}_{\rm inc} \cdot \boldsymbol{\sigma}_0|^2}{4|\boldsymbol{E}_{\rm inc}|^2}\right]. \qquad \text{(M9)}$$

For linearly polarized incident beams, differential cross sections for polarization in the plane of incidence (∥) and perpendicular to the plane of incidence (⊥) are found from Eq.(M9) to be

$$\frac{\mathrm{d}\mathcal{S}_\parallel}{\mathrm{d}\Omega} = R^6(\omega/c)^4(\cos\vartheta - ½)^2. \qquad \text{(M10)}$$

$$\frac{\mathrm{d}\mathcal{S}_\perp}{\mathrm{d}\Omega} = R^6(\omega/c)^4(1 - ½\cos\vartheta)^2. \qquad \text{(M11)}$$

Here, as before, $\vartheta$ is the deviation angle of $\boldsymbol{\sigma}_0$ from $\boldsymbol{\sigma}_{\rm inc}$.